\documentclass[a4paper]{llncs}

\input{Qcircuit}
\newcommand{\dw}[1][-1]{\ar @{--} [0,#1]}

\usepackage{algorithm}
\usepackage{hyperref}
\usepackage{braket}
\usepackage{bm}
\usepackage{graphicx}
\usepackage{caption}
\usepackage{subcaption}
\usepackage[margin=1in]{geometry}
\usepackage[affil-it]{authblk}
\usepackage{bbm}
\usepackage{bm}
\usepackage{mathtools}
\usepackage{amsfonts}
\usepackage{makecell}
\usepackage{booktabs}
\usepackage{dsfont}
\usepackage{cite}

\newcommand{\hgate}{{\sf H}}
\newcommand{\sgate}{{\sf S}}
\newcommand{\tgate}{{\sf T}}
\newcommand{\xgate}{{\sf X}}
\newcommand{\ygate}{{\sf Y}}
\newcommand{\zgate}{{\sf Z}}
\newcommand{\cnot}{{\sf CNOT}}
\newcommand{\cz}{{\sf CZ}}

\newcommand{\cP}{{\sf P}}

\newcommand{\cBPP}{{\sf BPP}}
\newcommand{\cNP}{{\sf NP}}

\newcommand{\cMA}{{\sf MA}}

\newcommand{\cBQP}{{\sf BQP}}

\newcommand{\cQMA}{{\sf QMA}}
\newcommand{\cQPIP}{{\sf QPIP}}

\newcommand{\cIP}{{\sf IP}}

\begin{document}
\sloppy

\title{Verification of quantum computation: \\ An overview of existing approaches}
\date{}

\author{Alexandru Gheorghiu\inst{1} \and Theodoros Kapourniotis\inst{2} \and Elham Kashefi\inst{1,3}}
\institute{School of Informatics, University of Edinburgh, UK
\and Department of Physics, University of Warwick, UK
\and CNRS LIP6, Universit\'{e} Pierre et Marie Curie, France}

\maketitle

\begin{abstract}
Quantum computers promise to efficiently solve not only problems believed to be intractable for classical computers, but also problems for which verifying the solution is also considered intractable. This raises the question of how one can check whether quantum computers are indeed producing correct results. This task, known as \emph{quantum verification},
has been highlighted as a significant challenge on the road to scalable quantum computing technology.
We review the most significant approaches to quantum verification and compare them in terms of structure, complexity and required resources. We also comment on the use of cryptographic techniques which, for many of the presented protocols, has proven extremely useful in performing verification. Finally, we discuss issues related to fault tolerance, experimental implementations and the outlook for future protocols.
\end{abstract}

%\tableofcontents

\section{Introduction}
Quantum computation is the subject of intense research due to the potential of quantum computers to efficiently solve problems which are believed to be intractable for classical computers.
The current focus of experiments, aiming to realize scalable quantum computation, is to demonstrate a \emph{quantum computational advantage}. In other words, this means performing a quantum computation in order to solve a problem which is proven to be classically intractable, based on plausible complexity-theoretic assumptions. Examples of such problems, suitable for near-term experiments, include \emph{boson sampling} \cite{bosonsampling}, \emph{instantaneous quantum polynomial time} (\textsf{IQP}) computations \cite{iqp} and others \cite{boixo2016characterizing,aaronson2016complexity,bermejovega2017}.
The prospect of achieving these tasks has ignited a flurry of experimental efforts \cite{tillmann2013experimental,spagnolo2014experimental,bentivegna2015experimental,lanyon2008experimental}.
However, while demonstrating a quantum computational advantage is an important milestone towards scalable quantum computing, it also raises a significant challenge:
\begin{quotation}
\emph{If a quantum experiment solves a problem which is proven to be intractable for classical computers, how can one verify the outcome of the experiment?}
 \end{quotation}
The first researcher who formalised the above ``paradox'' as a \emph{complexity theoretic} question was Gottesman, in a 2004 conference \cite{aaronsonchallenge}. It was then promoted, in 2007, as a complexity challenge by Aaronson who asked: ``\emph{If a quantum computer can efficiently solve a problem, can it also efficiently convince an observer that the solution is correct? More formally, does every language in the class of quantumly tractable problems ($\cBQP$) admit an interactive proof where the prover is in $\cBQP$ and the verifier is in the class of classically tractable problems ($\cBPP$)?}'' \cite{aaronsonchallenge}. Vazirani, then emphasized the importance of this question, not only from the perspective of complexity theory, but from a philosophical point of view \cite{vaziraniver}. In 2007, he raised the question of whether quantum mechanics is a \emph{falsifiable theory}, and suggested that a computational approach could answer this question.
This perspective was explored in depth by Aharonov and Vazirani in \cite{aharonov2013quantum}. They argued that although many of the predictions of quantum mechanics have been experimentally verified to a remarkable precision, all of them involved systems of low complexity. In other words, they involved few particles or few degrees of freedom for the quantum mechanical system.
But the same technique of ``\emph{predict and verify}'' would quickly become infeasible for systems of even a few hundred interacting particles due to the exponential overhead in classically simulating quantum systems.
And so what if, they ask, the predictions of quantum mechanics start to differ significantly from the real world in the high complexity regime? How would we be able to check this?
Thus, the fundamental question is whether there exists a verification procedure for quantum mechanical predictions which is efficient for arbitrarily large systems.

In trying to answer this question we return to complexity theory. The primary complexity class that we are interested in is $\cBQP$, which, as mentioned above, is the class of problems that can be solved efficiently by a quantum computer. The analogous class for classical computers, with randomness, is denoted $\cBPP$. Finally, concerning verification, we have the class $\cMA$, which stands for Merlin-Arthur. This consists of problems whose solutions can be verified by a $\cBPP$ machine when given a proof string, called a \emph{witness}\footnote{$\cBPP$ and $\cMA$ are simply the probabilistic versions of the more familiar classes $\cP$ and $\cNP$. Under plausible derandomization assumptions, $\cBPP = \cP$ and $\cMA = \cNP$ \cite{derand}.}. 
$\cBPP$ is contained in $\cBQP$, since any problem which can be solved efficiently on a classical computer can also be solved efficiently on a quantum computer. Additionally $\cBPP$ is contained in $\cMA$ since any $\cBPP$ problem admits a trivial empty witness. Both of these containments are believed to be strict, though this is still unproven. 

\begin{figure}[htbp!]
\centering
\includegraphics[scale=0.35]{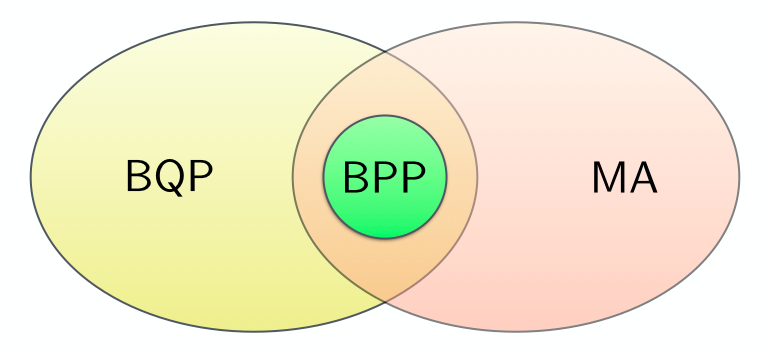}
\caption{Suspected relationship between $\cBQP$ and $\cMA$}
\label{fig:complexity}
\end{figure}

What about the relationship between $\cBQP$ and $\cMA$? Problems are known that are contained in both classes and are believed to be outside of $\cBPP$. One such example is \emph{factoring}. Shor's polynomial-time quantum algorithm for factoring demonstrates that the problem is in $\cBQP$ \cite{shor}. Additionally, for any number to be factored, the witness simply consists of a list of its prime factors, thus showing that the problem is also in $\cMA$.
In general, however, it is believed that $\cBQP$ is \emph{not} contained in $\cMA$ \cite{bv, watroussep}. The conjectured relationship between these complexity classes is illustrated in Figure~\ref{fig:complexity}.

What this tells us is that, very likely, there do not exist witnesses certifying the outcomes of general quantum experiments\footnote{Even if this were the case, i.e. $\cBQP \subseteq \cMA$, for this to be useful in practice one would require that computing the witness can also be done in $\cBQP$. In fact, there are candidate problems known to be in both $\cBQP$ and $\cMA$, for which computing the witness is believed to not be in $\cBQP$ (a conjectured example is \cite{childs2003exponential}).}. 
We therefore turn to a generalization of $\cMA$ known as an \emph{interactive-proof system}. This consists of two entities: a \emph{verifier} and a \emph{prover}. The verifier is a $\cBPP$ machine, whereas the prover has unbounded computational power. Given a problem for which the verifier wants to check a reported solution, the verifier and the prover interact for a number of rounds which is polynomial in the size of the input to the problem. At the end of this interaction, the verifier should accept a valid solution with high probability and reject, with high probability, otherwise. The class of problems which admit such a protocol is denoted $\cIP$\footnote{$\cMA$ can be viewed as an interactive-proof system where only one message is sent from the prover (Merlin) to the verifier (Arthur).}.
In contrast to $\cMA$, instead of having a single proof string for each problem, one has a transcript of back-and-forth communication between the verifier and the prover.

If we are willing to allow our notion of verification to include such interactive protocols, then one would like to know whether $\cBQP$ is contained in $\cIP$.
Unlike the relation between $\cBQP$ and $\cMA$, it is, in fact, the case that $\cBQP \subseteq \cIP$, which means that every problem which can be efficiently solved by a quantum computer admits an interactive-proof system.
One would be tempted to think that this solves the question of verification, however, the situation is more subtle. Recall that in $\cIP$, the prover is computationally unbounded, whereas for our purposes we would require the prover to be restricted to $\cBQP$ computations. Hence, the question that we would like answered and, arguably, the main open problem concerning quantum verification is the following:

\vspace{8pt}

 \textbf{Problem 1 (Verifiability of $\cBQP$ computations)}.  \label{prob:verification}
\emph{Does every problem in $\cBQP$ admit an interactive-proof system in which the prover is restricted to $\cBQP$ computations?}

\vspace{8pt}

\begin{figure}[htbp!]
    \centering
    \begin{subfigure}[b]{0.45\textwidth}
        \includegraphics[width=\textwidth]{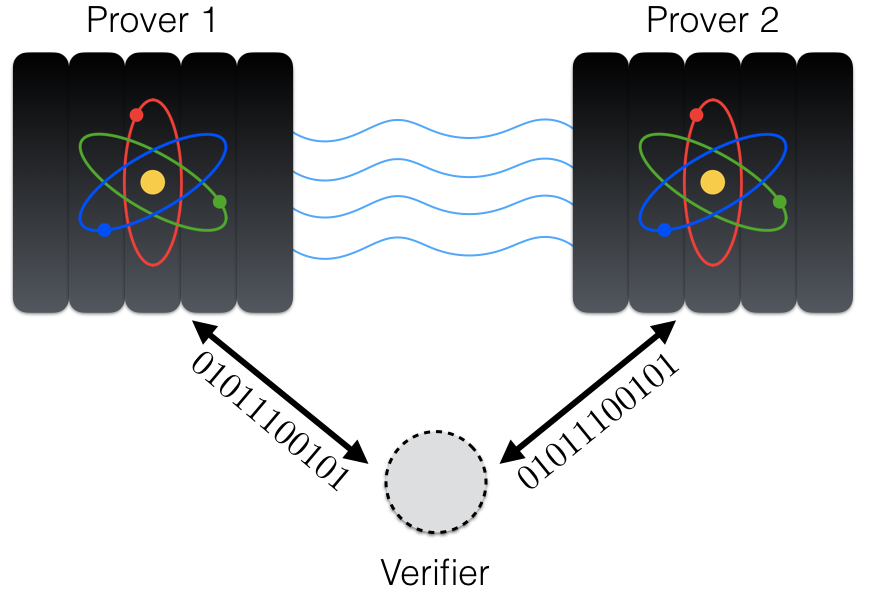}
        \caption{Classical verifier interacting with two entangled but non-communicating quantum provers}
    \end{subfigure}
    \hfill
    \begin{subfigure}[b]{0.45\textwidth}
        \includegraphics[width=\textwidth]{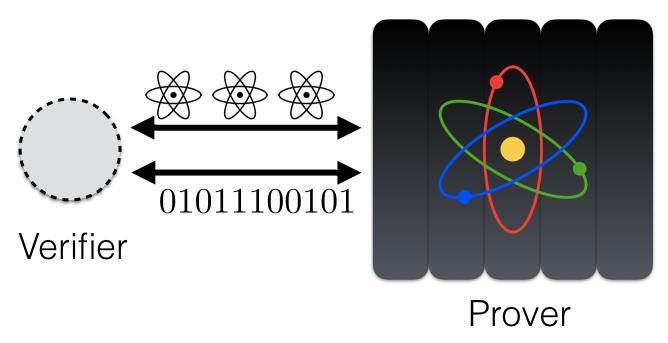}
        \caption{Verifier with the ability to prepare or measure constant-size quantum states interacting with a single quantum prover}
    \end{subfigure}
    \caption{Models for verifiable quantum computation}
    \label{fig:protos}
\end{figure}

As mentioned, this complexity theoretic formulation of the problem was considered by Gottesman, Aaronson and Vazirani \cite{aaronsonchallenge,vaziraniver} and, in fact, Aaronson has offered a $25\$$ prize for its resolution \cite{aaronsonchallenge}. While, as of yet, the question remains open, one does arrive at a positive answer through slight alterations of the interactive-proof system.
Specifically, if the verifier interacts with two or more $\cBQP$-restricted provers, instead of one, and the provers are not allowed to communicate with each other during the protocol, then it is possible to efficiently verify arbitrary $\cBQP$ computations \cite{ruv,gkw,hpdf,mckague,fh,nv,leash}.
Alternatively, in the single-prover setting, if we allow the verifier to have a constant-size quantum computer and the ability to send/receive quantum states to/from the prover then it is again possible to verify all polynomial-time quantum computations \cite{abe,abem,fk,broadbent,posthoc,hangleiter2017direct,monly,hypergraph,gwk}. Note that in this case, while the verifier is no longer fully ``classical'', its computational capability is still restricted to $\cBPP$ since simulating a constant-size quantum computer can be done in constant time.
These scenarios are depicted in Figure~\ref{fig:protos}.

The primary technique that has been employed in most, thought not all, of these settings, to achieve verification, is known as \emph{blindness}. This entails delegating a computation to the provers in such a way that they cannot distinguish this computation from any other of the same size, unconditionally\footnote{In other words, the provers would not be able to differentiate among the different computations even if they had unbounded computational power.}. Intuitively, verification then follows by having most of these computations be \emph{tests} or \emph{traps} which the verifier can check. If the provers attempt to deviate they will have a high chance of triggering these traps and prompt the verifier to reject.

In this paper, we review all of these approaches to verification. We broadly classify the protocols as follows:
\begin{enumerate}
\item \textbf{Single-prover prepare-and-send}. These are protocols in which the verifier has the ability to prepare quantum states and send them to the prover. They are covered in Section~\ref{sect:prepsend}.
\item \textbf{Single-prover receive-and-measure}. In this case, the verifier receives quantum states from the prover and has the ability to measure them. These protocols are presented in Section~\ref{sect:recvmeas}.
\item \textbf{Multi-prover entanglement-based}. In this case, the verifier is fully classical, however it interacts with more than one prover. The provers are not allowed to communicate during the protocol. Section~\ref{sect:entanglement} is devoted to these protocols.
\end{enumerate}
From the complexity-theoretic perspective, the protocols from the first two sections are classified as $\cQPIP$ (quantum prover interactive proofs) protocols, or protocols in which the verifier has a minimal quantum device and can send or receive quantum states. Conversely, the entanglement-based protocols are classified as $\sf{MIP^*}$ (multi prover interactive proofs with entanglement) protocols, in which the verifier is classical and interacting with provers that share entanglement\footnote{The definitions of these classes can be found in Subsection~\ref{subsect:complexity}.}.

After reviewing the major approaches to verification, in Section~\ref{sect:outlook}, we address a number of related topics. In particular, while all of the protocols from Sections~\ref{sect:prepsend}-\ref{sect:entanglement} are concerned with the verification of general $\cBQP$ computations, in Subsection~\ref{subsect:others} we mention \emph{sub-universal} protocols, designed to verify only a particular subclass of quantum computations. 
Next, in Subsection~\ref{sect:ft} we discuss an important practical aspect concerning verification, which is \emph{fault tolerance}. We comment on the possibility of making protocols resistant to noise which could affect any of the involved quantum devices. This is an important consideration for any realistic implementation of a verification protocol.
Finally, in Subsection~\ref{sect:exp} we outline some of the existing experimental implementations of these protocols.

Throughout the review, we are assuming familiarity with the basics of quantum information theory and some elements of complexity theory. However, we provide a brief overview of these topics as well as other notions that are used in this review (such as \emph{measurement-based quantum computing}) in the appendix, Section~\ref{sect:preliminaries}. Note also, that we will be referencing complexity classes such as $\cBQP$, $\cQMA$, $\cQPIP$ and $\sf{MIP^*}$. Definitions for all of these are provided in Subsection~\ref{subsect:complexity} of the appendix.
We begin with a short overview of \emph{blind quantum computing}.

\subsection{Blind quantum computing} \label{subsect:blind}
The concept of blind computing is highly relevant to quantum verification. Here, we simply give a succinct outline of the subject. For more details, see this review of blind quantum computing protocols by Fitzsimons \cite{Fitzsimons2017} as well as \cite{childs,ubqc,blindqc,blindqc2,blindqc3}. Note that, while the review of Fitzsimons covers all of the material presented in this section (and more), we restate the main ideas, so that our review is self-consistent and also in order to establish some of the notation that is used throughout the rest of the paper.

Blindness is related to the idea of \emph{computing on encrypted data} \cite{rad}. Suppose a client has some input $x$ and would like to compute a function $f$ of that input, however, evaluating the function directly is computationally infeasible for the client. Luckily, the client has access to a server with the ability to evaluate $f(x)$. The problem is that the client does not trust the server with the input $x$, since it might involve private or secret information (e.g. medical records, military secrets, proprietary information etc). The client does, however, have the ability to encrypt $x$, using some encryption procedure $\mathcal{E}$, to a ciphertext $y \leftarrow \mathcal{E}(x)$. 
As long as this encryption procedure hides $x$ sufficiently well, the client can send $y$ to the server and receive in return (potentially after some interaction with the server) a string $z$ which decrypts to $f(x)$.
In other words, $f(x) \leftarrow \mathcal{D}(z)$, where $\mathcal{D}$ is a decryption procedure that can be performed efficiently by the client\footnote{In the classical setting, computing on encrypted data culminated with the development of \emph{fully homomorphic encryption} (FHE), which is considered the ``\emph{holly grail}'' of the field \cite{gentry, fhe1, fhe2, fhe3}. Using FHE, a client can delegate the evaluation of \emph{any} polynomial-size classical circuit to a server, such that the input and output of the circuit are kept hidden from the server, based on reasonable computational assumptions. Moreover, the protocol involves only one round of back-and-forth interaction between client and server.}.
The encryption procedure can, roughly, provide two types of security: \emph{computational} or \emph{information-theoretic}. Computational security means that the protocol is secure as long as certain computational assumptions are true (for instance that the server is unable to invert one-way functions). Information-theoretic security (sometimes referred to as unconditional security), on the other hand, guarantees that the protocol is secure even against a server of unbounded computational power. See \cite{katz2014introduction} for more details on these topics.

In the quantum setting, the situation is similar to that of $\cQPIP$ protocols: the client is restricted to $\cBPP$ computations, but has some limited quantum capabilities, whereas the server is a $\cBQP$ machine. Thus, the client would like to delegate $\cBQP$ functions to the server, while keeping the input and the output hidden.
The first solution to this problem was provided by Childs \cite{childs}. His protocol achieves information-theoretic security but also requires the client and the server to exchange quantum messages for a number of rounds that is proportional to the size of the computation. This was later improved in a protocol by Broadbent, Fitzsimons and Kashefi \cite{ubqc}, known as \emph{universal blind quantum computing} (UBQC), which maintained information-theoretic security but reduced the quantum communication to a single message from the client to the server. UBQC still requires the client and the server to have a total communication which is proportional to the size of the computation, however, apart from the first quantum message, the interaction is purely classical. Let us now state the definition of perfect, or information-theoretic, blindness from \cite{ubqc}:

\begin{definition}[Blindness] \label{defn:blind}
Let \textsf{P} be a delegated quantum computation protocol involving a client and a server. The client draws the input from the random variable $X$. Let $L(X)$ be any function of this random variable. We say that the protocol is \emph{blind while leaking at most L(X)} if, on the client's input $X$, for any $l \in Range(L)$, the following two hold when given $l \leftarrow L(X)$:
\begin{enumerate}
\item \label{defn:blind-1} The distribution of the classical information obtained by the server in \textsf{P} is independent of $X$.
\item Given the distribution of classical information described in~\ref{defn:blind-1}, the state of the quantum system obtained by the server in \textsf{P} is fixed and independent of~$X$.
\end{enumerate}
\end{definition}

\noindent The definition is essentially saying that the server's ``view'' of the protocol should be independent of the input, when given the length of the input. This view consists, on the one hand, of the classical information he receives, which is independent of $X$, given $L(X)$. On the other hand, for any fixed choice of this classical information, his quantum state should also be independent of $X$, given $L(X)$. Note that the definition can be extended to the case of multiple servers as well. 
To provide intuition for how a protocol can achieve blindness, we will briefly recap the main ideas from \cite{childs, ubqc}. We start by considering the \emph{quantum one-time pad}.

\subsubsection{Quantum one-time pad.}
Suppose we have two parties, Alice and Bob, and Alice wishes to send one qubit, $\rho$, to Bob such that all information about $\rho$ is kept hidden from a potential eavesdropper, Eve. For this to work, we will assume that Alice and Bob share two classical random bits, denoted $b_1$ and $b_2$, that are known only to them.
Alice will then apply the operation $\xgate^{b_1} \zgate^{b_2}$ (the quantum one-time pad) to $\rho$, resulting in the state $\xgate^{b_1} \zgate^{b_2} \rho \zgate^{b_2} \xgate^{b_1}$, and send this state to Bob. If Bob then also applies $\xgate^{b_1} \zgate^{b_2}$ to the state he received, he will recover $\rho$. What happens if Eve intercepts the state that Alice sends to Bob?
Because Eve does not know the random bits $b_1$ and $b_2$, the state that she will intercept will be:
\begin{equation}
\frac{1}{4} \sum\limits_{b_1, b_2 \in \{0, 1\}} \xgate^{b_1} \zgate^{b_2} \rho \zgate^{b_2} \xgate^{b_1}
\end{equation}
However, it can be shown that for any single-qubit state $\rho$:
\begin{equation} \label{eqn:otp}
\frac{1}{4} \sum\limits_{b_1, b_2 \in \{0, 1\}} \xgate^{b_1} \zgate^{b_2} \rho \zgate^{b_2} \xgate^{b_1} = I/2
\end{equation}
In other words, the state that Eve intercepts is the totally mixed state, \emph{irrespective of the original state $\rho$}. 
But the totally mixed state is, by definition, the state of maximal uncertainty. Hence, Eve cannot recover any information about $\rho$, regardless of her computational power.
Note, that for this argument to work, and in particular for Equation~\ref{eqn:otp} to be true, Alice and Bob's shared bits must be \emph{uniformly random}.
If Alice wishes to send $n$ qubits to Bob, then as long as Alice and Bob share $2n$ random bits, they can simply perform the same procedure for each of the $n$ qubits. Equation~\ref{eqn:otp} generalizes for the multi-qubit case so that for an $n$-qubit state $\rho$ we have:
\begin{equation} 
\frac{1}{4^{n}} \sum\limits_{\mathbf{b_1}, \mathbf{b_2} \in \{0, 1\}^n} \xgate(\mathbf{b_1}) \zgate(\mathbf{b_2}) \rho \zgate(\mathbf{b_2}) \xgate(\mathbf{b_1}) = I/2^{n}
\end{equation}
Here, $\mathbf{b_1}$ and $\mathbf{b_2}$ are $n$-bit vectors, $\xgate(\mathbf{b}) = \bigotimes\limits_{i=1}^{n} \xgate^{\mathbf{b(i)}}$, $\zgate(\mathbf{b}) = \bigotimes\limits_{i=1}^{n} \zgate^{\mathbf{b(i)}}$ and $I$ is the $2^n$-dimensional identity matrix.

\subsubsection{Childs' protocol for blind computation.} \label{subsubsect:childs}
Now suppose Alice has some $n$-qubit state $\rho$ and wants a quantum circuit $\mathcal{C}$ to be applied to this state and the output to be measured in the computational basis. However, she only has the ability to store $n$ qubits, prepare qubits in the $\ket{0}$ state, swap any two qubits, or apply a Pauli $\xgate$ or $\zgate$ to any of the $n$ qubits. So in general, she will not be able to apply a general quantum circuit $\mathcal{C}$, or perform measurements. Bob, on the other hand, does not have these limitations as he is a $\cBQP$ machine and thus able to perform universal quantum computations.
How can Alice delegate the application of $\mathcal{C}$ to her state without revealing any information about it, apart from its size, to Bob?
The answer is provided by Childs' protocol \cite{childs}. Before presenting the protocol, recall that any quantum circuit, $\mathcal{C}$, can be expressed as a combination of Clifford operations and $\tgate$ gates. Additionally, Clifford operations normalise Pauli gates\footnote{In other words, for all Pauli operators $P$ and all Clifford operators $C$, there exists a Pauli operator $Q$ such that $CP = QC$.}. These notions are defined in the appendix, Subsection~\ref{subsubsect:qcomp}.

First, Alice will one-time pad her state and send the padded state to Bob. As mentioned, this will reveal no information to Bob about $\rho$. Next, Alice instructs Bob to start applying the gates in $\mathcal{C}$ to the padded state. Apart from the $\tgate$ gates, all other operations in $\mathcal{C}$ will be Clifford operations, which normalise the Pauli gates. Thus, if Alice's padded state is $\xgate(\mathbf{b_1}) \zgate(\mathbf{b_2}) \rho \zgate(\mathbf{b_2}) \xgate(\mathbf{b_1})$ and Bob applies the Clifford unitary $U_C$, the resulting state will be:
\begin{equation}
U_C \xgate(\mathbf{b_1}) \zgate(\mathbf{b_2}) \rho \zgate(\mathbf{b_2}) \xgate(\mathbf{b_1}) U_C^{\dagger} = 
\xgate(\mathbf{b'_1}) \zgate(\mathbf{b'_2}) U_C \rho U_C^{\dagger} \zgate(\mathbf{b'_2}) \xgate(\mathbf{b'_1})
\end{equation}
Here, $\mathbf{b'_1}$ and $\mathbf{b'_2}$ are linearly related to $\mathbf{b_1}$ and $\mathbf{b_2}$, meaning that Alice can compute them using only $xor$ operations. This gives her an updated pad for her state. If $\mathcal{C}$ consisted exclusively of Clifford operations then Alice would only need to keep track of the updated pad (also referred to as the \emph{Pauli frame}) after each gate. Once Bob returns the state, she simply undoes the one-time pad using the updated key, that she computed, and recovers $\mathcal{C} \rho \mathcal{C}^{\dagger}$.
Of course, this will not work if $\mathcal{C}$ contains $\tgate$ gates, since, up to an overall phase, we have that:
\begin{equation}
\tgate \xgate^{a} = \xgate^{a} \sgate^{a} \tgate
\end{equation}
where $\sgate = \tgate^2$ and is not a Pauli gate. In other words, if we try to commute the $\tgate$ operation with the one-time pad we will get an unwanted $\sgate$ gate applied to the state. Worse, the $\sgate$ will have a dependency on one of the secret pad bits for that particular qubit. This means that if Alice asks Bob to apply an $\sgate^a$ operation she will reveal one of her pad bits. Fortunately, as explained in \cite{childs}, there is a simple way to remedy this problem. 
After each $\tgate$ gate, Alice asks Bob to return the quantum state to her. Suppose that Bob had to apply a $\tgate$ on qubit $j$. Alice then applies a new one-time pad on that qubit. If the previous pad had no $\xgate$ gate applied to $j$, she will swap this qubit with a dummy state that does not take part in the computation\footnote{For instance, her initial state $\rho$ could contain a number of $\ket{0}$ qubits that is equal to the number of $\tgate$ gates in the circuit.}, otherwise she leaves the state unchanged. She then returns the state to Bob and asks him to apply an $\sgate$ gate to qubit $j$. Since this operation will always be applied, after a $\tgate$ gate, it does not reveal any information about Alice's pad. 
Bob's operation will therefore cancel the unwanted $\sgate$ gate when this appears and otherwise it will act on a qubit which does not take part in the computation.
The state should then be sent back to Alice so that she can undo the swap operation if it was performed.
Once all the gates in $\mathcal{C}$ have been applied, Bob is instructed to measure the resulting state in the computational basis and return the classical outcomes to Alice. Since the quantum output was one-time padded, the classical outcomes will also be one-time padded. Alice will then undo the pad an recover her desired output.

While Childs' protocol provides an elegant solution to the problem of quantum computing on encrypted data, it has significant requirements in terms of Alice's quantum capabilities. If Alice's input is fully classical, i.e. some state $\ket{x}$, where $x \in \{0,1\}^n$, then Alice would only require a constant-size quantum memory. Even so, the protocol requires Alice and Bob to exchange multiple quantum messages.
This, however, is not the case with UBQC which limits the quantum communication to one quantum message sent from Alice to Bob at the beginning of the protocol. Let us now briefly state the main ideas of that protocol.

\subsubsection{Universal Blind Quantum Computation (UBQC).} \label{subsubsect:ubqc}
In UBQC the objective is to not only hide the input (and output) from Bob, but also the circuit which will act on that input\footnote{This is also possible in Childs' protocol by simply encoding the description of the circuit $\mathcal{C}$ in the input and asking Bob to run a universal quantum circuit. The one-time padded input that is sent to Bob would then comprise of both the description of $\mathcal{C}$ as well as $x$, the input for $\mathcal{C}$.} \cite{ubqc}.
As in the previous case, Alice would like to delegate to Bob the application of some circuit $\mathcal{C}$ on her input (which, for simplicity, we will assume is classical). 
This time, however, we view $\mathcal{C}$ as an MBQC computation\footnote{For a brief overview of MBQC see Subsection~\ref{subsect:mbqc}.}. By considering some universal graph state, $\ket{G}$, such as the brickwork state (see Figure~\ref{fig:brickwork}), Alice can convert $\mathcal{C}$ into a description of $\ket{G}$ (the graph $G$) along with the appropriate measurement angles for the qubits in the graph state. By the property of the universal graph state, the graph $G$ would be the same for all circuits $\mathcal{C'}$ having the same number of gates as $\mathcal{C}$.
Hence, if she were to send this description to Bob, it would not reveal to him the circuit $\mathcal{C}$, merely an upper bound on its size. 
It is, in fact, the measurement angles and the ordering of the measurements (known as \emph{flow}) that uniquely characterise $\mathcal{C}$ \cite{flow}. But the measurement angles are chosen assuming all qubits in the graph state were initially prepared in the $\ket{+}$ state. Since these are $\xgate\ygate$-plane measurements, as explained in Subsection~\ref{subsect:qinfo}, the probabilities, for the two possible outcomes, depend only on the difference between the measurement angle and the preparation angle of the state, which is $0$, in this case\footnote{This remains true even if the qubits have been entangled with the $\cz$ operation.}.
Suppose instead that each qubit, indexed $i$, in the cluster state, were instead prepared in the state $\ket{+_{\theta_i}}$. Then, if the original measurement angle for qubit $i$ was $\phi_i$, to preserve the relative angles, the new value would be $\phi_i + \theta_i$.
If the values for $\theta_i$ are chosen at random, then they effectively act as a one-time pad for the original measurement angles $\phi_i$. This means that if Bob does not know the preparation angles of the qubits and were instructed to measure them at the updated angles $\phi_i + \theta_i$, to him, these angles would be indistinguishable from random, irrespective of the values of $\phi_i$. He would, however, learn the measurement outcomes of the MBQC computation. 
But there is a simple way to hide this information as well. One can flip the probabilities of the measurement outcomes for a particular state by performing a $\pi$ rotation around $\zgate$ axis. In other words, the updated measurement angles will be $\delta_i = \phi_i + \theta_i + r_i \pi$, where $r_i$ is sampled randomly from $\{0, 1\}$.

\begin{figure}[htbp!]
\centering
\includegraphics[scale=0.35]{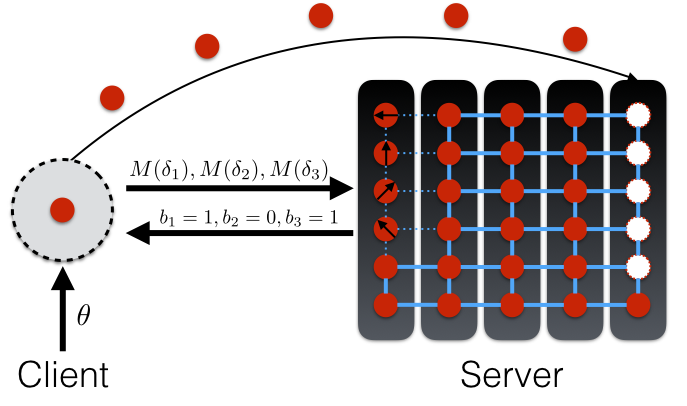}
\caption{Universal Blind Quantum Computation}
\label{fig:ubqc}
\end{figure}

To recap, UBQC works as follows:
\begin{enumerate}\addtolength{\itemsep}{+0.5\baselineskip}
\item[\textbf{(1)}] Alice chooses an input $x$ and a quantum computation $\mathcal{C}$ that she would like Bob to perform on $\ket{x}$.
\item[\textbf{(2)}] She converts $x$ and $\mathcal{C}$ into a pair $(G, \{\phi_i\}_i)$, where $\ket{G}$ is an $N$-qubit universal graph state (with an established ordering for measuring the qubits), $N = O(|\mathcal{C}|)$ and $\{\phi_i\}_i$ is the set of computation angles allowing for the MBQC computation of $\mathcal{C}\ket{x}$.
\item[\textbf{(3)}] She picks, uniformly at random, values $\theta_i$, with $i$ going from $1$ to $N$, from the set $\{0, \pi/4, 2\pi/4, ... 7\pi/4\}$ as well as values $r_i$ from the set $\{0, 1\}$.
\item[\textbf{(4)}] She then prepares the states $\ket{+_{\theta_i}}$ and sends them to Bob, who is instructed to entangle them, using $\cz$ operations, according to the graph structure $G$.
\item[\textbf{(5)}] Alice then asks Bob to measure the qubits at the angles $\delta_i = \phi'_i + \theta_i + r_i \pi$ and return the measurement outcomes to her. Here, $\phi'_i$ is an updated version of $\phi_i$ that incorporates corrections resulting from previous measurements, as in the description of MBQC given in Subsection~\ref{subsect:mbqc}.
\item[\textbf{(6)}] After all the measurements have been performed, Alice undoes the $r_i$ one-time padding of the measurement outcomes, thus recovering the true outcome of the computation.
\end{enumerate}
The protocol is illustrated schematically in Figure~\ref{fig:ubqc}, reproduced from \cite{cubqc} (the variables $b_1$, $b_2$, $b_3$ indicate measurement outcomes).

We can see that as long as Bob does not know the values of the $\theta_i$ and $r_i$ variables, the measurements he is asked to perform, as well as their outcomes, will appear totally random to him. The reason why Bob cannot learn the values of $\theta_i$ and $r_i$ from the qubits prepared by Alice is due to the limitation, in quantum mechanics, that one cannot distinguish between non-orthogonal states. In fact, a subsequent paper by Dunjko and Kashefi shows that Alice can utilize any two non-overlapping, non-orthogonal states in order to perform UBQC \cite{dunjkokashefi}.

\section{Prepare-and-send protocols} \label{sect:prepsend}
We start by reviewing $\cQPIP$ protocols in which the only quantum capability of the verifier is to prepare and send constant-size quantum states to the prover (no measurement). 
The verifier must use this capability in order to delegate the application of some $\cBQP$ circuit, $\mathcal{C}$, on an input $\ket{\psi}$\footnote{This input can be a classical bit string $\ket{x}$, though it can also be more general.}.
Through interaction with the prover, the verifier will attempt to certify that the correct circuit was indeed applied on her input, with high probability, aborting the protocol otherwise.

There are three major approaches that fit this description and we devote a subsection to each of them:
\begin{enumerate}
\item \textbf{Subsection~\ref{subsect:abem}}: two protocols based on quantum authentication, developed by Aharonov, Ben-Or, Eban and Mahadev \cite{abe, abem}.
\item \textbf{Subsection~\ref{subsect:fk}}: a trap-based protocol, developed by Fitzsimons and Kashefi \cite{fk}.
\item \textbf{Subsection~\ref{subsect:howtoverify}}: a scheme based on repeating indistinguishable runs of tests and computations, developed by Broadbent \cite{broadbent}.
\end{enumerate}

In the context of prepare-and-send protocols, it is useful to provide more refined notions of completeness and soundness than the ones in the definition of a $\cQPIP$ protocol. This is because, apart from knowing that the verifier wishes to delegate a $\cBQP$ computation to the prover, we also know that it prepares a particular quantum state and sends it to the prover to act with some unitary operation on it (corresponding to the quantum circuit associated with the $\cBQP$ computation).
This extra information allows us to define \emph{$\delta$-correctness} and \emph{$\epsilon$-verifiability}. 
We start with the latter:

\begin{definition}[$\epsilon$-verifiability]  \label{def:verifiability}
Consider a delegated quantum computation protocol between a verifier and a prover and let the verifier's quantum state be $\Ket{\psi} \Ket{flag}$, where $\Ket{\psi}$ is the input state to the protocol and $\ket{flag}$ is a flag state denoting whether the verifier accepts ($\ket{flag} = \ket{acc}$) or rejects ($\ket{flag} = \ket{rej}$) at the end of the protocol.
Consider also the quantum channel $Enc_s$ (encoding), acting on the verifier's state, where $s$ denotes a private random string, sampled by the verifier from some distribution $p(s)$.
Let $\mathcal{P}_{honest}$ denote the CPTP map corresponding to the honest action of the prover in the protocol (i.e. following the instructions of the verifier) acting on the verifier's state.
Additionally, define:
\begin{equation}
P_{incorrect}^s = (I - \ket{\Psi^{s}_{out}}\bra{\Psi^{s}_{out}}) \otimes \ket{acc^s}\bra{acc^s}
\end{equation}
as a projection onto the orthogonal complement of the correct output:
\begin{equation}
\ket{\Psi^{s}_{out}}\bra{\Psi^{s}_{out}} = Tr_{flag} (\mathcal{P}_{honest} (Enc_s(\Ket{\psi}\bra{\psi} \otimes \Ket{acc} \Bra{acc})))
\end{equation}
and on acceptance for the flag state:
\begin{equation}
\ket{acc^s}\bra{acc^s} = Tr_{input} (\mathcal{P}_{honest} (Enc_s(\Ket{\psi}\bra{\psi} \otimes \Ket{acc} \Bra{acc})))
\end{equation}

We say that such a protocol is $\epsilon$-verifiable (with $0\leq \epsilon \leq 1$), if for any action $\mathcal{P}$, of the prover, we have that\footnote{An alternative to Equation~\ref{eqn:verifiability} is: $TD(\rho_{out}, p \ket{\Psi^{s}_{out}}\bra{\Psi^{s}_{out}} \otimes \ket{acc^s}\bra{acc^s} + (1 -p) \rho \otimes \ket{rej^s}\bra{rej^s}) \leq \epsilon$, for some $0 \leq p \leq 1$ and some density matrix $\rho$, where $TD$ denotes trace distance. In other words, the output state of the protocol, $\rho_{out}$, is close to a state which is a mixture of the correct output state with acceptance and an arbitrary state and rejection.
This definition can be more useful when one is interested in a \emph{quantum output} for the protocol (i.e. the prover returns a quantum state to the verifier). Such a situation is particularly useful when composing verification protocols \cite{dunjko2014composable, garbled, kapourniotis2015optimising, gkw}.}:
\begin{equation} \label{eqn:verifiability}
Tr \left( \sum_{s} p(s)  P_{incorrect}^s \mathcal{P}(Enc_s(\ket{\psi}\bra{\psi} \otimes \ket{acc}\bra{acc})) \right) \leq \epsilon
\end{equation}
\end{definition}

\noindent Essentially, this definition says that the probability for the output of the protocol to be incorrect \emph{and} the verifier accepting, should be bounded by $\epsilon$. As a simple mathematical statement we would write this as the joint distribution:
\begin{equation}
Pr( incorrect, accept ) \leq \epsilon
\end{equation}
One could also ask whether $Pr(incorrect | accept)$ should also be upper bounded. Indeed, it would seem like this conditional distribution is a better match for our intuition regarding the ``\emph{probability of accepting an incorrect outcome}''. However, giving a sensible upper bound for the conditional distribution can be problematic. To understand why, note that we can express the conditional distribution as:
\begin{equation}
Pr( incorrect | accept ) = \frac{Pr( incorrect, accept )}{Pr(accept)}
\end{equation}
Now, it is true that if $Pr(accept)$ is close to $1$ \emph{and} the joint distribution is upper bounded, then the conditional distribution will also be upper bounded. 
Suppose however that $Pr(accept) = 2^{-O(|\mathcal{C}|)}$. In other words, the probability of acceptance is exponentially small in the size of the delegated computation\footnote{One could imagine this happening if, for instance, the prover provides random responses to the verifier instead of performing the desired computation $\mathcal{C}$.}. 
In this case, to upper bound the conditional distribution, it must be that the joint probability is also inverse exponential in the size of the computation. But this is a highly unusual condition, for it would mean that the prover is more likely to deceive the verifier for smaller computations, rather than for larger ones. Moreover, as we will see with the presented protocols, it is typical for the joint probability to be upper bounded by a quantity that is independent of the size of the computation.
For this reason, approaches to verification will either bound $Pr(incorrect, accept)$, or provide a bound for $Pr( incorrect | accept )$ conditioned on the fact that $Pr(accept)$ is close to $1$ (see \cite{ruv} for an example of this).

We now define $\delta$-correctness:
\begin{definition}[$\delta$-correctness]  \label{def:correctness}
Consider a delegated quantum computation protocol between a verifier and a prover. Using the notation from Definition~\ref{def:verifiability}, and letting:
\begin{equation}
P_{correct}^s =  \ket{\Psi^{s}_{out}}\bra{\Psi^{s}_{out}} \otimes \ket{acc^s}\bra{acc^s}
\end{equation}
be the projection onto the correct output and on acceptance for the flag state, we say that such a protocol is $\delta$-correct (with $0\leq \delta \leq 1$), if for all strings $s$ we have that:
\begin{equation}
Tr \left( P_{correct}^s \mathcal{P}_{honest}(Enc_s(\ket{\psi}\bra{\psi} \otimes \ket{acc}\bra{acc})) \right) \geq \delta
\end{equation}
\end{definition}

\noindent This definition says that when the prover behaves honestly, the verifier obtains the correct outcome, with high probability, for any possible choice of its secret parameters.

If a prepare-and-send protocol has both $\delta$-correctness and $\epsilon$-verifiability, for some $\delta > 0$, $\epsilon < 1$, it will also have completeness $\delta (1/2 + 1/poly(n))$ and soundness $\epsilon$ as a $\cQPIP$ protocol, where $n$ is the size of the input. 
The reason for the asymmetry in completeness and soundness is that in the definition of $\delta$-correctness we require that the output quantum state of the protocol is $\delta$-close to the output quantum state of the desired computation. But the computation outcome is dictated by a measurement of this state, which succeeds with probability at least $1/2 + 1/poly(n)$, from the definition of $\cBQP$. Combining these facts leads to $\delta (1/2 + 1/poly(n))$ completeness.
It follows that for this to be a valid $\cQPIP$ protocol it must be that $\delta(1/2 + 1/poly(n)) - \epsilon \geq 1/poly(n)$, for all inputs. 
For simplicity, we will instead require $\delta/2 - \epsilon \geq 1/poly(n)$, which implies the previous inequality.
As we will see, for all prepare-and-send protocols $\delta = 1$. This condition is easy to achieve by simply designing the protocol so that the honest behaviour of the prover leads to the correct unitary being applied to the verifier's quantum state. Therefore, the main challenge with these protocols will be to show that $\epsilon \leq 1/2 - 1/poly(n)$.

\subsection{Quantum authentication-based verification} \label{subsect:abem}
This subsection is dedicated to the two protocols presented in \cite{abe,abem} by Aharonov et al. These protocols are extensions of \emph{Quantum Authentication Schemes} (QAS), a security primitive introduced in \cite{barnum02qas} by Barnum et al. 
A QAS is a scheme for transmitting a quantum state over an insecure quantum channel and being able to indicate whether the state was corrupted or not. 
More precisely, a QAS involves a \emph{sender} and a \emph{receiver}. The sender has some quantum state $\ket{\psi}\ket{flag}$ that it would like to send to the receiver over an insecure channel. The state $\ket{\psi}$ is the one to be authenticated, while $\ket{flag}$ is an indicator state used to check whether the authentication was performed successfully. We will assume that $\ket{flag}$ starts in the state $\ket{acc}$. It is also assumed that the sender and the receiver share some classical key $k$, drawn from a probability distribution $p(k)$.
To be able to detect the effects of the insecure channel on the state, the sender will first apply some encoding procedure $Enc_k$ thus obtaining $\rho = \sum_k p(k) Enc_k(\ket{\psi}\ket{acc})$. This state is then sent over the quantum channel where it can be tampered with by an eavesdropper resulting in a new state $\rho'$.
The receiver, will then apply a decoding procedure to this state, resulting in $Dec_k(\rho')$ and decide whether to accept or reject by measuring the flag subsystem\footnote{The projectors for the measurement are assumed to be $P_{acc} = \ket{acc}\bra{acc}$, for acceptance and $P_{rej} = I - \ket{acc}\bra{acc}$ for rejection.}.
Similar to verification, this protocol must satisfy two properties:
\begin{enumerate}
\item 
\textbf{$\delta$-correctness}. Intuitively this says that if the state sent through the channel was not tampered with, then the receiver should accept with high probability (at least $\delta$), irrespective of the used keys. More formally, for $0 \leq \delta \leq 1$, let:
\begin{equation*}
P_{correct} = \ket{\psi}\bra{\psi} \otimes \ket{acc}\bra{acc}
\end{equation*}
be the projector onto the correct state $\ket{\psi}$ and on acceptance for the flag state. Then, it must be the case that for all keys $k$:
\begin{equation*}
Tr \left( P_{correct} Dec_k(Enc_k( \ket{\psi}\bra{\psi} \otimes \ket{acc}\bra{acc} )) \right) \geq \delta
\end{equation*}
\item 
\textbf{$\epsilon$-security}. This property states that for any deviation that the eavesdropper applies on the sent state, the probability that the resulting state is far from ideal \emph{and} the receiver accepts is small. Formally, for $0 \leq \epsilon \leq 1$, let:
\begin{equation*}
P_{incorrect} = (I - \ket{\psi}\bra{\psi}) \otimes \ket{acc}\bra{acc}
\end{equation*}
be the projector onto the orthogonal complement of the correct state $\ket{\psi}$, and on acceptance, for the flag state. Then, it must be the case that for any CPTP action, $\mathcal{E}$, of the eavesdropper, we have:
\begin{equation*}
Tr \left( P_{incorrect} \sum_k p(k) Dec_k (\mathcal{E}(Enc_k(  \ket{\psi}\bra{\psi} \otimes \ket{acc}\bra{acc} ))) \right) \leq \epsilon
\end{equation*}
\end{enumerate}

\begin{figure}
  \centering
  \includegraphics[width=0.7\linewidth]{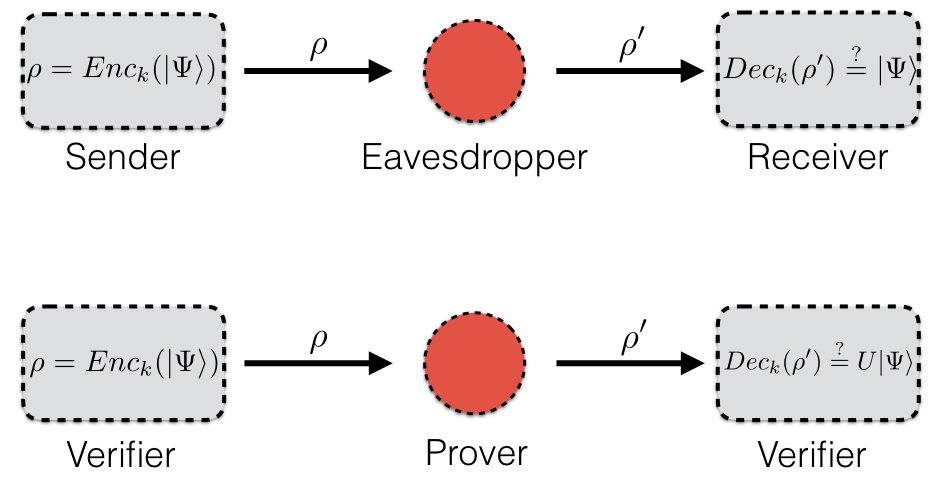}
  \caption{QAS-based verification}
  \label{fig:authver}
\end{figure}

To make the similarities between QAS and prepare-and-send protocols more explicit, suppose that, in the above scheme, the receiver were trying to authenticate the state $U\ket{\psi}$ instead of $\ket{\psi}$, for some unitary $U$. In that case, we could view the sender as the verifier at the beginning of the protocol, the eavesdropper as the prover and the receiver as the verifier at the end of the protocol. This is illustrated in Figure~\ref{fig:authver}, reproduced from \cite{cubqc}.
If one could therefore augment a QAS scheme with the ability of applying a quantum circuit on the state, while keeping it authenticated, then one would essentially have a prepare-and-send verification protocol. This is what is achieved by the two protocols of Aharonov et al.

\subsubsection{Clifford-QAS VQC.}
The first protocol, named \emph{Clifford QAS-based Verifiable Quantum Computing} (Clifford-QAS VQC) is based on a QAS which uses Clifford operations in order to perform the encoding procedure. 
Strictly speaking, this protocol is not a prepare-and-send protocol, since, as we will see, it involves the verifier performing measurements as well. However, it is a precursor to the second protocol from \cite{abe,abem}, which is a prepare-and-send protocol. Hence, why we review the Clifford-QAS VQC protocol here.

Let us start by explaining the authentication scheme first. As before, let $\Ket{\psi}\ket{flag}$ be the state that the sender wishes to send to the receiver and $k$ be their shared random key. We will assume that $\ket{\psi}$ is an $n$-qubit state, while $\ket{flag}$ is an $m$-qubit state. Let $t = n + m$ and $\mathfrak{C}_t$ be the set of $t$-qubit Clifford operations\footnote{Note that:
\begin{equation}
\mathfrak{C}_t = \{ U \in U(2^t) | \sigma \in \mathds{P}_t \implies U \sigma U^{\dagger} \in \mathds{P}_t \}
\end{equation}
where:
 \begin{equation}
\mathds{P}_t = \{ \alpha \; \sigma_1 \otimes ... \otimes \sigma_t | \alpha \in \{+1, -1, +i, -i\}, \sigma_i \in \{ I, \xgate, \ygate, \zgate \} \}
\end{equation}
is the $t$-qubit Pauli group. See Subsection~\ref{subsubsect:qcomp} for more details.}
We also assume that each possible key, $k$, can specify a unique $t$-qubit Clifford operation, denoted $C_k$\footnote{Hence $|k| = O(log(|\mathfrak{C}_t|))$.}.
The QAS works as follows:
\begin{enumerate}\addtolength{\itemsep}{+0.5\baselineskip}
\item[\textbf{(1)}] The sender performs the encoding procedure $Enc_k$. This consists of applying the Clifford operation $C_k$ to the state $\ket{\psi}\ket{acc}$.
\item[\textbf{(2)}] The state is sent through the quantum channel.
\item[\textbf{(3)}] The receiver applies the decoding procedure $Dec_k$ which consists of applying $C_k^{\dagger}$ to the received state.
\item[\textbf{(4)}] The receiver measures the $flag$ subsystem and accepts if it is in the $\ket{acc}$ state. 
\end{enumerate}

We can see that this protocol has correctness $\delta=1$, since, the sender and receiver's operations are exact inverses of each other and, when there is no intervention from the eavesdropper, they will perfectly cancel out.
It is also not too difficult to show that the protocol achieves security $\epsilon=2^{-m}$. 
We will include a sketch proof of this result as all other proofs of security, for prepare-and-send protocols, rely on similar ideas. Aharonov et al start by using the following lemma:
\begin{lemma}[Clifford twirl] \label{lemma:cliff}
Let $P_1$, $P_2$ be two operators from the $n$-qubit Pauli group, such that $P_1 \neq P_2$\footnote{Technically, what is required here is that $|P_1| \neq |P_2|$, since global phases are ignored.}. For any $n$-qubit density matrix $\rho$ it is the case that:
\begin{equation}
\sum_{C \in \mathfrak{C}_n} C^{\dagger} P_1 C \rho C^{\dagger} P_2 C = 0
\end{equation}
\end{lemma}

To see how this lemma is applied, recall that any CPTP map admits a Kraus decomposition, so we can express the eavesdropper's action as:
\begin{equation}
\mathcal{E}(\rho) = \sum_i K_i \rho K_i^{\dagger}
\end{equation}
where, $\{ K_i \}_i$ is the set of Kraus operators, satisfying:
\begin{equation}
\sum_i K_i^{\dagger} K_i = I
\end{equation}
Additionally, recall that the $n$-qubit Pauli group is a basis for all $2^n \times 2^n$ matrices, which means that we can express each Kraus operator as:
\begin{equation}
K_i = \sum_j \alpha_{ij} P_j
\end{equation}
where $j$ ranges over all indices for $n$-qubit Pauli operators and $\{ \alpha_{ij} \}_{i,j}$ is a set of complex numbers such that:
\begin{equation}
\sum_{ij} \alpha_{ij} \alpha^*_{ij} = 1
\end{equation}
For simplicity, assume that the phase information of each Pauli operator, i.e. whether it is $+1$, $-1$, $+i$ or $-i$, is absorbed in the $\alpha_{ij}$ terms.
One can then re-express the eavesdropper's deviation as:
\begin{equation} \label{eqn:paulidecomp}
\mathcal{E}(\rho) = \sum_{ijk} \alpha_{ij} \alpha^*_{ik} P_j \rho P_k
\end{equation}

We would now like to use Lemma~\ref{lemma:cliff} to see how this deviation affects the encoded state. Given that the encoding procedure involves applying a random Clifford operation to the initial state, which we will denote $\ket{\Psi_{in}} = \Ket{\psi}\ket{acc}$, the state received by the eavesdropper will be:
\begin{equation}
\rho = \frac{1}{|\mathfrak{C}_t|} \sum_l C_l \ket{\Psi_{in}}\bra{\Psi_{in}} C_l^{\dagger}
\end{equation}

\noindent Acting with $\mathcal{E}$ on this state and using Equation~\ref{eqn:paulidecomp} yields:
\begin{equation}
\mathcal{E}(\rho) = \frac{1}{|\mathfrak{C}_t|} \sum_{ijkl}  \alpha_{ij} \alpha^*_{ik} P_j C_l \ket{\Psi_{in}}\bra{\Psi_{in}} C_l^{\dagger} P_k
\end{equation}
The receiver takes this state and applies the decoding operation, which involves inverting the Clifford that was applied by the sender. This will produce the state:
\begin{equation}
\frac{1}{|\mathfrak{C}_t|} \sum_{ijkl} \alpha_{ij} \alpha^*_{ik} C_l^{\dagger} P_j C_l \ket{\Psi_{in}}\bra{\Psi_{in}} C_l^{\dagger} P_k C_l
\end{equation}
Finally, using Lemma~\ref{lemma:cliff} we can see that all terms which act with different Pauli operations on both sides (i.e. $j \neq k$) will vanish, resulting in:
\begin{equation}
\sigma = \frac{1}{|\mathfrak{C}_t|} \sum_{ijl} \alpha_{ij} \alpha^*_{ij} C_l^{\dagger} P_j C_l \ket{\Psi_{in}}\bra{\Psi_{in}} C_l^{\dagger} P_j C_l
\end{equation}

Let us take a step back and understand what happened. We saw that any general map can be expressed as a combination of Pauli operators acting on both sides of the target state, $\rho$. Importantly, the Pauli operators on both sides needed not be equal.
However, if the target state is an equal mixture of Clifford terms acting on some other state (in our case $\Ket{\Psi_{in}}\Bra{\Psi_{in}}$), which are then ``undone'' by the decoding procedure, the Clifford twirl lemma makes all non-equal Pauli terms vanish. 
In the resulting state, $\sigma$, we notice that each Pauli term is conjugated by Clifford operators from the set $\mathfrak{C}_t$. 
We know that conjugating a Pauli matrix by a Clifford operator results in a new Pauli matrix. 
Moreover, we know that for all $j$ such that $P_j \neq I$ it is the case that:
\begin{equation}
\sum_l C_l^{\dagger} P_j C_l = \frac{|\mathfrak{C}_t|}{4^t - 1}\sum_{P \in \mathds{P}_t \backslash \{I\}} P
\end{equation}
In other words, averaging a non-identity Pauli operator over the Clifford group results in an equal mixture of all non-identity Pauli operations\footnote{In \cite{abem} this is referred to as \emph{Clifford decoherence}.}.
From this and since $\alpha_{ij}\alpha^*_{ij} = |\alpha_{ij}|^2$ is a positive real number and $\sum_{ij} \alpha_{ij}\alpha^*_{ij} = 1$, the resulting state is a convex combination of the ideal state and a state that is a \emph{uniform} mixture of Pauli deviations on the ideal state.
Mathematically, this means:
\begin{equation}
\sigma = \beta \Ket{\Psi_{in}}\Bra{\Psi_{in}} + \frac{1 - \beta}{4^t - 1} \sum_{i, P_i \neq I}  P_i \ket{\Psi_{in}}\bra{\Psi_{in}} P_i
\end{equation}
where $0 \leq \beta \leq 1$.

The last element in the proof is to compute $Tr(P_{incorrect} \sigma)$. Since the first term in the mixture is the ideal state, we will be left with:
\begin{equation}
Tr(P_{incorrect} \sigma) = \frac{1 - \beta}{4^t - 1} \sum_{i, P_i \neq I}  Tr(P_{incorrect} P_i \ket{\Psi_{in}}\bra{\Psi_{in}} P_i)
\end{equation}
The terms in the summation will be non-zero whenever $P_i$ acts as identity on the flag subsystem. The number of such terms can be computed to be exactly $4^n 2^m - 1$ and using the fact that $t = m + n$ and $1 - \beta \leq 1$, we have:
\begin{equation}
Tr(P_{incorrect} \sigma) \leq (1 - \beta) \frac{4^n 2^m - 1}{4^{m+n}} \leq \frac{1}{2^m}
\end{equation}
concluding the proof.

As mentioned, in all prepare-and-send protocols we assume that the verifier will prepare some state $\ket{\psi}$ on which it wants to apply a quantum circuit denoted $\mathcal{C}$. Since we are assuming that the verifier has a constant-size quantum device, the state $\ket{\psi}$ will be a product state, i.e. $\ket{\psi} = \ket{\psi_1} \otimes \ket{\psi_2} \otimes ... \otimes \ket{\psi_n}$. For simplicity, assume each $\ket{\psi_i}$ is one qubit, though any constant number of qubits is allowed.
In Clifford-QAS VQC the verifier will use the prover as an untrusted quantum storage device.
Specifically, each $\ket{\psi_i}$, from $\ket{\psi}$, will be paired with a constant-size flag system in the accept state, $\ket{acc}$, resulting in a \emph{block} of the form $\ket{block_i} = \ket{\psi_i}\ket{acc}$. Each block will be encoded, by having a random Clifford operation applied on top of it. The verifier prepares these blocks, one at a time, for all $i \in \{1,... n\}$, and sends them to the prover. The prover is then asked to return pairs of blocks to the verifier so that she may apply gates from $\mathcal{C}$ on them (after undoing the Clifford operations). The verifier then applies new random Clifford operations on the blocks and sends them back to the prover. The process continues until all gates in $\mathcal{C}$ have been applied.

But what if the prover corrupts the state or deviates in some way? This is where the QAS enters the picture. Since each block has a random Clifford operation applied, the idea is to have the verifier use the Clifford QAS scheme to ensure that the quantum state remains authenticated after each gate in the quantum circuit is applied. In other words, if the prover attempts to deviate at any point resulting in a corrupted state, this should be detected by the authentication scheme.
Putting everything together, the protocol works as follows:
\begin{enumerate}\addtolength{\itemsep}{+0.5\baselineskip}
\item[\textbf{(1)}] Suppose the input state that the verifier intends to prepare is $\ket{\psi} = \ket{\psi_1} \otimes \ket{\psi_2} \otimes ... \otimes \ket{\psi_n}$, where each $\ket{\psi_i}$ is a one qubit state\footnote{This can simply be the state $\ket{x}$, if the verifier wishes to apply $\mathcal{C}$ on the classical input $x$. However, the state can be more general which is why we are not restricting it to be $\ket{x}$.}. Also let $\mathcal{C}$ be quantum circuit that the verifier wishes to apply on $\ket{\psi}$. The verifier prepares (one block at a time) the state $\ket{\psi}\ket{flag} = \ket{block_1} \otimes \ket{block_2} \otimes ... \otimes \ket{block_n}$, where $\ket{block_i} = \ket{\psi_i}\ket{acc}$ and each $\ket{acc}$ state consists of a constant number $m$ of qubits. Additionally let the size of each block be $t = m + 1$.

\item[\textbf{(2)}] The verifier applies a random Clifford operation, from the set $\mathfrak{C}_t$ on each block and sends it to the prover.

\item[\textbf{(3)}] The verifier requests a pair of blocks, $(\ket{block_i}, \ket{block_j})$, from the prover, in order to apply a gate from $\mathcal{C}$ on the corresponding qubits, $(\ket{\psi_i}, \ket{\psi_j})$. Once the blocks have been received, the verifier undoes the random Clifford operations and measures the flag registers, aborting if these are not in the $\ket{acc}$ state. Otherwise, the verifier performs the gate from $\mathcal{C}$, applies new random Clifford operations on each block and sends them back to the prover. This step repeats until all gates in $\mathcal{C}$ have been performed. 

\item[\textbf{(4)}] Once all gates have been performed, the verifier requests all the blocks (one by one) in order to measure the output. As in the previous step, the verifier will undo the Clifford operations first and measure the flag registers, aborting if any of them are not in the $\ket{acc}$ state.
\end{enumerate}

We can see that the security of this protocol reduces to the security of the Clifford QAS. Moreover, it is also clear that if the prover behaves honestly, then the verifier will obtain the correct output state exactly. Hence:
\begin{theorem}
For a fixed constant $m > 0$, Clifford-QAS VQC is a prepare-and-send $\cQPIP$ protocol having correctness $\delta=1$ and verifiability $\epsilon = 2^{-m}$.
\end{theorem}

\subsubsection{Poly-QAS VQC.} \label{subsect:polyqas}
The second protocol in \cite{abe, abem}, is referred to as \emph{Polynomial QAS-based Verifiable Quantum Computing} (Poly-QAS VQC).
It improves upon the previous protocol by removing the interactive quantum communication between the verifier and the prover, reducing it to a single round of quantum messages sent at the beginning of the protocol. 
To encode the input, this protocol uses a specific type of quantum error correcting code known as a \emph{polynomial CSS code} \cite{aharonov2008fault}. 
We will not elaborate on the technical details of these codes as that is beyond the scope of this review. We only mention a few basic characteristics which are necessary in order to understand the Poly-QAS VQC protocol.
The polynomial CSS codes operate on \emph{qudits} instead of qubits. A $q$-qudit is simply a quantum state in a $q$-dimensional Hilbert space. The generalized computational basis for this space is given by $\{ \ket{i} \}_{i \leq q}$. The code takes a $q$-qudit, $\ket{i}$, as well as $\ket{0}$ states, and encodes them into a state of $t = 2d + 1$ qudits as follows:
\begin{equation}
E \ket{i}\ket{0}^{\otimes t - 1} = \sum_{p, deg(p) \leq d, p(i) = 0} \ket{p(\alpha_1)}\ket{p(\alpha_2)} ... \ket{p(\alpha_t)}
\end{equation}
where $E$ is the encoding unitary, $p$ ranges over polynomials of degree less than $d$ over the field $F_q$ of integers $mod \; q$, and $\{ \alpha_j \}_{j \leq t}$ is a fixed set of $m$ \emph{non-zero} values from $F_q$ (it is assumed that $q > t$).
The code can detect errors on at most $d$ qudits and can correct errors on up to $\lfloor \frac{d}{2} \rfloor$ qudits (hence $\lfloor \frac{d}{2} \rfloor$ is the weight of the code).
Importantly, the code is transversal for Clifford operations.
Aharonov et al consider a slight variation of this scheme called a \emph{signed polynomial code}, which allows one to randomize over different polynomial codes. The idea is to have the encoding (and decoding) procedure also depend on a key $k \in \{-1, +1\}^t$ as follows:
\begin{equation} \label{eqn:signedencode}
E_k \ket{i}\ket{0}^{\otimes t - 1} = \sum_{p, deg(p) \leq d, p(i) = 0} \ket{k_1 p(\alpha_1)}\ket{k_2 p(\alpha_2)} ... \ket{k_t p(\alpha_t)}
\end{equation}

The signed polynomial CSS code can be used to create a simple authentication scheme having security $\epsilon = 2^{-d}$. 
This works by having the sender encode the state $\ket{\Psi_{in}} = \ket{\psi}\ket{0}^{\otimes t - 1}$, where $\ket{\psi}$ is a qudit to be authenticated, in the signed code and then one-time padding the encoded state. Note that the $\ket{0}^{\otimes t - 1}$ part of the state is acting as a flag system. We are assuming that the sender and the receiver share both the sign key of the code and the key for the one-time padding. The one-time padded state is then sent over the insecure channel.
The receiver undoes the pad and applies the inverse of the encoding operation. It then measures the last $t - 1$ qudits, accepting if and only if they are all in the $\ket{0}$ state.
Proving security is similar to the Clifford QAS and relies on two results:
\begin{lemma}[Pauli twirl] \label{lemma:pauli}
Let $P_1$, $P_2$ be two operators from the $n$-qudit Pauli group, denoted $\mathds{P}_n$, such that $P_1 \neq P_2$. For any $n$-qudit density matrix $\rho$ it is the case that:
\begin{equation}
\sum_{Q \in \mathds{P}_n} Q^{\dagger} P_1 Q \rho Q^{\dagger} P_2 Q = 0
\end{equation}
\end{lemma}
This result is identical to the Clifford twirl lemma, except the Clifford operations are replaced with Pauli operators\footnote{Note that by abuse of notation we assume $\mathds{P}_n$ refers to the group of generalized Pauli operations over qudits, whereas, typically, one uses this notation to refer to the Pauli group of qubits.}. The result is also valid for qubits.
\begin{lemma}[Signed polynomial code security] \label{lemma:signedpoly}
Let $\rho = \ket{\psi}\bra{\psi} \otimes \ket{0}\bra{0}^{\otimes t - 1}$, be a state which will be encoded in the signed polynomial code, $P = (I - \ket{\psi}\bra{\psi}) \otimes \ket{0}\bra{0}^{\otimes t - 1}$, be a projector onto the orthogonal complement of $\ket{\psi}$ and on $\ket{0}^{t-1}$, and $Q \in \mathds{P}_t \backslash \{I\}$ be a non-identity Pauli operation on $t$ qudits. Then it is the case that: 
\begin{equation}
\frac{1}{2^t} \sum_{k \in \{-1, +1\}^t} Tr \left( P \;\; E_k^{\dagger} Q E_k \; \rho \; E_k^{\dagger} Q E_k \right)
\leq \frac{1}{2^{t-1}}
\end{equation}
\end{lemma}

Using these two results, and the ideas from the Clifford QAS scheme, it is not difficult to prove the security of the above described authentication scheme. As before, the eavesdropper's map is decomposed into Kraus operators which are then expanded into Pauli operations. Since the sender's state is one-time padded (and the receiver will undo the one-time pad), the Pauli twirl lemma will turn the eavesdropper's deviation into a convex combination of Pauli deviations:
\begin{equation}
\frac{1}{2^t} \sum_{k \in \{-1, +1\}^t} \sum_{Q \in \mathds{P}_t} \beta_Q \; Q E_k \ket{\Psi_{in}}\bra{\Psi_{in}} E_k^{\dagger} Q^{\dagger}
\end{equation}
which can be split into the identity and non-identity Pauli terms:
\begin{equation}
\frac{1}{2^t} \sum_{k \in \{-1, +1\}^t} \left( \beta_{I} E_k \ket{\Psi_{in}}\bra{\Psi_{in}} E_k^{\dagger} + \sum_{Q \in \mathds{P}_t \backslash \{I\}} \beta_Q \; Q E_k \ket{\Psi_{in}}\bra{\Psi_{in}} E_k^{\dagger} Q^{\dagger} \right)
\end{equation}
where $\beta_Q$ are positive real coefficients satisfying:
\begin{equation}
\sum_{Q \in \mathds{P}_n} \beta_Q = 1
\end{equation}
The receiver takes this state and applies the inverse encoding operation, resulting in:
\begin{equation}
\rho = \frac{1}{2^t} \sum_{k \in \{-1, +1\}^t} \left( \beta_{I} \ket{\Psi_{in}}\bra{\Psi_{in}}  + \sum_{Q \in \mathds{P}_t \backslash \{I\}} \beta_Q \; Q E_k \ket{\Psi_{in}}\bra{\Psi_{in}} E_k^{\dagger} Q^{\dagger} \right)
\end{equation}
But now we know that $\epsilon = Tr(P_{incorrect} \rho)$, and using Lemma~\ref{lemma:signedpoly} together with the facts that $Tr(P_{incorrect} \ket{\Psi_{in}}\bra{\Psi_{in}}) = 0$ and that the $\beta_Q$ coefficients sum to $1$ we end up with:
\begin{equation}
\epsilon \leq \frac{1}{2^{t-1}} \leq \frac{1}{2^d}
\end{equation}

There are two more aspects to be mentioned before giving the steps of the Poly-QAS VQC protocol. The first is that the encoding procedure for the signed polynomial code is implemented using the following interpolation operation:
\begin{equation}
D_k \ket{i} \ket{k_2 p(\alpha_2)} ... \ket{k_{d+1} p(\alpha_{d+1})} \ket{0}^{\otimes d} = \ket{k_1 p(\alpha_1)} ... \ket{k_t p(\alpha_t)}
\end{equation}
The inverse operation $D_k^{\dagger}$ can be though of as a decoding of one term from the superposition in Equation~\ref{eqn:signedencode}. Akin to Lemma~\ref{lemma:signedpoly}, the signed polynomial code has the property that, when averaging over all sign keys, $k$, if such a term had a non-identity Pauli applied to it, when decoding it with $D_k^{\dagger}$, the probability that its last $d$ qudits are not $\ket{0}$ states is upper bounded by $2^{-d}$.

The second aspect is that, as mentioned, the signed polynomial code is transversal for Clifford operations. However, in order to apply non-Clifford operations it is necessary to measure encoded states together with so-called \emph{magic states} (which will also be encoded). This manner of performing gates is known as \emph{gate teleportation} \cite{Gottesman1999}. The target state, on which we want to apply a non-Clifford operation, and the magic state are first entangled using a Clifford operation and then the magic state is measured in the computational basis. The effect of the measurement is to have a non-Clifford operation applied on the target state, along with Pauli errors which depend on the measurement outcome. For the non-Clifford operations, Aharonov et al use Toffoli gates\footnote{See Subsection~\ref{subsubsect:qcomp} for the definition of the Toffoli gate.}.

\begin{figure}[htbp!]
  \centering
  \includegraphics[width=0.7\linewidth]{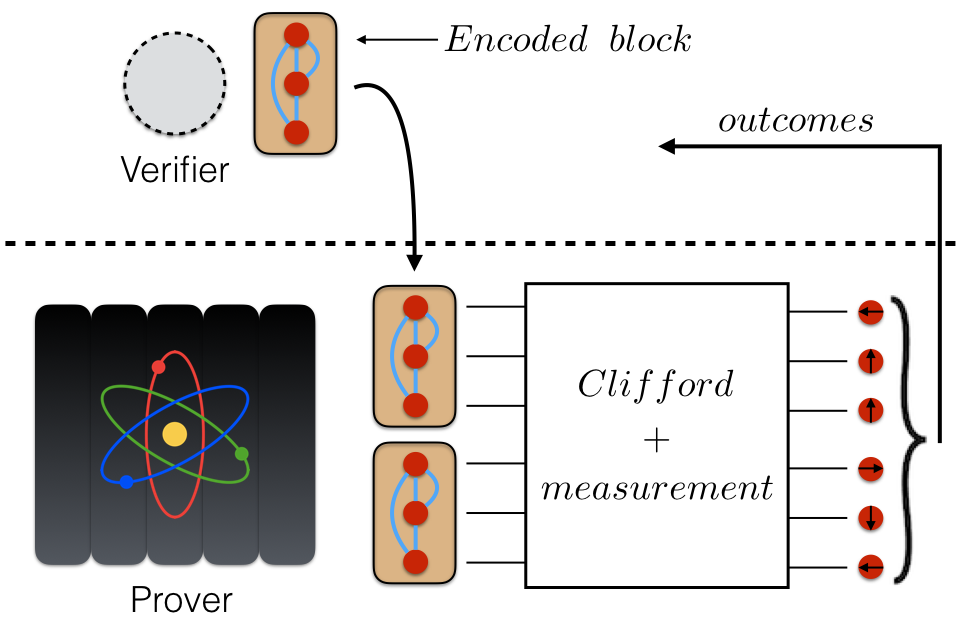}
  \caption{Poly-QAS VQC}
  \label{fig:polyqas}
\end{figure}

Given all of these, the Poly-QAS VQC protocol works as follows:
\begin{enumerate}\addtolength{\itemsep}{+0.5\baselineskip}
\item[\textbf{(1)}] Suppose the input state that the verifier intends to prepare is $\ket{\psi} = \ket{\psi_1} \otimes \ket{\psi_2} \otimes ... \otimes \ket{\psi_n}$, where each $\ket{\psi_i}$ is a $q$-qudit. 
Also suppose that the verifier wishes to apply the quantum circuit $\mathcal{C}$ on $\ket{\psi}$, which contains $L$ Toffoli gates.
The verifier prepares the state $\ket{\Psi_{in}} = \ket{\psi_1}\ket{0}^{t-1} \otimes \ket{\psi_2}\ket{0}^{t-1} \otimes ... \otimes \ket{\psi_n}\ket{0}^{t-1} \otimes \ket{M_1}\ket{0}^{3t-3} \otimes ... \otimes \ket{M_L}\ket{0}^{3t-3}$, where $t = 2d + 1$ and each $\ket{M_i}$ is a $3$-qudit magic state, used for performing Toffoli gates. Groups of $t$ qubits will comprise a block as follows. The first $n$ blocks are simply $\ket{block_i} =\ket{\psi_i}\ket{0}^{t-1}$, with $i \in \{1, ..., n\}$. Next, we have the states of the form $\ket{M_i}\ket{0}^{3t-3}$ which consist of $3$ blocks, each. Each block, from such a state, will comprise of one qudit from $\ket{M_i}$ and a $\ket{0}^{t-1}$ state. Note that we can no longer represent these blocks as pure states, since the $3$ qudits of a $\ket{M_i}$ state are entangled. So, to summarize, each block contains one qudit from either the state $\ket{\psi}$ or a magic state $\ket{M_i}$, together with a flag system, $\ket{0}^{t-1}$.

\item[\textbf{(2)}] The verifier encodes each block in a signed polynomial code with a randomly chosen key $k \in \{-1, +1\}^t$ (the same key for each block) and then quantum one-time pads each block (using different keys for the padding of each block). 
The blocks are prepared and encoded in sequence (the verifier has the ability to process $3$ blocks, or $3t$ qudits, at a time) and then sent to the prover.

\item[\textbf{(3)}] When applying Clifford operations, the verifier simply asks the prover to apply the gates in a transversal fashion. Since Clifford operations normalise Pauli operators, the verifier then updates the one-time pad keys similar to Childs' protocol (see Subsection~\ref{subsubsect:childs}).

\item[\textbf{(4)}] When applying a Toffoli gate, the verifier asks the prover to measure $3$ blocks, comprising a magic state, in the computational basis and report the measurement outcomes. It is assumed that the magic state was entangled, using a Clifford operation, with $3$ target blocks on which the Toffoli gate is to be applied. The verifier undoes the (classical) one-time padding of the measurement outcomes and expects each of the $3$ groups of measurement outcomes (associated with each of the $3$ blocks) to be of the form $[ k_1 p(\alpha_1), ..., k_t p(\alpha_t)]$. The verifier then takes these classical strings and turns them into states of the form $\ket{\phi} = \ket{k_1 p(\alpha_1)} ... \ket{k_t p(\alpha_t)}$ (using her constant-sized quantum computer)\footnote{Note that no actual quantum state was returned to the verifier by the prover. Instead, she locally prepared a quantum state from the classical outcomes reported by the prover.}. She then applies $D_k^{\dagger}$ on each of these $\ket{\phi}$ states and checks that the last $d$ qudits, of each state, are $\ket{0}$, aborting otherwise.
Assuming not-abort, the verifier instructs the prover to perform the appropriate Pauli corrections resulting from the gate teleportation.

\item[\textbf{(5)}] Once all gates have been performed, the verifier instructs the prover to measure all blocks in the computational basis. As in step $4$, the verifier will then de-one-time pad the outcomes, apply $D_k^{\dagger}$ to each state of the form $\ket{\phi}$ (prepared from these outcomes), and check that the last $d$ qudits are $\ket{0}$, aborting otherwise.
\end{enumerate}
\noindent The protocol is schematically illustrated in Figure~\ref{fig:polyqas}.

As with the previous protocol, the security is based on the security of the authentication scheme. However, there is a significant difference. In the Clifford-QAS VQC protocol, one could always assume that the state received by the verifier was the correctly encoded state with a deviation on top that was independent of this encoding.
However, in the Poly-QAS VQC protocol, the quantum state is never returned to the verifier and, moreover, the prover's instructed actions on this state are adaptive based on the responses of the verifier. Since the prover is free to deviate at any point throughout the protocol, if we try to commute all of his deviations to the end (i.e. view the output state as the correct state resulting from an honest run of the protocol, with a deviation on top that is independent of the secret parameters), we find that the output state will have a deviation on top which depends on the verifier's responses. Since the verifier's responses depend on the secret keys, we cannot directly use the security of the authentication scheme to prove that the protocol is $2^{-d}$-verifiable.

The solution, as explained in \cite{abem}, is to consider the state of the entire protocol comprising of the prover's system, the verifier's system \emph{and} the transcript of all classical messages exchanged during the protocol. 
For a fixed interaction transcript, the prover's attacks can be commuted to the end of the protocol. This is because, if the transcript is fixed, there is no dependency of the prover's operations on the verifier's messages. We simply view all of his operations as unitaries acting on the joint system of his private memory, the input quantum state and the transcript.
One can then use Lemma~\ref{lemma:pauli} and Lemma~\ref{lemma:signedpoly} to bound the projection of this state onto the incorrect subspace with acceptance.
The whole state, however, will be a mixture of all possible interaction transcripts, but since each term is bounded and the probabilities of the terms in the mixture must add up to one, it follows that the protocol is $2^{-d}$-verifiable:

\begin{theorem}
For a fixed constant $d > 0$, Poly-QAS VQC is a prepare-and-send $\cQPIP$ protocol having correctness $\delta=1$ and verifiability $\epsilon = 2^{-d}$.
\end{theorem}

Let us briefly summarize the two protocols in terms of the verifier's resources. In both protocols, if one fixes the security parameter, $\epsilon$, the verifier must have a $O(log(1/\epsilon))$-size quantum computer. Additionally, both protocols are interactive with the total amount of communication (number of messages times the size of each message) being upper bounded by $O(|\mathcal{C}| \cdot log(1/\epsilon))$, where $\mathcal{C}$ is the quantum circuit to be performed\footnote{To be precise, the communication in the Poly-QAS VQC scheme is $O((n + L) \cdot log(1/\epsilon))$, where $n$ is the size of the input and $L$ is the number of Toffoli gates in $\mathcal{C}$.}. 
However, in Clifford-QAS VQC, this communication is quantum whereas in Poly-QAS VQC only one quantum message is sent at the beginning of the protocol and the rest of the interaction is classical.

Before ending this subsection, we also mention the result of Broadbent et al. from \cite{broadbent2013quantum}. This result generalises the use of quantum authentication codes for achieving verification of delegated quantum computation (not limited to decision problems). 
Moreover, the authors prove the security of these schemes in the \emph{universal composability framework}, which allows for secure composition of cryptographic protocols and primitives \cite{univcomp}.

\subsection{Trap-based verification} \label{subsect:fk}
In this subsection we discuss \emph{Verifiable Universal Blind Quantum Computing} (VUBQC), which was developed by Fitzsimons and Kashefi in \cite{fk}. 
The protocol is written in the language of MBQC and relies on two essential ideas. The first is that an MBQC computation can be performed blindly, using UBQC, as described in Subsection~\ref{subsect:blind}. The second is the idea of embedding checks or \emph{traps} in a computation in order to verify that it was performed correctly. Blindness will ensure that these checks remain hidden and so any deviation by the prover will have a high chance of triggering a trap.
Notice that this is similar to the QAS-based approaches where the input state has a flag subsystem appended to it in order to detect deviations and the whole state has been encoded in some way so as to hide the input and the flag subsystem.
This will lead to a similar proof of security. However, as we will see, the differences arising from using MBQC and UBQC lead to a reduction in the quantum resources of the verifier. In particular, in VUBQC the verifier requires only the ability to prepare single qubit states, which will be sent to the prover, in contrast to the QAS-based protocols which required the verifier to have a constant-size quantum computer. 

Recall the main steps for performing UBQC. The client, Alice, sends qubits of the form $\ket{+_{\theta_i}}$ to Bob, the server, and instructs him to entangle them according to a graph structure, $G$, corresponding to some universal graph state. She then asks him to measure qubits in this graph state at angles $\delta_i = \phi'_i + \theta_i + r_i \pi$, where $\phi'_i$ is the corrected computation angle and $r_i \pi$ acts a random $\zgate$ operation which flips the measurement outcome.
Alice will use the measurement outcomes, denoted $b_i$, provided by Bob to update the computation angles for future measurements.
Throughout the protocol, Bob's perspective is that the states, measurements and measurement outcomes are indistinguishable from random. Once all measurements have been performed, Alice will undo the $r_i$ padding of the final outcomes and recover her output.
Of course, UBQC does not provide any guarantee that the output she gets is the correct one, since Bob could have deviated from her instructions.

Transitioning to VUBQC, we will identify Alice as the verifier and Bob as the prover.
To augment UBQC with the ability to detect malicious behaviour on the prover's part, the verifier will introduce traps in the computation. How will she do this? Recall that the qubits which will comprise $\ket{G}$ need to be entangled with the $\cz$ operation. Of course, for $\xgate\ygate$-plane states $\cz$ does indeed entangle the states. However, if either qubit, on which
$\cz$ acts, is $\ket{0}$ or $\ket{1}$, then no entanglement is created. So suppose that we have a $\ket{+_{\theta}}$ qubit whose neighbours, according to $G$, are computational basis states. Then, this qubit will remain disentangled from the rest of the qubits in $\ket{G}$.
This means that if the qubit is measured at its preparation angle, the outcome will be deterministic. The verifier can exploit this fact to certify that the prover is performing the correct measurements.
Such states are referred to as \emph{trap qubits}, whereas the $\ket{0}$, $\ket{1}$ neighbours are referred to as \emph{dummy qubits}.
Importantly, as long as $G$'s structure remains that of a universal graph state\footnote{Note that adding dummy qubits into the graph will have the effect of disconnecting qubits that would otherwise have been connected. It is therefore important that the chosen graph state allows for the embedding of traps and dummies so that the desired computation can still be performed. For instance, the brickwork state from Subsection~\ref{subsect:mbqc} allows for only one trap qubit to be embedded, whereas other graph states allows for multiple traps. See \cite{fk, dottedtriplegraph} for more details.} and as long as the dummy qubits and the traps are chosen at random, adding these extra states as part of the UBQC computation will not affect the blindness of the protocol. The implication of this is that the prover will be completely unaware of the positions of the traps and dummies.
The traps effectively play a role that is similar to that of the flag subsystem in the authentication-based protocols. The dummies, on the other hand, are there to ensure that the traps do not get entangled with the rest of qubits in the graph state. They also serve another purpose. When a dummy is in a $\ket{1}$ state, and a $\sf{CZ}$ acts on it and a trap qubit, in the state $\ket{+_{\theta}}$, the effect is to ``flip'' the trap to $\ket{-_{\theta}}$ (alternatively $\ket{-_{\theta}}$ would have been flipped to $\ket{+_{\theta}}$). This means that if the trap is measured at its preparation angle, $\theta$, the measurement outcome will also be flipped, with respect to the initial preparation. Conversely, if the dummy was initially in the state $\ket{0}$, then no flip occurs. Traps and dummies, therefore, serve to also certify that the prover is performing the $\cz$ operations correctly. Thus, by using the traps (and the dummies), the verifier can check both the prover's measurements and his entangling operations and hence verify his MBQC computation.

We are now ready to present the steps of VUBQC:
\begin{enumerate}\addtolength{\itemsep}{+0.5\baselineskip}
\item[\textbf{(1)}] The verifier chooses an input $x$ and a quantum computation $\mathcal{C}$ that she would like the prover to perform on $\ket{x}$\footnote{As in the previous protocols, this need not be a classical input and the verifier could prepare an input of the form $\ket{\psi} = \ket{\psi_1} \otimes ... \otimes \ket{\psi_n}$.}.
\item[\textbf{(2)}] She converts $x$ and $\mathcal{C}$ into a pair $(G, \{\phi_i\}_i)$, where $\ket{G}$ is an $N$-qubit universal graph state (with an established ordering for measuring the qubits), which admits an embedding of $T$ traps and $D$ dummies. We therefore have that $N = T + D + Q$, where $Q = O(|\mathcal{C}|)$ is the number of computation qubits used for performing $\mathcal{C}$ and $\{\phi_i\}_{i \leq Q}$ is the associated set of computation angles\footnote{Note that the number of traps, $T$, and the number of dummies, $D$, are related, since each trap should have only dummy neighbours in $\ket{G}$.}.
\item[\textbf{(3)}] Alice picks, uniformly at random, values $\theta_i$, with $i$ going from $1$ to $T+Q$, from the set $\{0, \pi/4, 2\pi/4, ... 7\pi/4\}$ as well as values $r_i$ from the set $\{0, 1\}$ for the trap and computation qubits.
\item[\textbf{(4)}] She then prepares the $T+Q$ states $\ket{+_{\theta_i}}$, as well as $D$ dummy qubits which are states chosen at random from $\{ \ket{0}, \ket{1} \}$. All these states are sent to Bob, who is instructed to entangle them, using $\cz$ operations, according to the graph structure $G$.
\item[\textbf{(5)}] Alice then asks Bob to measure the qubits as follows: computation qubits will be measured at $\delta_i = \phi'_i + \theta_i + r_i \pi$, where $\phi'_i$ is an updated version of $\phi_i$ that incorporates corrections resulting from previous measurements; trap qubits will be measured at $\delta_i = \theta_i + r_i \pi$; dummy qubits are measured at randomly chosen angles from $\{0, \pi/4, 2\pi/4, ... 7\pi/4\}$. This step is interactive as Alice needs to update the angles of future measurements based on past outcomes. The number of rounds of interaction is proportional to the depth of $\mathcal{C}$. If any of the trap measurements produce incorrect outcomes, Alice will abort upon completion of the protocol.
\item[\textbf{(6)}] Assuming all trap measurements succeeded, after all the measurements have been performed, Alice undoes the $r_i$ one-time padding of the measurement outcomes, thus recovering the outcome of the computation.
\end{enumerate}

\begin{figure}[htbp!]
  \centering
  \includegraphics[width=0.7\linewidth]{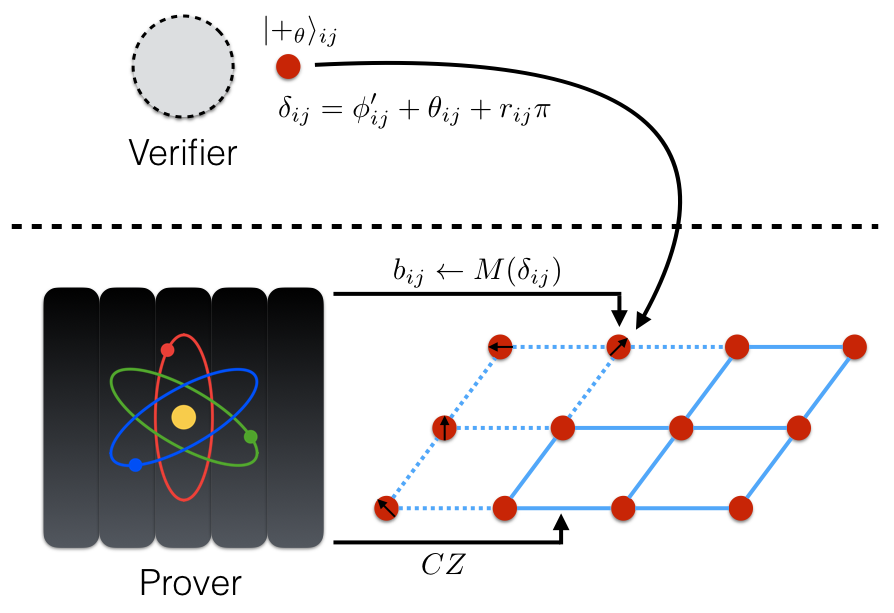}
  \caption{Verifiable Universal Blind Quantum Computing}
  \label{fig:fk}
\end{figure}

The protocol is illustrated schematically in Figure~\ref{fig:fk}, where all the parameters have been labelled by their position, $(i, j)$, in a rectangular cluster state.

One can see that VUBQC has correctness $\delta=1$, since if the prover behaves honestly then all trap measurements will produce the correct result and the computation will have been performed correctly. What about verifiability?
We will first answer this question for the case where there is a single trap qubit ($T=1$) at a uniformly random position in $\ket{G}$, denoted $\ket{+_{\theta_t}}$.
Adopting a similar notation to that from \cite{fk}, we let:
\begin{equation}
\mathcal{B}_j (\nu) = \sum_{\mathbf{s}} p_{\nu,j}(\mathbf{s}) \ket{\mathbf{s}}\bra{\mathbf{s}} \otimes \rho_{\nu, j}^{\mathbf{s}}
\end{equation}
denote the outcome density operator of \emph{all classical and quantum messages} exchanged between the verifier and the prover throughout the protocol, excluding the last round of measurements (which corresponds to measuring the output of the computation).
Additionally, $\nu$ denotes the set of secret parameters of Alice (i.e. the positions of the traps and dummies as well as the sets $\{ \phi_i \}_i$, $\{ \theta_i \}_i$ and $\{ r_i \}_i$); $j$ ranges over the possible strategies of the prover\footnote{Since the prover is unbounded and is free to choose any of the \emph{uncountably} many CPTP strategies, $j$ should be thought more of as a symbolic parameter indicating that there is a dependence on the prover's strategy and whether or not this strategy is the ideal one.} with $j=0$ corresponding to the honest strategy; $\mathbf{s}$ is a binary vector which ranges over all possible \emph{corrected} values of the measurement outcomes sent by the prover; lastly, $\rho_{\nu, j}^{\mathbf{s}}$ is the state of the unmeasured qubits, representing the output state of the computation (prior to the final measurement).
To match Definition~\ref{def:verifiability}, one also considers:
\begin{equation}
P^{\nu}_{incorrect} = (I - \mathcal{C} \ket{x}\bra{x} \mathcal{C}^{\dagger}) \otimes \ket{+^{\nu}_{\theta_t}} \bra{+^{\nu}_{\theta_t}}
\end{equation}
to be the projection onto the orthogonal complement of the correct output together with the trap state being projected onto acceptance. The dependence on $\nu$, for the trap qubit, arises because the acceptance outcome depends on the states of the dummy neighbors for that qubit. This is because if one of the dummies is $\ket{1}$, the $\cz$ operation has the effect of flipping $\ket{+_{\theta_t}}$ to $\ket{-_{\theta_t}}$. 
Additionally, $\nu$ also encodes the position of this trap, in the graph state, as well as the $\zgate$ flip specified by the $r_i$ parameter, for $i=t$.
One then needs to find an $\epsilon$ such that:
\begin{equation}
Tr \left( \sum_{\nu} p(\nu)P^{\nu}_{incorrect} \mathcal{B}_j(\nu) \right) \leq \epsilon
\end{equation}
This is done in a manner similar to the proof of security for the Poly-QAS VQC scheme of the previous section\footnote{Note that the security proof for Poly-QAS VQC was in fact inspired from that of the VUBQC protocol, as mentioned in \cite{abem}.}.
Specifically, one fixes the interaction transcript for the protocol.
This just means fixing the measurement angles $\delta_i$, and then considering all possible transcripts compatible with the fixed angles. One can do this because UBQC guarantees that the prover learns nothing from the interaction except for, at most, an upper bound on $|\mathcal{C}|$. This means that there will be multiple transcripts compatible with the same values for the $\delta_i$ angles.
It also means that any deviation that the prover performs is independent of the secret parameters of the verifier (though it can depend on the $\delta_i$ angles) and can therefore be commuted to the end of the protocol. The outcome density operator $\mathcal{B}_j(\nu)$ can then be expressed as the ideal outcome with a CPTP deviation, $\mathcal{E}_j$, on top, that is independent of $\nu$:
\begin{equation}
\mathcal{B}_j(\nu) = \mathcal{E}_j(\mathcal{B}_0(\nu))
\end{equation}
The deviation $\mathcal{E}_j$ is then decomposed into Kraus operators which, in turn, are decomposed into Pauli operators leading to:
\begin{equation}
\mathcal{B}_j(\nu) = \sum_{k,l,m} \alpha_{kl}(j)\alpha^*_{km}(j) \; P_l \mathcal{B}_0(\nu) P_m
\end{equation}
where $\alpha_{kl}(j)$ (and their conjugates) are the complex coefficients for the Pauli operators.
This summation can be split into the terms that act as identity on $\mathcal{B}_0(\nu)$ and those that do not. Suppose the terms that act trivially have weight $0 \leq \beta \leq 1$, we then have:
\begin{equation}
\mathcal{B}_j(\nu) = \beta\mathcal{B}_0(\nu) + (1 - \beta)\sum_{k,l,m} \alpha_{kl}(j)\alpha^*_{km}(j) \; P_l \mathcal{B}_0(\nu) P_m
\end{equation}
where the second term is summing over Pauli operators that act non-trivially. We use this to compute the probability of accepting an incorrect outcome, noting that $P_{incorrect}^{\nu} \mathcal{B}_0(\nu) = 0$:
\begin{equation}
Tr \left( \sum_{\nu} p(\nu)P^{\nu}_{incorrect} \mathcal{B}_j(\nu) \right) = 
(1 - \beta) Tr \left( \sum_{\nu} \sum_{k,l,m} p(\nu)P^{\nu}_{incorrect} ( \alpha_{kl}(j) \alpha^*_{km}(j) \; P_l \mathcal{B}_0(\nu) P_m) \right)
\end{equation}
We now use the fact that $P^{\nu}_{incorrect} = (I - \mathcal{C} \ket{x}\bra{x} \mathcal{C}^{\dagger}) \otimes \ket{+^{\nu}_{\theta_t}} \bra{+^{\nu}_{\theta_t}}$ and keep only the projection onto the trap qubit. The projection onto the space orthogonal to the correct state is a trace decreasing operation and also $(1 - \beta) \leq 1$ hence:
\begin{equation}
Tr \left( \sum_{\nu} p(\nu)P^{\nu}_{incorrect} \mathcal{B}_j(\nu) \right) \leq Tr \left( \sum_{\nu} p(\nu) \ket{+^{\nu}_{\theta_t}} \bra{+^{\nu}_{\theta_t}} \; \sum_{k,l,m} \alpha_{kl}(j)\alpha^*_{km}(j) \; P_l \mathcal{B}_0(\nu) P_m \right)
\end{equation}
The summation over $\nu$ can be broken into two summations: one over the position of the trap (and the dummies) and one over the remaining parameters. This latter sum makes the reduced state appear totally mixed to the prover (a fact which is ensured by UBQC). The above expression then becomes:
\begin{equation} \label{eqn:fksoundness}
Tr \left( \sum_{\nu^t} p(\nu^t) \ket{+^{\nu^t}_{\theta_t}} \bra{+^{\nu^t}_{\theta_t}} \; \sum_{k,l,m} \alpha_{kl}(j) \alpha^*_{km}(j) \; P_l (\ket{+_{\theta_t}^{\nu^t}} \bra{+_{\theta_t}^{\nu^t}} \otimes (I/Tr(I))) P_m \right)
\end{equation}
where $\nu^t$ denotes the secret parameters for the trap qubit and consists of $\theta_t$, $r_t$ and the position of the trap in the graph.
But notice that, on the identity system, the terms in which $l \neq m$ will have no contribution to the summation. This is because at least one of the Pauli terms (either $P_l$ or $P_m$) will act on the identity system. Since Pauli operators are traceless, when taking the trace these terms will be zero.
For the trap system we will have:
\begin{equation}
Tr \left(\sum_{\nu^t} p(\nu^t) \ket{+^{\nu^t}_{\theta_t}} \bra{+^{\nu^t}_{\theta_t}} \; P_l \ket{+_{\theta_t}^{\nu^t}} \bra{+_{\theta_t}^{\nu^t}} P_m \right) =
 \sum_{\nu^{t}} p(\nu^{t}) \bra{+^{\nu^{t}}_{\theta_t}} P_l \ket{+^{\nu^{t}}_{\theta_t}} \bra{+^{\nu^{t}}_{\theta_t}} P_m \ket{+^{\nu^{t}}_{\theta_t}}
\end{equation}
Note that we are taking $p(\nu^{t})$ to be the uniform distribution over these parameters. By summing over $\theta_t$ and $r_t$, the above expression becomes zero, whenever $l \neq m$. This is a result of the Pauli twirl Lemma~\ref{lemma:pauli}.
Thus, only terms in which $l = m$ will remain. Substituting this back into expression~\ref{eqn:fksoundness} leads to:
\begin{equation}
Tr \left( \sum_{\nu^t} p(\nu^t) \ket{+^{\nu^t}_{\theta_t}} \bra{+^{\nu^t}_{\theta_t}} \; \sum_{k,l} |\alpha_{kl}(j)|^2 \; P_l (\ket{+_{\theta_t}^{\nu^t}} \bra{+_{\theta_t}^{\nu^t}} \otimes (I/Tr(I))) P_l \right)
\end{equation}
In other words, the resulting state is a convex combination of Pauli deviations.
The position of the trap is completely randomised so that it is equally likely that any of the $N$ qubits is the trap. Therefore, in the above summation, there will be $N$ terms (corresponding to the $N$ possible positions of the trap), one of which will be zero (the one in which the non-trivial Pauli deviations act on the trap qubit). Hence:
\begin{equation}
Tr \left( \sum_{\nu} p(\nu)P^{\nu}_{incorrect} \mathcal{B}_j(\nu) \right) \leq \frac{N-1}{N} = 1 - \frac{1}{N}
\end{equation}
We have found that for the case of a single trap qubit, out of the total $N$ qubits, one has $\epsilon = 1 - \frac{1}{N}$.

If however, there are multiple trap states, the bound improves. Specifically, for a type of resource state called \emph{dotted-triple graph}, the number of traps can be a constant fraction of the total number of qubits, yielding $\epsilon = 8/9$. 
If the protocol is then repeated a constant number of times, $d$, with the verifier aborting if any of these runs gives incorrect trap outcomes, it can be shown that $\epsilon = (8/9)^{d}$ \cite{dottedtriplegraph}.
Alternatively, if the input state and computation are encoded in an error correcting code of distance $d$, then one again obtains $\epsilon = (8/9)^{d}$. This is useful if one is interested in a quantum output, or a classical bit string output. If, instead, one would only like a single bit output (i.e. the outcome of the decision problem) then sequential repetition and taking the majority outcome is sufficient. The fault tolerant encoding need not be done by the verifier. Instead, the prover will simply be instructed to prepare a larger resource state which also offers topological error-correction. See \cite{fk, rhg1, rhg2} for more details.
An important observation, however, is that the fault tolerant encoding, just like in the Poly-QAS VQC protocol, is used \emph{only to boost security} and not for correcting deviations arising from faulty devices. This latter case is discussed in Section~\ref{sect:ft}.
To sum up:

\begin{theorem}
For a fixed constant $d > 0$, VUBQC is a prepare-and-send $\cQPIP$ protocol having correctness $\delta=1$ and verifiability $\epsilon = (8/9)^d$.
\end{theorem}

It should be noted that in the original construction of the protocol, the fault tolerant encoding, used for boosting security, required the use of a resource state having $O(|\mathcal{C}|^2)$ qubits. The importance of the dotted-triple graph construction is that it achieves the same level of security while keeping the number of qubits linear in $|\mathcal{C}|$.
The same effect is achieved by a composite protocol which combines the Poly-QAS VQC scheme, from the previous section, with VUBQC \cite{kapourniotis2015optimising}.
This works by having the verifier run small instances of VUBQC in order to prepare the encoded blocks used in the Poly-QAS VQC protocol. Because of the blindness property, the prover does not learn the secret keys used in the encoded blocks. The verifier can then run the Poly-QAS VQC protocol with the prover, using those blocks.
This hybrid approach illustrates how composition can lead to more efficient protocols. In this case, the composite protocol maintains a single qubit preparation device for the verifier (as opposed to a $O(log(1/\epsilon))$-size quantum computer) while also achieving linear communication complexity. We will encounter other composite protocols when reviewing entanglement-based protocols in Section~\ref{sect:entanglement}.

Lastly, let us explicitly state the resources and overhead of the verifier throughout the VUBQC protocol. As mentioned, the verifier requires only a single-qubit preparation device, capable of preparing states of the form $\ket{+_{\theta}}$, with $\theta \in \{0, \pi/4, 2\pi/4, ... 7\pi/4 \}$, and $\ket{0}$, $\ket{1}$. The number of qubits needed is on the order of $O(|\mathcal{C}|)$. After the qubits have been sent to the prover, the two interact classically and the size of the communication is also on the order of $O(|\mathcal{C}|)$.

\subsection{Verification based on repeated runs} \label{subsect:howtoverify}
The final prepare-and-send protocol we describe is the one defined by Broadbent in \cite{broadbent}. While the previous approaches relied on hiding a flag subsystem or traps in either the input or the computation, this protocol has the verifier alternate between different runs designed to either test the behaviour of the prover or perform the desired quantum computation. We will refer to this as the \emph{Test-or-Compute} protocol.
From the prover's perspective, the possible runs are indistinguishable from each other, thus making him unaware if he is being tested or performing the verifier's chosen computation.
Specifically, suppose the verifier would like to delegate the quantum circuit $\mathcal{C}$ to be applied on the $\ket{0}^{\otimes n}$ state\footnote{The preparation of a specific input $\ket{x}$ can be done as part of the circuit $\mathcal{C}$.}, where $n$ is the size of the input. The verifier then chooses randomly between three possible runs:
\begin{itemize}
\item \textbf{Computation run}. The verifier delegates $\mathcal{C} \ket{0}^{\otimes n}$ to the prover. 
\item \textbf{$\xgate$-test run}. The verifier delegates the identity computation on the $\ket{0}^{\otimes n}$ state to the prover. 
\item \textbf{$\zgate$-test run}. The verifier delegates the identity computation on the $\ket{+}^{\otimes n}$ state to the prover.
\end{itemize}
It turns out that this suffices in order to test against any possible malicious behavior of the prover, with high probability.

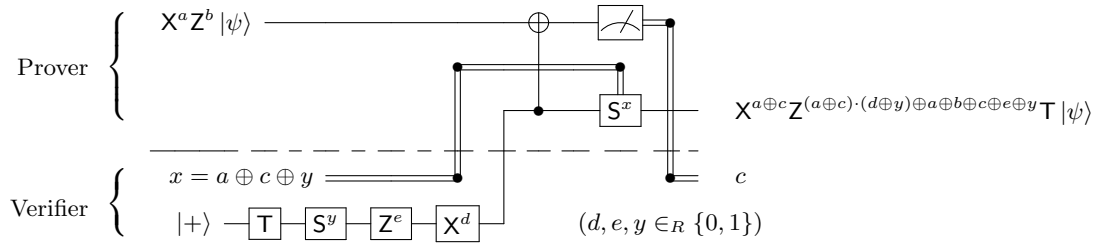
\begin{figure}[H]
\centerline{
 \Qcircuit @C=1em @R=1em  {
 %wire 1
&&&&&&\lstick{\xgate^a \zgate^b\ket{\psi}} & \qw & \qw & \qw & \qw & \targ &\qw    &  \meter & \control \cw \cwx[4]  &    &   \\
%wire 2
\lstick{\text{Prover}}&&& & && &&& \control  & \cw    & \cw
&   \cw & \control \cw  & &&\\
%wire 3
&&&& & & && &&& \ctrl{-2} &\qw    & \gate{\sgate^x} \cwx &  \qw  &
\qw& \rstick{\xgate^{a\oplus c}  \zgate^{(a \oplus c)\cdot (d \oplus y)   \oplus a \oplus b \oplus c \oplus e  \oplus y}  \tgate \ket{\psi}}\\
%wire dashed
%cheat!
&&&\dw&\dw&\dw&\dw& \dw & \dw & \dw &\dw & \dw
 &  \dw & \dw  & \dw &\dw&\\
%wire 4
\lstick{\raisebox{-1cm}{\text{Verifier}}}&& &&&& &\lstick{x = a\oplus c \oplus y } &\cw & \control \cw \cwx[-3] & &
&   & & \control &  \cw& \rstick{c} \\
%wire 5
&&&&& \lstick{\ket{+}}   & \gate{\tgate} & \gate{\sgate^y} & \gate{\zgate^{e}} &\gate{\xgate^d}&
\qw \qwx[-3]
 &   & & &\mbox{($d, e, y \in_R \{0,1\} $)} & &&
 \rstick{}
 \gategroup{1}{2}{3}{2}{0.7em}{\{}
 \gategroup{5}{2}{6}{2}{0.7em}{\{}
 }
 }
 \caption{Gadget for performing $\tgate$ on one-time padded state, reproduced from \cite{broadbent}}
 \label{fig:gadget}
\end{figure}

In more detail, the protocol uses a technique for quantum computing on encrypted data, described in \cite{qcompenc}, which is similar to Childs' protocol from Subsection~\ref{subsubsect:childs}, except it does not involve two-way quantum communication. The verifier will one-time pad either the $\ket{0}^{\otimes n}$ state or the $\ket{+}^{\otimes n}$ state and send the qubits to the prover. The prover is then instructed to apply the circuit $\mathcal{C}$, which consists of the gates $\xgate$, $\zgate$, $\hgate$, $\tgate$, $\cnot$. As we know, the Clifford operations commute with (or normalise) the one-time pad, so the verifier would only need to appropriately update the one-time pad to account for this. However, $\tgate$ gates do not commute with the pad. In particular, commuting them past the $\xgate$ gates introduces unwanted $\sgate$ operations. To resolve this issue, the verifier will use a particular gadget which will allow the prover to apply $\tgate$ and correct for $\sgate$ at the same time. This gadget is shown in Figure~\ref{fig:gadget}, reproduced from \cite{broadbent}.

The gadget works in a gate teleportation fashion. For each qubit, labelled $j$, on which the prover should apply a $\tgate$ gate, the verifier sends a qubit of the form $\xgate^d \zgate^c \sgate^y \tgate \ket{+}$, as well as the classical bit $x = a \oplus c \oplus y$, where $a$ is the $\xgate$ padding of qubit $j$ and $c$, $d$ and $y$ are chosen at random. The verifier then instructs the prover to apply a $\cnot$ between the sent qubit and qubit $j$, effectively entangling them, and then measure qubit $j$ in the computational basis. Lastly, the verifier instructs the prover to apply an $\sgate^x$ gate to the sent qubit. The end result is that this qubit will be the same as the de-one-time padded qubit $j$ but with a $\tgate$ and a new one-time pad acting on it.
Importantly, the new pad is kept secret from the prover.

The $\tgate$ gate gadget allows the verifier to control the application of either a non-Clifford operation or a Clifford operation through gate teleportation. 
For note that if the verifier does not apply a $\tgate$ gate on the qubit sent to the prover, the resulting computation is Clifford. This is what allows the verifier to switch between the computation run and the two test runs. The prover cannot distinguish between the two cases, since his side of the gadget is identical in both instances. Thus, in a test run, the computation the prover performs will be Clifford and the verifier can simply update the one-time pad, of the input, accordingly. There is, however, one complication. In an $\xgate$-test run, the input is $\ket{0}^{\otimes n}$ and should remain this way until the end of the circuit, up to qubit flips resulting from the one-time pad. But any Hadamard gate in the circuit will map $\ket{0}$ to $\ket{+}$. The same is true for $\zgate$-test runs, where $\ket{+}$ states can be mapped to $\ket{0}$.
To resolve this issue, Broadbent uses the following identities:
\begin{equation} \label{eqn:h}
\hgate \tgate^2 \hgate \tgate^2 \hgate \tgate^2 \hgate = \hgate
\end{equation}
\begin{equation} \label{eqn:id}
\hgate \hgate \hgate \hgate = I
\end{equation}
The idea is to have the prover implement each Hadamard operation in $\mathcal{C}$ by applying four $\hgate$ gates alternating with $\sgate = \tgate^2$ gates. Each $\tgate^2$ operation is performed by using the $\tgate$ gate gadget twice. When the verifier chooses a computation run, she will apply the $\tgate$ gates in the gadget and therefore, via Equation~\ref{eqn:h}, this leads to a Hadamard operation. Conversely, in a rest run, no $\tgate$ gates are applied, hence, from Equation~\ref{eqn:id}, no Hadamard operation will act on the target qubit.
Since the output is always measured, by the prover, in the computational basis, in an $\xgate$-test run the verifier simply checks that the de-one-time padded output is $\ket{0}^{\otimes n}$.

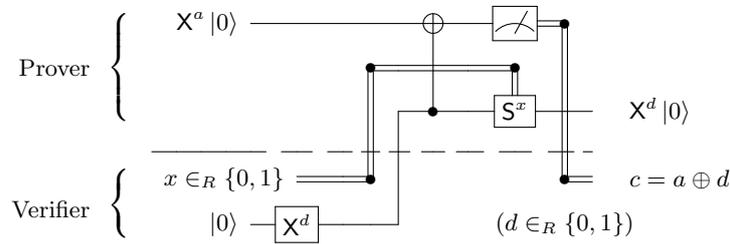
\begin{figure}[H]
\centerline{
 \Qcircuit @C=1em @R=1em  {
 %wire 1
&&&&&&\lstick{\xgate^a\ket{0}} & \qw & \qw & \qw & \qw & \targ &\qw    &  \meter & \control \cw \cwx[4]  &    &   \\
%wire 2
\lstick{\text{Prover}}&&& & && &&& \control  & \cw    & \cw
&   \cw & \control \cw  & &&\\
%wire 3
&&&& & & && &&& \ctrl{-2} &\qw    & \gate{\sgate^x} \cwx &  \qw  &
\qw& \rstick{\xgate^d \ket{0}}\\
%wire dashed
%cheat!
&&&\dw&\dw&\dw&\dw& \dw & \dw & \dw &\dw & \dw
 &  \dw & \dw  & \dw &\dw&\\
%wire 4
\lstick{\raisebox{-1cm}{\text{Verifier}}}&& &&&& &\lstick{x \in_R \{0,1\}} &\cw & \control \cw \cwx[-3] & &
&   & & \control &  \cw& \rstick{c = a \oplus d} \\
%wire 5
&&&&& &\lstick{\ket{0}}   & \gate{\xgate^d} & \qw & \qw &
\qw \qwx[-3]
 &   & & &\mbox{($d \in_R \{0,1\} $)} & &&
 \rstick{}
 \gategroup{1}{2}{3}{2}{0.7em}{\{}
 \gategroup{5}{2}{6}{2}{0.7em}{\{}
 }
 }
 \caption{$\tgate$ gate gadget in $\xgate$-test run, reproduced from \cite{broadbent}}
 \label{fig:xgadget}
\end{figure}

There is, in fact, an additional testing step being performed during an $\xgate$-test run. Consider the $\tgate$ gadget for such a run in Figure~\ref{fig:xgadget}, reproduced from \cite{broadbent}.

Note that the measurement bit, $c$, provided by the prover to the verifier should be an $xor$ of the original $\xgate$ padding of the input and the updated $\xgate$ padding of the input. Checking the value of this bit allows the verifier to test that the gadget was applied correctly.

What about the $\zgate$-test run? In that case, the output should be the $\ket{+}^{\otimes n}$ which, upon measurement, should collapse with equal probability into any of the $2^n$ possible binary strings.
The verifier does not test for this. Instead, each $\hgate$ gate that should be applied by the prover has the effect of switching between the $\zgate$-test run and the $\xgate$-test run. Thus, whenever a Hadamard is applied to a qubit, during a $\zgate$-test run, the verifier switches to an $\xgate$-test run until the next Hadamard operation. In the $\xgate$-test runs, the verifier will use the $\tgate$ gate gadget from Figure~\ref{fig:xgadget}. These are the only checks that are performed in the $\zgate$-test run.

Putting everything together, the protocol starts by having the verifier choose an input $x$ and a quantum computation $\mathcal{C}$ to act on this input (that includes preparing the input from the $\ket{0}^{\otimes n}$ state). Then, the verifier will randomly choose to perform one of the following:
\begin{itemize}\addtolength{\itemsep}{+0.5\baselineskip}
\item \textbf{Computation run}. The verifier one-time pads the $\ket{0}^{\otimes n}$ state and sends it to the prover. The prover is then instructed to apply $\mathcal{C}$ on this state, such that for each $\tgate$ gate in the circuit the prover and the verifier interact in order to perform the $\tgate$ gadget. Additionally, any $\hgate$ in $\mathcal{C}$ is performed as in Equation~\ref{eqn:h}.
For Clifford operations, the verifier updates the one-time pad of the state accordingly. The prover is instructed to measure the output state of the circuit in the computational basis and return the outcome to the verifier. The verifier undoes the padding of this outcome and accepts if the output of the circuit indicates acceptance.
\item \textbf{$\xgate$-test run}. The verifier one-time pads the $\ket{0}^{\otimes n}$ state and sends it to the prover. As in the computation run, for each $\tgate$, the verifier and the prover will interact to run the $\tgate$ gate gadget. In this case, however, the verifier will use the $\tgate$ gate gadget from Figure~\ref{fig:xgadget}, making the circuit effectively act as identity and checking that the prover is performing these gadgets correctly (rejecting otherwise). 
Additionally, the $\hgate$ gates in $\mathcal{C}$ will also act as identity, from Equation~\ref{eqn:id}, as described previously.
The verifier updates the one-time padding of the state accordingly for all gates in the circuit. Once the circuit is finished, the prover is instructed to measure the output in the computational basis and report the outcome to the verifier. The verifier accepts if the de-one-time padded output is $\ket{0}^{\otimes n}$.
\item \textbf{$\zgate$-test run}. The verifier one-time pads the $\ket{+}^{\otimes n}$ state and sends it to the prover. As in the $\xgate$-test run, the $\tgate$ gate gadgets will act as identity. 
The $\hgate$ operations that the prover performs will temporarily switch the $\zgate$-test run into an $\xgate$-test run, in which the verifier uses the gadget from Figure~\ref{fig:xgadget} to check that prover implemented it correctly. Any subsequent $\hgate$ will switch back to a $\zgate$-test run.
Additionally, the verifier updates the one-time padding of the state accordingly for all gates in the circuit. The prover is instructed to measure the output in the computational basis and report the outcome to the verifier, however in this case the verifier discards the output.
\end{itemize}

The asymmetry between the $\xgate$-test run and the $\zgate$-test run stems from the fact that the output is always measured in the computational basis. This means that an incorrect output is one which has been bit-flipped. In turn, this implies that only $\xgate$ and $\ygate$ operations on the output will act as deviations, since $\zgate$ effectively acts as identity on computational basis states. If the circuit $\mathcal{C}$ does not contain any Hadamard gates and hence, the computation takes place entirely in the computational basis, then the $\xgate$-test is sufficient for detecting such deviations. 
However, when Hadamard gates are present, this is no longer the case since deviations can occur in the conjugate basis, $(\ket{+}, \ket{-})$, as well. This is why the $\zgate$-test is necessary. Its purpose is to check that the prover's operations are performed correctly when switching to the conjugate basis. For this reason, a Hadamard gate will switch a $\zgate$-test run into an $\xgate$-test run which provides verification using the $\tgate$ gate gadget.

In terms of the correctness of the protocol, we can see that if the prover behaves honestly then the correct outcome is obtained in the computation run and the verifier will accept the test runs, hence $\delta = 1$\footnote{However, note that if the verifier chooses a test run, in the case where the prover is honest, this will lead to acceptance irrespective of the outcome of the decision problem. This is in contrast to the previous protocols in which the testing is performed at the same time as the computation and, when the test succeeds, the verifier outputs the result of the computation.}. 
For verifiability, the analysis is similar to the previous protocols. Suppose that $\ket{\psi}$ is either the $\ket{0}^{\oplus n}$ or the $\ket{+}^{\oplus n}$ state representing the input. Additionally, assuming there are $t$ $\tgate$ gates in $\mathcal{C}$ (including the ones used for performing the Hadamards), let $\ket{\phi}$ be the state of the $t$ qubits that the verifier sends for the $\tgate$ gate gadgets. Then, the one-time padded state that the prover receives is:
\begin{equation}
\frac{1}{4^{n+t}}\sum_{Q \in \mathds{P}_{n+t}} Q \ket{\psi}\bra{\psi} \otimes \ket{\phi}\bra{\phi} Q^{\dagger}
\end{equation}
The prover is then instructed to follow the steps of the protocol in order to run the circuit $\mathcal{C}$. Note that all of the operations that he is instructed to perform are Clifford operations. This is because any non-Clifford operation from $\mathcal{C}$ is performed with the $\tgate$ gate gadgets (which require only Clifford operations) and the states from $\ket{\phi}$, prepared by the verifier.
In following the notation from \cite{broadbent}, we denote the honest action of the protocol as $C$. As in the previous protocols, the prover's deviation can be commuted to the end of the protocol, so that it acts on top of the correct state. After expressing the deviation map in terms of Pauli operators one gets:
\begin{equation}
\frac{1}{4^{n+t}}\sum_{P_i,P_j,Q \in \mathds{P}_{n+t}} \alpha_i \alpha_j^* \; P_i Q C\ket{\psi}\bra{\psi} \otimes \ket{\phi}\bra{\phi}C^{\dagger} Q^{\dagger} P_j^{\dagger}
\end{equation}
Note that we have also commuted $C$ past the one-time pad so that it acts on the state $\ket{\psi}\ket{\phi}$, rather than on the one-time padded versions of these states. This is possible precisely because $C$ is a Clifford operation and therefore normalises Pauli operations.
One can then assume that the verifier performs the decryption of the padding before the final measurement, yielding:
\begin{equation}
\frac{1}{4^{n+t}}\sum_{P_i,P_j,Q \in \mathds{P}_{n+t}} \alpha_i \alpha_j^* \; Q^{\dagger} P_i Q C\ket{\psi}\bra{\psi} \otimes \ket{\phi}\bra{\phi}C^{\dagger} Q^{\dagger} P_j^{\dagger} Q
\end{equation}
We now use the Pauli twirl from Lemma~\ref{lemma:pauli} to get:
\begin{equation}
\frac{1}{4^{n+t}}\sum_{P_i \in \mathds{P}_{n+t}} |\alpha_i|^2 \; P_i C\ket{\psi}\bra{\psi} \otimes \ket{\phi}\bra{\phi}C^{\dagger} P_i^{\dagger}
\end{equation}
which is a convex combination of Pauli attacks acting on the correct output state.
If we now denote $M$ to be the set of non-benign Pauli attacks (i.e. attacks which do not act as identity on the output of the computation), then one of the test runs will reject with probability:
\begin{equation}
\sum_{P_i \in M} |\alpha_i|^2
\end{equation}
This is because non-benign Pauli $\xgate$ or $\ygate$ operations are detected by the $\xgate$-test run, whereas non-benign Pauli $\zgate$ operations are detected by the $\zgate$-test run. Since either test occurs with probability $1/3$, it follows that, the probability of the verifier accepting an incorrect outcome is at most $2/3$, hence $\epsilon = 2/3$.

Note that when discussing the correctness and verifiability of the Test-or-Compute protocol, we have slightly abused the terminology, since this protocol does not rigorously match the established definitions for correctness and verifiability that we have used for the previous protocols. The reason for this is the fact that in the Test-or-Compute protocol there is no additional flag or trap subsystem to indicate failure. Rather, the verifier detects malicious behaviour by alternating between different runs. 
It is therefore more appropriate to view the Test-or-Compute protocol simply as a $\cQPIP$ protocol having a constant gap between completeness and soundness:

\begin{theorem}
Test-or-Compute is a prepare-and-send $\cQPIP$ protocol having completeness $8/9$ and soundness $7/9$.
\end{theorem}

\noindent In terms of the verifier's quantum resources, we notice that, as with the VUBQC protocol, the only requirement is the preparation of single qubit states. All of these states are sent in the first round of the protocol, the rest of the interaction being completely classical.

\subsection{Summary of prepare-and-send protocols}
The protocols, while different, have the common feature that they all use blindness or have the potential to be blind protocols. Out of the five presented protocols, only the Poly-QAS VQC and the Test-or-Compute protocols are not explicitly blind since, in both cases, the computation is revealed to the server. However, it is relatively easy to make the protocols blind by encoding the circuit into the input (which is one-time padded). Hence, one can say that all protocols achieve blindness.

This feature is essential in the proof of security for these protocols. Blindness combined with either the Pauli twirl Lemma~\ref{lemma:pauli} or the Clifford twirl Lemma~\ref{lemma:cliff} have the effect of reducing any deviation of the prover to a convex combination of Pauli attacks. Each protocol then has a specific way of detecting such an attack. 
In the Clifford-QAS VQC protocol, the convex combination is turned into a uniform combination and the attack is detected by a flag subsystem associated with a quantum authentication scheme. A similar approach is employed in the Poly-QAS VQC protocol, using a quantum authentication scheme based on a special type of quantum error correcting code. The VUBQC protocol utilizes trap qubits and either sequential repetition or encoding in an error correcting code to detect Pauli attacks.
Finally, the Test-or-Compute protocol uses a hidden identity computation acting on either the $\ket{0}^{\otimes n}$ or $\ket{+}^{\otimes n}$ states, in order to detect the malicious behavior of the prover.

\begin{table}[htb]
\centering
\begin{tabular}{l*{4}{c}}
\toprule
 \bfseries Protocol  \hspace{0.1in} & \bfseries Verifier resources \hspace{0.1in} & \bfseries Communication \hspace{0.1in} & \bfseries 2-way quantum comm. \\
\midrule
Clifford-QAS VQC      & $O(log(1/\epsilon))$               &  $O(N \cdot log(1/\epsilon))$             & Y               \\
Poly-QAS VQC          & $O(log(1/\epsilon))$               &  $O((n + L) \cdot log(1/\epsilon))$             & N               \\
VUBQC                 & $O(1)$                             &  $O(N \cdot log(1/\epsilon))$         & N               \\
Test-or-Compute            & $O(1)$                             &  $O((n + T) \cdot log(1/\epsilon))$    & N          \\ \bottomrule
\end{tabular}\par
\bigskip
\captionof{table}{Comparison of prepare-and-send protocols. 
If we denote as $\mathcal{C}$ the circuit that the verifier wishes to delegate to the prover, and as $x$ the input to this circuit, then $n = |x|$, $N = |\mathcal{C}|$. Additionally, $T$ denotes the number of $\tgate$ gates in $\mathcal{C}$, $L$ denotes the number of Toffoli gates in $\mathcal{C}$ and $\epsilon$ denotes the verifiability of the protocols. The second column refers to the verifier's \emph{quantum} resources. The third column quantifies the \emph{total} communication complexity, both classical and quantum, of the protocols (i.e. number of messages times the size of a message).} \label{tab:prepsendcomp}
\end{table}

Because of these differences, each protocol will have different ``quantum requirements'' for the verifier. For instance, in the authentication-based protocols, the verifier is assumed to be a quantum computer operating on a quantum memory of size $O(log(1/\epsilon))$, where $\epsilon$ is the desired verifiability of the protocol. 
In VUBQC and Test-or-Compute, however, the verifier only requires a device capable of preparing single-qubit states.
Additionally, out of all of these protocols, only Clifford-QAS VQC requires $2$-way quantum communication, whereas the other three require the verifier to send only one quantum message at the beginning of the protocol, while the rest of the communication is classical.
These facts, together with the communication complexities of the protocols are shown in Table~\ref{tab:prepsendcomp}.

As mentioned, if we want to make the Poly-QAS VQC and Test-or-Compute protocols blind, the verifier will hide her circuit by incorporating it into the input. The input would then consist of an encoding of $\mathcal{C}$ and an encoding of $x$. The prover would be asked to perform controlled operations from the part of the input containing the description of $\mathcal{C}$, to the part containing $x$, effectively acting with $\mathcal{C}$ on $x$.
We stress that in this case, the protocols would have a communication complexity of $O(|\mathcal{C}| \cdot log(1/\epsilon))$, just like VUBQC and Clifford-QAS VQC\footnote{Technically, the complexity should be $O((|x| + |\mathcal{C}|) \cdot log(1/\epsilon))$, however we are assuming that $\mathcal{C}$ acts non-trivially on $x$ (i.e. there are at least $|x|$ gates in $\mathcal{C}$).}.

\section{Receive-and-measure protocols} \label{sect:recvmeas}
\begin{figure}[htbp!]
  \centering
  \includegraphics[width=0.7\linewidth]{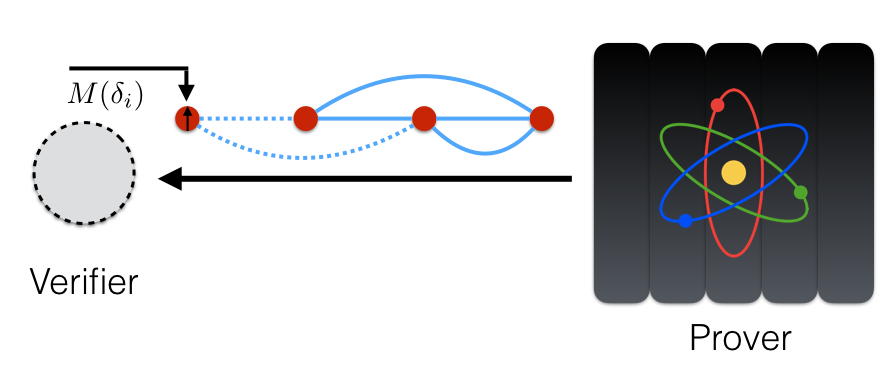}
  \caption{Receive-and-measure protocols}
  \label{fig:recv}
\end{figure}

The protocols presented so far have utilized a verifier with a trusted preparation device (and potentially a trusted quantum memory) interacting with a prover having the capability of storing and performing operations on arbitrarily large quantum systems. In this section, we explore protocols in which the verifier possesses a trusted measurement device. The point of these protocols is to have the prover prepare a specific quantum state and send it to the verifier. The verifier's measurements have the effect of either performing the quantum computation or extracting the outcome of the computation.
An illustration of receive-and-measure protocols is shown in Figure~\ref{fig:recv}.

For prepare-and-send protocols we saw that blindness was an essential feature for achieving verifiability. While most of the receive-and-measure protocols are blind as well, we will see that it is possible to perform verification without hiding any information about the input or computation, from the prover.
Additionally, while in prepare-and-send protocols the verifier was sending an encoded or encrypted quantum state to the prover, in receive-and-measure protocols, the quantum state received by the verifier is not necessarily encoded or encrypted. Moreover, this state need not contain a flag or a trap subsystem.
For this reason, we can no longer consistently define $\epsilon$-verifiability and $\delta$-correctness, as we did for prepare-and-send protocols. Instead, we will simply view receive-and-measure protocols as $\cQPIP$ protocols.

The protocols presented in this section are:
\begin{enumerate}
\item \textbf{Subsection~\ref{subsect:monly}}: a \emph{measurement-only} protocol developed by Morimae and Hayashi that employs ideas from MBQC in order to perform verification \cite{monly}.
\item \textbf{Subsection~\ref{subsect:posthoc}}: a \emph{post hoc} verification protocol, developed by Morimae and Fitzsimons \cite{posthoc,posthocpublished} (and independently by Hangleiter et al \cite{hangleiter2017direct}).
\end{enumerate}

There is an additional receive-and-measure protocol by Gheorghiu, Wallden and Kashefi \cite{gwk} which we refer to as \emph{Steering-based VUBQC}. That protocol, however, is similar to the entanglement-based GKW protocol from Subsection~\ref{subsect:GKW}. We will therefore review Steering-based VUBQC in that subsection by comparing it to the entanglement-based protocol.

\subsection{Measurement-only verification} \label{subsect:monly}
In this section we discuss the measurement-only protocol from \cite{monly}, which we shall simply refer to as the \emph{measurement-only protocol}. 
This protocol uses MBQC to perform the quantum computation, like the VUBQC protocol from Subsection~\ref{subsect:fk}, however the manner in which verification is performed is more akin to Broabdent's Test-or-Compute protocol, from Subsection~\ref{subsect:howtoverify}. This is because, just like in the Test-or-Compute protocol, the measurement-only approach has the verifier alternate between performing the computation or testing the prover's operations.

The key idea for this protocol, is the fact that graph states can be completely specified by a set of stabilizer operators. This fact is explained in Subsection~\ref{subsect:mbqc}. To reiterate the main points, recall that for a graph $G$, with associated graph state $\ket{G}$, if we denote as $V(G)$ the set of vertices in $G$ and as $N_G(v)$ the set of neighbours for a given vertex $v$, then the generators for the stabilizer group of $\ket{G}$ are:
\begin{equation}
K_v = \xgate_v \prod_{w \in N_G(v)} \zgate_w
\end{equation}
for all $v \in V(G)$. In other words, the $K_v$ operators generate the entire group of operators, $O$, such that $O\ket{G} = \ket{G}$.

When viewed as observables, stabilizers allow one to test that an unknown quantum state is in fact a particular graph state $\ket{G}$, with high probability. This is done by measuring random stabilizers of $\ket{G}$ on multiple copies of the unknown quantum state. If all measurements return the $+1$ outcome, then, the unknown state is close in trace distance to $\ket{G}$. This is related to a concept known as \emph{self-testing}, which is the idea of determining whether an unknown quantum state and an unknown set of observables are close to a target state and observables, based on observed statistics. We postpone a further discussion of this topic to the next section, since self-testing is ubiquitous in entanglement-based protocols.

\begin{figure}[htbp!]
    \centering
    \begin{subfigure}[b]{0.45\textwidth}
        \includegraphics[width=\textwidth]{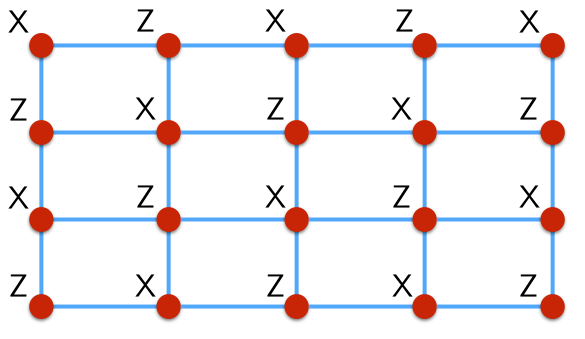}
        \caption{$\xgate\zgate$ group measurement operators}
        \label{fig:xz}
    \end{subfigure}
    \hfill
    \begin{subfigure}[b]{0.45\textwidth}
        \includegraphics[width=\textwidth]{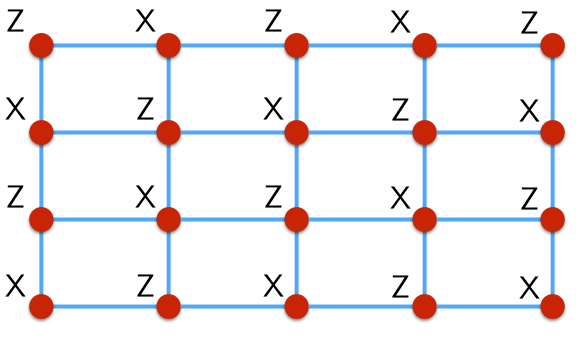}
        \caption{$\zgate\xgate$ group measurement operators}
        \label{fig:zx}
    \end{subfigure}
    \caption{Stabilizer measurements}\label{fig:stabs}
\end{figure}

As mentioned, the measurement-only protocol involves a testing phase and a computation phase. The prover will be instructed to prepare multiple copies of a 2D cluster state, $\ket{G}$, and send them, qubit by qubit, to the verifier. The verifier will then randomly use one of these copies to perform the MBQC computation, whereas the other copies are used for testing that the correct cluster state was prepared\footnote{This is very much in the spirit of a cryptographic technique known as \emph{cut-and-choose} \cite{crepeau2011cut}, which has also been used in the context of testing quantum states \cite{qcutandchoose}.}.
This testing phase will involve checking all possible stabilizers of $\ket{G}$. In particular, the verifier will divide the copies to be tested into two groups, which we shall refer to as the $\xgate\zgate$ group and the $\zgate\xgate$ group.
In the $\xgate\zgate$ group of states, the verifier will measure the qubits according to the 2D cluster structure, starting with an $\xgate$ operator in the upper left corner of the lattice and then alternating between $\xgate$ and $\zgate$.
In the $\zgate\xgate$ group, she swaps $\xgate$ and $\zgate$ and once again measures the qubits according to the 2D cluster structure. The two cases are illustrated in Figure~\ref{fig:stabs}.

Together, the measurement outcomes of the two groups can be used to infer outcomes of all stabilizer measurements defined by the $K_v$ operators.
For instance, given that the measurement outcomes for the qubits take values $\pm 1$, to compute the outcome of a $K_v$ measurement, for some node $v$ that is measured with $\mathsf{X}$, the verifier simply takes product of the measurement outcomes for all nodes in $\{v\} \cup N_v$.
These tests allow the verifier to certify that the prover is indeed preparing copies of the state $\ket{G}$. She can then use one of these copies to run the computation. Since the prover does not know which state the verifier will use for the computation, any deviation he implements has a high chance of being detected by one of the verifier's tests.
Hence, the protocol works as follows:
\begin{enumerate}\addtolength{\itemsep}{+0.5\baselineskip}
\item[\textbf{(1)}] The verifier chooses an input $x$ and a quantum computation $\mathcal{C}$.
\item[\textbf{(2)}] She instructs the prover to prepare $2k + 1$ copies of a 2D cluster state, $\ket{G}$, for some constant $k$, and send all of the qubits, one at a time, to the verifier.
\item[\textbf{(3)}] The verifier randomly picks one copy to run the computation of $\mathcal{C}$ on $x$ in an MBQC fashion. The remaining $2k$ copies are randomly divided into the $\xgate\zgate$ groups and the $\zgate\xgate$ group and measured, as described above, so as to check the stabilizers of $\ket{G}$.
\item[\textbf{(4)}] If all stabilizer measurement outcomes are successful (i.e. produced the outcome $+1$), then the verifier accepts the outcome of the computation, otherwise she rejects.
\end{enumerate}

As with all protocols, completeness follows immediately, since if the prover behaves honestly, the verifier will accept the outcome of the computation. In the case of soundness, Hayashi and Morimae treat the problem as a \emph{hypothesis test}. 
In other words, in the testing phase of the protocol the verifier is checking the hypothesis that the prover prepared $2k+1$ copies of the state $\ket{G}$. 
They then prove the following theorem:
\begin{theorem}
Let $1/(2k+1) \leq \alpha \leq 1$ be the verifier's confidence level in the testing phase of the measurement-only protocol. Then, the state used by the verifier for the computation, denoted $\rho$, satisfies:
\begin{equation} \label{eqn:hyp}
\bra{G} \rho \ket{G} \geq 1 - \frac{1}{\alpha(2k+1)}
\end{equation}
\end{theorem}
This theorem is essentially showing that as the number of copies of $\ket{G}$, requested by the verifier, increases, and the verifier accepts in the testing phase, one gets that the state $\rho$, used by the verifier for the computation, is close in trace distance to the ideal state, $\ket{G}$.
The confidence level, $\alpha$, represents the maximum acceptance probability for the verifier, such that the computation state, $\rho$, \emph{does not} satisfy Equation~\ref{eqn:hyp}. 
Essentially this represents the probability for the verifier to accept a computation state that is far from ideal.
Hayashi and Morimae argue that the lower bound, $\alpha \geq 1/(2k+1)$, is tight, because if the prover corrupts one of the $2k + 1$ states sent to the verifier, there is a $1/(2k+1)$ chance that that state will not be tested and the verifier accepts.

If one now denotes with $C$ the POVM that the verifier applies on the computation state in order to perform the computation of $\mathcal{C}$, then it is the case that:
\begin{equation}
\lvert Tr( C \rho) - Tr(C \ket{G}\bra{G}) \rvert \leq \frac{1}{\sqrt{\alpha(2k+1)}}
\end{equation}
What this means is that the distribution of measurement outcomes for the state $\rho$, sent by the prover in the computation run, is almost indistinguishable from the distribution of measurement outcomes for the ideal state $\ket{G}$.
The soundness of the protocol is therefore upper bounded by $\frac{1}{\sqrt{\alpha(2k+1)}}$. This implies that to achieve soundness below $\epsilon$, for some $\epsilon > 0$, the number of copies that the prover would have to prepare scales as $O \left( \frac{1}{\alpha} \cdot \frac{1}{\epsilon^2} \right)$.

In terms of the quantum capabilities of the verifier, she only requires a single qubit measurement device capable of measuring the observables: $\xgate, \ygate, \zgate, (\xgate + \ygate)/\sqrt{2}, (\xgate - \ygate)/\sqrt{2}$.
Recently, however, Morimae, Takeuchi and Hayashi have proposed a similar protocol which uses \emph{hypergraph states} \cite{hypergraph}. These states have the property that one can perform universal quantum computations by measuring only the Pauli observables ($\xgate$, $\ygate$ and $\zgate$).
Hypergraph states are generalizations of graph states in which the vertices of the graph are linked by hyperedges, which can connect more than two vertices. Hence, the entangling of qubits is done with a generalized $\cz$ operation involving multiple qubits. The protocol itself is similar to the one from \cite{monly}, as the prover is required to prepare many copies of a hypergraph state and send them to the verifier. The verifier will then test all but one of these states using stabilizer measurements and use the remaining one to perform the MBQC computation. 
For a computation, $\mathcal{C}$, the protocol has completeness lower bounded by $1 - |\mathcal{C}| e^{-|\mathcal{C}|}$ and soundness upper bounded by $1/\sqrt{|\mathcal{C}|}$.
The communication complexity is higher than the previous measurement-only protocol, as the prover needs to send $O(|\mathcal{C}|^{21})$ copies of the $O(|\mathcal{C}|)$-qubit graph state, leading to a total communication cost of $O(|\mathcal{C}|^{22})$.
We end with the following result:

\begin{theorem}
The measurement-only protocols are receive-and-measure $\cQPIP$ protocols having an inverse polynomial gap between completeness and soundness.
\end{theorem}

\subsection{Post hoc verification} \label{subsect:posthoc}
The protocols we have reviewed so far have all been based on cryptographic primitives. There were reasons to believe, in fact, that any quantum verification protocol would have to use some form of encryption or hiding. This is due to the parallels between verification and authentication, which were outlined in Section~\ref{sect:prepsend}. However, it was shown that this is not the
 case when Morimae and Fitzsimons, and independently Hangleiter et al, proposed a protocol for \emph{post hoc} quantum verification \cite{posthoc, hangleiter2017direct}. The name ``post hoc'' refers to the fact that the protocol is not interactive, requiring a single round of back and forth communication between the prover and the verifier. Moreover, verification is performed after the computation has been carried out.
It should be mentioned that the first post hoc protocol was proposed in \cite{fh}, by Fitzsimons and Hajdu{\v{s}}ek,
however, that protocol utilizes multiple quantum provers, and we review it in Subsection~\ref{subsect:entposthoc}.

In this section, we will present the post hoc verification approach, refered to as \emph{1S-Post-hoc}, from the perspective of the Morimae and Fitzsimons paper \cite{posthoc}. The reason for choosing their approach, over the Hangleiter et al one, is that the entanglement-based post hoc protocols, from Subsection~\ref{subsect:entposthoc}, are also described using similar terminology to the Morimae and Fitzsimons paper.
The protocol of Hangleiter et al is essentially identical to the Morimae and Fitzsimons one, except it is presented from the perspective of certifying the ground state of a gapped, local Hamiltonian. Their certification procedure is then used to devise a verification protocol for a class of quantum simulation experiments, with the purpose of demonstrating a quantum computational advantage \cite{hangleiter2017direct}.

The starting point is the complexity class $\sf{QMA}$, for which we have stated the definition in Subsection~\ref{subsect:complexity}.
Recall, that one can think of $\sf{QMA}$ as the class of problems for which the solution can be checked by a $\cBQP$ verifier receiving a quantum state $\ket{\psi}$, known as a witness, from a prover.
We also stated the definition of the $k$-local Hamiltonian problem, a complete problem for the class $\cQMA$, in Definition~\ref{def:LH}.
We mentioned that for $k=2$ the problem is $\cQMA$-complete \cite{kempe2006complexity}. For the post hoc protocol, Morimae and Fitzsimons consider a particular type of $2$-local Hamiltonian known as an $\xgate\zgate$-Hamiltonian.

To define an $\xgate\zgate$-Hamiltonian we introduce some helpful notation. Consider an $n$-qubit operator $S$, which we shall refer to as $\xgate\zgate$-term, such that $S = \bigotimes_{j=1}^{n} P_j$, with $P_j \in \{I, \xgate, \zgate\}$. Denote $w_X(S)$ as the $\xgate$-weight of $S$, representing the total number of $j$'s for which $P_j = \xgate$. Similarly denote $w_Z(S)$ as the $\zgate$-weight for $S$.
An $\xgate\zgate$-Hamiltonian is then a $2$-local Hamiltonian of the form $H = \sum_i a_i S_i$, where the $a_i$'s are real numbers and the $S_i$'s are $\xgate\zgate$-terms having $w_X(S_i) + w_Z(S_i) \leq 2$. 

The 1S-Post-hoc protocol starts with the observation that $\sf{BQP} \subseteq \sf{QMA}$. This means that any problem in $\sf{BQP}$ can be viewed as an instance of the $2$-local Hamiltonian problem.
Therefore, for any language $L \in \sf{BQP}$ and input $x$, there exists an $\xgate\zgate$-Hamiltonian, $H$, such that the smallest eigenvalue of $H$ is less than $a$ when $x \in L$ or larger than $b$, when $x \not\in L$, where $a$ and $b$ are a pair of numbers satisfying $b - a \geq 1/poly(|x|)$.
Hence, the lowest energy eigenstate of $H$ (also referred to as \emph{ground state}), denoted $\ket{\psi}$, is a quantum witness for $x \in L$. In a $\cQMA$ protocol, the prover would be instructed to send this state to the verifier. The verifier then performs a measurement on $\ket{\psi}$ to estimate its energy, accepting if the estimate is below $a$ and rejecting otherwise.
However, we are interested in a verification protocol for $\cBQP$ problems where the verifier has minimal quantum capabilities. This means that there will be two requirements: the verifier can only perform single-qubit measurements; the prover is restricted to $\cBQP$ computations.
The 1S-Post-hoc protocol satisfies both of these constraints.

The first requirement is satisfied because estimating the energy of a quantum state, $\ket{\psi}$, with respect to an $\xgate\zgate$-Hamiltonian $H$, can be done by measuring one of the observables $S_i$ on the state $\ket{\psi}$.
Specifically, it is shown in \cite{morimae2016quantum} that if one chooses the local term $S_i$ according to a probability distribution given by the normalized terms $|a_i|$, and measures $\ket{\psi}$ with the $S_i$ observables, this provides an estimate for the energy of $\ket{\psi}$.
Since $H$ is an $\xgate\zgate$-Hamiltonian, this entails performing at most two measurements, each of which can be either an $\xgate$ measurement or a $\zgate$ measurement.
This implies that the verifier need only perform single-qubit measurements.

For the second requirement, one needs to show that for any $\cBQP$ computation, there exists an $\xgate\zgate$-Hamiltonian such that the ground state can be prepared by a polynomial-size quantum circuit.
Suppose the computation that the verifier would like to delegate is denoted as $\mathcal{C}$ and the input for this computation is $x$. Given what we have mentioned above, regarding the local Hamiltonian problem, it follows that there exists an $\xgate\zgate$-Hamiltonian $H$ and numbers $a$ and $b$, with $b - a \geq 1/poly(|x|)$, such that if $\mathcal{C}$ accepts $x$ with high probability then the ground state of $H$ has energy below $a$, otherwise it has energy above $b$.
It was shown in \cite{kitaev2002classical,kempe2006complexity,2local}, that starting from $\mathcal{C}$ and $x$ one can construct an $\xgate\zgate$-Hamiltonian satisfying this property and which also has a ground state that can be prepared by a $\cBQP$ machine.
The ground state is known as the \emph{Feynman-Kitaev clock state}. 
To describe this state, suppose the circuit $\mathcal{C}$ has $T$ gates (i.e. $T=|\mathcal{C}|$) and that these gates, labelled in the order in which they are applied, are denoted $\{U_i\}_{i=0}^T$. For $i=0$ we assume $U_0 = I$. The Feynman-Kitaev state is the following:
\begin{equation}
\ket{\psi} = \frac{1}{\sqrt{T + 1}} \sum\limits_{t=0}^{T} U_t U_{t-1} ... U_0 \ket{x} \ket{1^{t} 0^{T - t}}
\end{equation}
This is essentially a superposition over all time steps, $t$, of the time evolved state in the circuit $\mathcal{C}$. Hence, the state can be prepared by a $\cBQP$ machine.
The $\xgate\zgate$-Hamiltonian is then a series of $2$-local constraints that are all simultaneously satisfied by this state.

We can now present the steps of the 1S-Post-hoc protocol:
\begin{enumerate}\addtolength{\itemsep}{+0.5\baselineskip}
\item[\textbf{(1)}] The verifier chooses a quantum circuit, $\mathcal{C}$, and an input $x$ to delegate to the prover.
\item[\textbf{(1)}] The verifier computes the terms $a_i$ of the $\xgate\zgate$-Hamiltonian, $H = \sum_i a_i S_i $, having as a ground state the Feynman-Kitaev state associated with $\mathcal{C}$ and $x$, denoted $\ket{\psi}$.
\item[\textbf{(2)}] The verifier instructs the prover to send her $\ket{\psi}$, qubit by qubit.
\item[\textbf{(4)}] The verifier chooses one of the $\xgate\zgate$-terms $S_i$, according to the normalized distribution $\{|a_i|\}_i$,  and measures it on $\ket{\psi}$. She accepts if the measurement indicates the energy of $\ket{\psi}$ is below $a$.
\end{enumerate}
\noindent Note that the protocol is not blind, since the verifier informs the prover about both the computation $\mathcal{C}$ and the input $x$.

As mentioned, the essential properties that any $\cQPIP$ protocol should satisfy are completeness and soundness. For the post hoc protocol, these follow immediately from the local Hamiltonian problem. Specifically, we know that there exist $a$ and $b$ such that $b - a \geq 1/poly(|x|)$. When $\mathcal{C}$ accepts $x$ with high probability, the state $\ket{\psi}$ will be an eigenstate of $H$ having eigenvalue smaller than $a$. Otherwise, any state, when measured under the $H$ observable, will have an energy greater than $b$. Of course, the verifier is not computing the exact energy $\ket{\psi}$ under $H$, merely an estimate. This is because she is measuring only one local term from $H$. However, it is shown in \cite{posthoc} that the precision of her estimate is also inverse polynomial in $|x|$. Therefore:

\begin{theorem}
1S-Post-hoc is a receive-and-measure $\cQPIP$ protocol having an inverse polynomial gap between completeness and soundness.
\end{theorem}

The only quantum capability of the verifier is the ability to measure single qubits in the computational and Hadamard bases (i.e. measuring the $\zgate$ and $\xgate$ observables). The protocol, as described, suggests that it is sufficient for the verifier to measure only two qubits. However, since the completeness-soundness gap decreases with the size of the input, in practice one would perform a sequential repetition of this protocol in order to boost this gap. It is easy to see that, for a protocol with a completeness-soundness gap of $1/p(|x|)$, for some polynomial $p$, in order to achieve a constant gap of at least $1 - \epsilon$, where $\epsilon > 0$, the protocol needs to be repeated $O(p(|x|) \cdot log(1/\epsilon))$ times. It is shown in \cite{hangleiter2017direct,gap} that $p(|x|)$ is $O(|\mathcal{C}|^2)$, hence the protocol should be repeated $O(|\mathcal{C}|^2 \cdot log(1/\epsilon))$ times and this also gives us the total number of measurements for the verifier\footnote{As a side note, the total number of measurements is not the same as the communication complexity for this protocol, since the prover would have to send $O(|\mathcal{C}|^3 \cdot log(1/\epsilon))$ qubits in total. This is because, for each repetition, the prover sends a state of $O(|\mathcal{C}|)$ qubits, but the verifier only measures $2$ qubits from each such state.}. 
Note, however, that this assumes that each run of the protocol is independent of the previous one (in other words, that the states sent by the prover to the verifier in each run are uncorrelated). Therefore, the $O(|\mathcal{C}|^2 \cdot log(1/\epsilon))$ overhead should be taken as an i.i.d. (independent and identically distributed states) estimate.
This is, in fact, mentioned explicitly in the Hangleiter et al result, where they explain that the prover should prepare ``\emph{a number of independent and identical copies of a quantum state}'' \cite{hangleiter2017direct}.
Thus, when considering the most general case of a malicious prover that does not obey the i.i.d. constraint, one requires a more thorough analysis involving non-independent runs, as is done in the measurement-only protocol \cite{monly} or the steering-based VUBQC protocol \cite{gwk}.

\subsection{Summary of receive-and-measure protocols}
Receive-and-measure protocols are quite varied in the way in which they perform verification. The measurement-only protocols use stabilizers to test that the prover prepared a correct graph state and then has the verifier use this state to perform an MBQC computation. The 1S-Post-hoc protocol relies on the entirely different approach of estimating the ground state energy of a local Hamiltonian. Lastly, the steering-based VUBQC protocol, which we detail in Subsection~\ref{subsect:GKW}, is different from these other two approaches by having the verifier remotely prepare the VUBQC states on the prover's side and then doing trap-based verification.
Having such varied techniques leads to significant differences in the total number of measurements performed by the verifier, as we illustrate in Table~\ref{tab:recvandmeas}.

\begin{table}[htb] 
\centering
\begin{tabular}{l*{3}{c}} 
\toprule
 \bfseries Protocol  \hspace{0.1in} & \bfseries Measurements \hspace{0.1in} & \bfseries Observables \hspace{0.1in} &  \bfseries Blind \\
\midrule
Measurement-only                          & $O(N \cdot 1/\alpha \cdot 1/\epsilon^2)$         &  5                  & Y          \\
Hypergraph measurement-only               & $O(max(N, 1/\epsilon^2)^{22})$         &  3                                               & Y          \\
1S-Post-hoc                                & $O(N^2 \cdot log(1/\epsilon))$                   &  2                  & N          \\
Steering-based VUBQC \hspace{0.1in}       & $O(N^{13} log(N) \cdot log(1/\epsilon))$         &  5                  & Y          \\ \bottomrule
\end{tabular}
\begin{center}
\caption{Comparison of receive-and-measure protocols. We denote $N = |\mathcal{C}|$ to be the size of the delegated quantum computation. The number of measurements is computed for a target gap between completeness and soundness of $1 - \epsilon$ for some constant $\epsilon > 0$. For the first measurement-only protocol, $\alpha$ denotes the confidence level of the verifier in the hypothesis test.}
\label{tab:recvandmeas}
\end{center}
\end{table}

Of course, the number of measurements is not the only metric we use in comparing the protocols. Another important aspect is
how many observables the verifier should be able to measure. The 1S-Post-hoc protocol is optimal in that sense, since the verifier need only measure $\xgate$ and $\zgate$ observables. Next is the hypergraph state measurement-only protocol which requires all three Pauli observables. Lastly, the other two protocols require the verifier to be able to measure the $\xgate\ygate$-plane observables $\xgate$, $\ygate$, $(\xgate+\ygate)/\sqrt{2}$ and $(\xgate-\ygate)/\sqrt{2}$ plus the $\zgate$ observable.

Finally, we compare the protocols in terms of blindness, which we have seen plays an important role in prepare-and-send protocols. For receive-and-measure protocols, the 1S-Post-hoc protocol is the only one that is not blind. 
While this is our first example of a verification protocol that does not hide the computation and input from the prover, it is not the only one. In the next section, we review two other post hoc protocols that are also not blind.

\section{Entanglement-based protocols} \label{sect:entanglement}
The protocols discussed in the previous sections have been either prepare-and-send or receive-and-measure protocols. Both types employ a verifier with some minimal quantum capabilities interacting with a single $\cBQP$ prover.
In this section we explore protocols which utilize multiple non-communicating provers that share entanglement and a fully classical verifier. The main idea will be for the verifier to distribute a quantum computation among many provers and verify its correct execution from correlations among the responses of the provers. 

We classify the entanglement-based approaches as follows:
\begin{enumerate}
\item \textbf{Subsection~\ref{subsect:ruv}} three protocols which make use of the CHSH game, the first one developed by Reichardt, Unger and Vazirani \cite{ruv}, the second by Gheorghiu, Kashefi and Wallden \cite{gkw} and the third by Hajdu{\v{s}}ek, P{\'e}rez-Delgado and Fitzsimons.
\item \textbf{Subsection~\ref{subsect:mckague}} a protocol based on self-testing graph states, developed by McKague \cite{mckague}.
\item \textbf{Subsection~\ref{subsect:entposthoc}} two post hoc protocols, one developed by Fitzsimons and Hajdu{\v{s}}ek \cite{posthoc} and another by Natarajan and Vidick \cite{nv}.
\end{enumerate}

Unlike the previous sections where, for the most part, each protocol was based on a different underlying idea for performing verification, entanglement-based protocols are either based on some form of rigid self-testing or on testing local Hamiltonians via the post hoc approach. In fact, as we will see, even the post hoc approaches employ self-testing. Of course, there are distinguishing features within each of these broad categories, but due to their technical specificity, we choose to label the protocols in this section by the initials of the authors.

Since self-testing plays such a crucial role in entanglement-based protocols, let us provide a brief description of the concept.
The idea of self-testing was introduced by Mayers and Yao in \cite{mayersyao}, and is concerned with characterising the shared quantum state and observables of $n$ non-communicating players in a \emph{non-local game}. A non-local game is one in which a referee (which we will later identify with the verifier) will ask questions to the $n$ players (which we will identify with the provers) and, based on their responses, decide whether they win the game or not. Importantly, we are interested in games where there is a quantum strategy that outperforms a classical strategy. By a classical strategy, we mean that the players can only produce local correlations\footnote{To define local correlations, consider a setting with two players, Alice and Bob. Each player receives an input, $x$ for Alice and $y$ for Bob and produces an output, denoted $a$ for Alice and $b$ for Bob. We say that the players' responses are locally correlated if: $Pr(a,b|x,y) = \sum_{\lambda} Pr(a | x, \lambda) Pr(b | y, \lambda) Pr(\lambda)$. Where $\lambda$ is known as a \emph{hidden variable}. In other words, given this hidden variable, the players' responses depend only on their local inputs.}. Conversely, in a quantum strategy, the players are allowed to share entanglement in order to produce non-local correlations and achieve a higher win rate. Even so, there is a limit to how well the players can perform in the game. In other words, the optimal quantum strategy has a certain probability of winning the game, which may be less than $1$.
Self-testing results are concerned with non-local games in which the optimal quantum strategy is \emph{unique}, up to local isometries on the players' systems. This means that if the referee observes a near maximal win rate for the players, in the game, she can conclude that they are using the optimal strategy and can therefore characterise their shared state and their observables, up to a local isometries.
More formally, we give the definition of self-testing, adapted from \cite{coladangelo2017separation} and using notation similar to that of \cite{nv}:

\begin{definition}[Self-testing]
Let $G$ denote a game involving $n$ non-communicating players denoted $\{ P_i \}_{i=1}^n$. Each player will receive a question from a set, $Q$ and reply with an answer from a set $A$. There exists some condition establishing which combinations of answers to the questions constitutes a win for the game. Let $\omega^*(G)$ denote the maximum winning probability of the game for players obeying quantum mechanics.

The players provide their responses by implementing a measurement strategy $S = (\ket{\psi}, \{ O_i^j \}_{ij})$ consisting of a state $\ket{\psi}$ shared among the $n$ players and local observables $\{ O_i^j \}_j$, for each player $P_i$.
We say that the game $G$ \textbf{self-tests} the strategy $S$, with robustness $\epsilon = \epsilon(\delta)$, for some $\delta > 0$, if, for any strategy $\tilde{S} = (\ket{\tilde{\psi}}, \{ \tilde{O}_i^j \}_{ij})$ achieving winning probability $\omega^*(G) - \epsilon$ there exists a local isometry $\Phi = \bigotimes_{i=1}^{n} \Phi_i$ and a state $\ket{junk}$ such that:
\begin{equation}
TD(\Phi(\ket{\tilde{\psi}}), \ket{junk}\ket{\psi}) \leq \delta
\end{equation}
and for all $j$:
\begin{equation}
TD \left( \Phi \left( \bigotimes_{i=1}^{n} \tilde{O}_i^j \ket{\tilde{\psi}}   \right), \ket{junk} \bigotimes_{i=1}^{n} O_i^j \ket{\psi}  \right) \leq \delta
\end{equation}
\end{definition}
Note that $TD$ denotes trace distance, and is defined in Subsection~\ref{subsubsect:td}.

\subsection{Verification based on CHSH rigidity} \label{subsect:ruv}
\subsubsection{RUV protocol.} \label{subsect:RUV}
In \cite{tsirelson}, Tsirelson gave an upper bound for the total amount of non-local correlations shared between two non-communicating parties, as predicted by quantum mechanics. In particular, consider a two-player game consisting of Alice and Bob. Alice is given a binary input, labelled $a$, and Bob is given a binary input, labelled $b$. They each must produce a binary output and we label Alice's output as $x$ and Bob's output as $y$. Alice and Bob win the game iff $a \cdot b = x \oplus y$.
The two are \emph{not} allowed to communicate during the game, however they are allowed to share classical or quantum correlations (in the form of entangled states).
This defines a non-local game known as the \emph{CHSH game} \cite{chsh}. The optimal \emph{classical} strategy for winning the game achieves a success probability of $75\%$, whereas, what Tsirelson proved, is that any \emph{quantum} strategy achieves a success probability of at most $cos^2(\pi/8) \approx 85.3\%$. This maximal winning probability, in the quantum case, can in fact be achieved by having Alice and Bob do the following. First, they will share the state $\Ket{\Phi_+} = (\Ket{00} + \Ket{11}) / \sqrt{2}$. If Alice receives input $a=0$, then she will measure the Pauli $\xgate$ observable on her half of the $\Ket{\Phi_+}$ state, otherwise (when $a=1$) she measures the Pauli $\zgate$ observable. Bob, on input $b=0$ measures $(\xgate+\zgate)/\sqrt{2}$, on his half of the Bell pair, and on input $b=1$, he measures $(\xgate-\zgate)/\sqrt{2}$. 
We refer to this strategy as the \emph{optimal quantum strategy} for the CHSH game.

\begin{figure}[htbp!]
\centering
\includegraphics[scale=0.4]{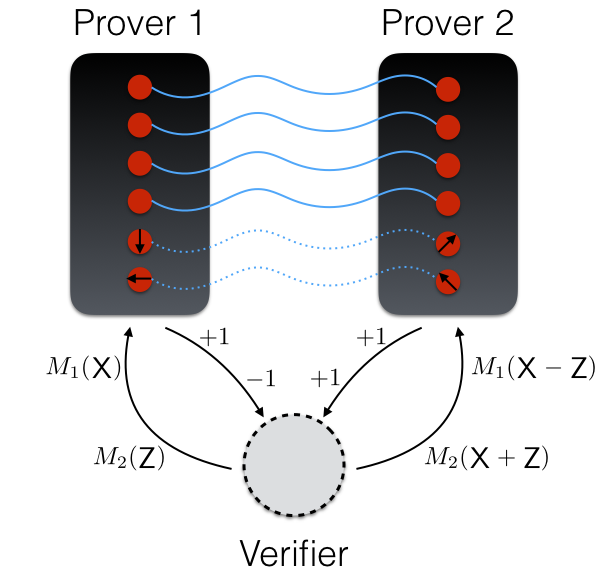}
\caption{Ideal CHSH game strategy}
\label{fig:ruv}
\end{figure}

McKague, Yang and Scarani proved a converse of Tsierlson's result, by showing that if one observes two players winning the CHSH game with a near $cos^2(\pi/8)$ probability, then it can be concluded that the players' shared state is close to a Bell pair and their observables are close to the ideal observables of the optimal strategy (Pauli $\xgate$ and $\zgate$, for Alice, and $(\xgate + \zgate)/\sqrt{2}$ and $(\xgate - \zgate)/\sqrt{2}$, for Bob) \cite{mckague2012robust}. This is effectively a self-test for a Bell pair.
Reichardt, Unger and Vazirani then proved a more general result for self-testing a \emph{tensor product} of multiple Bell states as well as the observables acting on these states \cite{ruv}\footnote{Note that the McKague, Yang and Scarani result could also be used to certify a tensor product of Bell pairs, by repeating the self-test of a single Bell pair multiple times. However, this would require each repetition to be independent of the previous one. In other words the states shared by Alice and Bob, as well as their measurement outcomes, should be independent and identically distributed (i.i.d.) in each repetition. The Reichardt, Unger and Vazirani result makes no such assumption.}. It is this latter result that is relevant for the RUV protocol so we give a more formal statement for it:
\begin{theorem} \label{thm:rigidity}
Suppose two players, Alice and Bob, are instructed to play $n$ sequential CHSH games.
Let the inputs, for Alice and Bob, be given by the $n$-bit strings $\mathbf{a}, \mathbf{b} \in \{0, 1\}^n$.
Additionally, let $S = (\ket{\tilde{\psi}}, \tilde{A}(\mathbf{a}), \tilde{B}(\mathbf{b}))$ be the strategy employed by Alice and Bob in playing the $n$ CHSH games, where $\ket{\tilde{\psi}}$ is their shared state and $\tilde{A}(\mathbf{a})$ and $\tilde{B}(\mathbf{b})$ are their respective observables, for inputs $\mathbf{a}, \mathbf{b}$. 

Suppose Alice and Bob win at least $n(1-\epsilon)cos^2(\pi/8)$ games, with $\epsilon = poly(\delta, 1/n)$ for some $\delta > 0$, such that $\epsilon \rightarrow 0$ as $\delta \rightarrow 0$ or $n \rightarrow \infty$. Then, there exist a local isometry $\Phi = \Phi_A \otimes \Phi_B$ and a state $\ket{junk}$ such that:
\begin{equation}
TD(\Phi(\ket{\tilde{\psi}}), \ket{junk} \ket{\Phi_+}^{\otimes n}) \leq \delta
\end{equation}
and:
\begin{equation}
TD \left( \Phi \left( \tilde{A}(\mathbf{a}) \otimes \tilde{B}(\mathbf{b}) \ket{\tilde{\psi}} \right), \ket{junk} A(\mathbf{a}) \otimes B(\mathbf{b}) \ket{\Phi_+}^{\otimes n} \right) \leq \delta
\end{equation}
where $A(\mathbf{a}) = \bigotimes\limits_{i=1}^{n} P(\mathbf{a(i)})$, $B(\mathbf{b}) = \bigotimes\limits_{i=1}^{n} Q(\mathbf{b(i)})$ and $P(0)=\xgate$, $P(1)=\zgate$, $Q(0)=(\xgate+\zgate)/\sqrt{2}$, $Q(1)=(\xgate-\zgate)/\sqrt{2}$.
\end{theorem}

What this means is that, up to a local isometry, the players share a state which is close in trace distance to a tensor product of Bell pairs and their measurements are close to the ideal measurements.
This result, known as \emph{CHSH game rigidity}, is the key idea for performing multi-prover verification using a classical verifier. We will refer to the protocol in this section as the \emph{RUV protocol}.

Before giving the description of the protocol, let us first look at an example of \emph{gate teleportation}, which we also mentioned when presenting the Poly-QAS VQC protocol of Subsection~\ref{subsect:polyqas}. Suppose two parties, Alice and Bob, share a Bell state $\ket{\Phi_+}$. Bob applies a unitary $U$ on his share of the entangled state so that the joint state becomes $(I \otimes U) \ket{\Phi_+}$. Alice now takes an additional qubit, labelled $\ket{\psi}$ and measures this qubit and the one from the $\ket{\Phi_+}$ state in the Bell basis given by the states:
\begin{equation*}
\ket{\Phi_+} = \frac{\ket{00} + \ket{11}}{\sqrt{2}} \;\;\;\;
\ket{\Phi_-} = \frac{\ket{00} - \ket{11}}{\sqrt{2}}
\end{equation*}
\begin{equation*}
\ket{\Psi_+} = \frac{\ket{01} + \ket{10}}{\sqrt{2}} \;\;\;\;
\ket{\Psi_-} = \frac{\ket{01} - \ket{10}}{\sqrt{2}}
\end{equation*}
The outcome of this measurement will be two classical bits which we label $b_1$ and $b_2$. After the measurement, the state on Bob's system will be $U \xgate^{b_1} \zgate^{b_2} \ket{\psi}$. Essentially, Bob has a one-time padded version of $\ket{\psi}$ with the $U$ gate applied.

We now describe the RUV protocol. It uses two quantum provers but can be generalized to any number of provers greater than two. Suppose that Alice and Bob are the two provers. They are allowed to share an unbounded amount of quantum entanglement but are not allowed to communicate during the protocol. A verifier will interact classically with both of them in order to delegate and check an arbitrary quantum computation specified by the quantum circuit $\mathcal{C}$. The protocol consists in alternating randomly between four sub-protocols:

\begin{itemize}\addtolength{\itemsep}{+0.5\baselineskip}
\item \textbf{CHSH games.} In this subprotocol, the verifier will simply play CHSH games with Alice and Bob. To be precise, the verifier will repeatedly instruct Alice and Bob to perform the ideal measurements of the CHSH game. She will collect the answers of the two provers (which we shall refer to as CHSH statistics) and after a certain number of games, will compute the win rate of the two provers. The verifier is interested in the case when Alice and Bob win close to the maximum number of games as predicted by quantum mechanics. Thus, at the start of the protocol she takes $\epsilon = poly(1/|\mathcal{C}|)$ and accepts the statistics produced by Alice and Bob if and only if they win at least a fraction $(1 - \epsilon)cos^2(\pi/8)$ of the total number of games. Using the rigidity result, this implies that Alice and Bob share a state which is close to a tensor product of perfect Bell states (up to a local isometry). This step is schematically illustrated in Figure~\ref{fig:ruv}.
\item \textbf{State tomography.} This time the verifier will instruct Alice to perform the ideal CHSH game measurements, as in the previous case. However, she instructs Bob to measure his halves of the entangled states so that they collapse to a set of \emph{resource states} which will be used to perform gate teleportation. The resource states are chosen so that they are universal for quantum computation. 
Specifically, in the RUV protocol, the following resource states are used: $\{ \mathsf{P}\Ket{0},  (\mathsf{HP})_2 \Ket{\Phi_+}, (\mathsf{GY})_2 \Ket{\Phi_+},
\cnot_{2,4}\mathsf{P}_2 \mathsf{Q}_4 (\Ket{\Phi_+} \otimes \Ket{\Phi_+}) : \mathsf{P}, \mathsf{Q} \in \{\xgate, \ygate, \zgate, I \}  \}$, where $\mathsf{G} = exp \left( -i \frac{\pi}{8} \ygate\right)$ and the subscripts indicate on which qubits the operators act. 
Assuming Alice and Bob do indeed share Bell states, Bob's measurements will collapse Alice's states to the same resource states (up to a one-time padding known to the verifier). Alice's measurements on these states are used to check Bob's preparation, effectively performing state tomography on the resource states.
\item \textbf{Process tomography.} This subprotocol is similar to the state tomography one, except the roles of Alice and Bob are reversed. The verifier instructs Bob to perform the ideal CHSH game measurements. Alice, on the other hand, is instructed to perform Bell basis measurements on pairs of qubits. As in the previous subprotocol, Bob's measurement outcomes are used to tomographically check that Alice is indeed performing the correct measurements. 
\item \textbf{Computation.} The final subprotocol combines the previous two. Bob is asked to perform the resource preparation measurements, while Alice is asked to perform Bell basis measurements. This effectively makes Alice perform the desired computation through repeated gate teleportation.
\end{itemize}

An important aspect, in proving the correctness of the protocol, is the local similarity of pairs of subprotocols. For instance, Alice cannot distinguish between the CHSH subprotocol and the state tomography one, or between the process tomography one and computation. This is because, in those situations, she is asked to perform the same operations on her side, while being unaware of what Bob is doing. Moreover, since the verifier can test all but the computation part, if Alice deviates there will be a high probability of her deviation being detected. The same is true for Bob. In this way, the verifier can, essentially, enforce that the two players behave honestly and thus perform the correct quantum computation.
Note, that this is not the same as the blindness property, discussed in relation to the previous protocols. The RUV protocol does, however, posses that property as well. This follows from a more involved argument regarding the way in which the computation by teleportation is performed.

It should be noted that there are only two constraints imposed on the provers: that they cannot communicate once the protocol has commenced and that they produce close to quantum optimal win-rates for the CHSH games.
Importantly, there are no constraints on the quantum systems possessed by the provers, which can be arbitrarily large. Similarly, there are no constraints on what measurements they perform or what strategy they use in order to respond to the verifier.
In spite of this, the rigidity result shows that for the provers to produce statistics that are accepted by the verifier, they must behave according to the ideal strategy (up to local isometry).
Having the ability to fully characterise the prover's shared state and their strategies in this way is what allows the verifier to check the correctness of the delegated quantum computation.
This approach, of giving a full characterisation of the states and observables of the provers, is a powerful technique which is employed by all the other entanglement-based protocols, as we will see.

In terms of practically implementing such a protocol, there are two main considerations: the amount of communication required between the verfier and the provers and the required quantum capabilities of the provers.
For the latter, it is easy to see that the RUV protocol requires both provers to be universal quantum computers (i.e. $\cBQP$ machines), having the ability to store multiple quantum states and perform quantum circuits on these states. 
In terms of the communication complexity, since the verifier is restricted to $\cBPP$, the amount of communication must scale polynomially with the size of the delegated computation.
It was computed in \cite{gkw}, that this communication complexity is of the order $O(|\mathcal{C}|^c)$, with $c > 8192$. Without even considering the constant factors involved, this scaling is far too large for any sort of practical implementation in the near future\footnote{However, with added assumptions (such as i.i.d. states and measurement statistics for the two provers), the scaling can become small enough that experimental testing is possible. A proof of concept experiment of this is realized in \cite{huang2017}.}.

There are essentially two reasons for the large exponent in the scaling of the communication complexity. The first, as mentioned by the authors, is that the bounds derived in the rigidity result are not tight and could possibly be improved.
The second and, arguably more important reason, stems from the rigidity result itself. In Theorem~\ref{thm:rigidity}, notice that $\epsilon = poly(\delta, 1/n)$ and $\epsilon \rightarrow 0$ as $n \rightarrow \infty$. We also know that the provers need to win a fraction $(1-\epsilon)cos^2(\pi/8)$ of CHSH games, in order to pass the verifier's checks. 
Thus, the completeness-soundness gap of the protocol will be determined by $\epsilon$. But since, for fixed $\delta$, $\epsilon$ is essentially inverse polynomial in $n$, the completeness-soundness gap will also be inverse polynomial in $n$. Hence, one requires polynomially many repetition in order to boost the gap to constant.

\noindent We conclude with:
\begin{theorem}
The RUV protocol is an $\mathsf{MIP^*}$ protocol achieving an inverse polynomial gap between completeness and soundness.
\end{theorem}

\subsubsection{GKW protocol.} \label{subsect:GKW}
As mentioned, in the RUV protocol the two quantum provers must be universal quantum computers. One could ask whether this is a necessity or whether there is a way to reduce one of the provers to be non-universal. In a paper by Gheorghiu, Kashefi and Wallden it was shown that the latter option is indeed possible. This leads to a protocol which we shall refer to as the \emph{GKW protocol}.
The protocol is based on the observation that one could use the state tomography subprotocol of RUV in such a way so that one prover is remotely preparing single qubit states for the other prover. The preparing prover would then only be required to perform single qubit measurements and, hence, not need the full capabilities of a universal quantum computer. The specific single qubit states that are chosen, can be the ones used in the VUBQC protocol of Subsection~\ref{subsect:fk}. This latter prover can then be instructed to perform the VUBQC protocol with these states. Importantly, because the provers are not allowed to communicate, this would preserve the blindness requirement of VUBQC.
We will refer to the preparing prover as the \emph{sender} and the remaining prover as the \emph{receiver}. Once again, we assume the verifier wishes to delegate to the provers the evaluation of some quantum circuit $\mathcal{C}$.

\begin{figure}[htbp!]
    \centering
    \begin{subfigure}[b]{0.45\textwidth}
        \includegraphics[width=\textwidth]{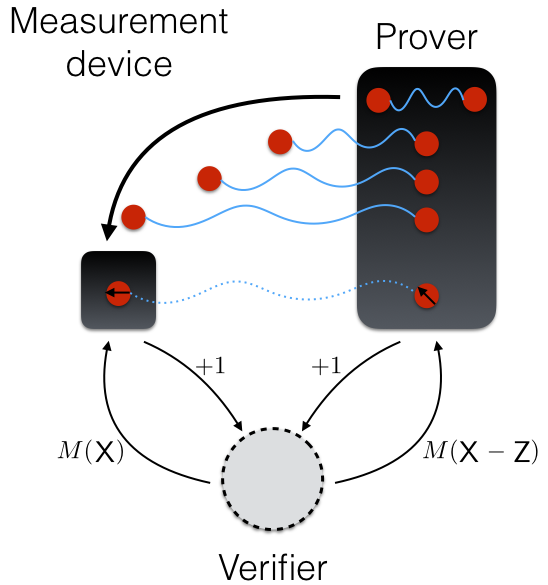}
        \caption{Testing preparation}
        \label{fig:test}
    \end{subfigure}
    \hfill
    \begin{subfigure}[b]{0.433\textwidth}
        \includegraphics[width=\textwidth]{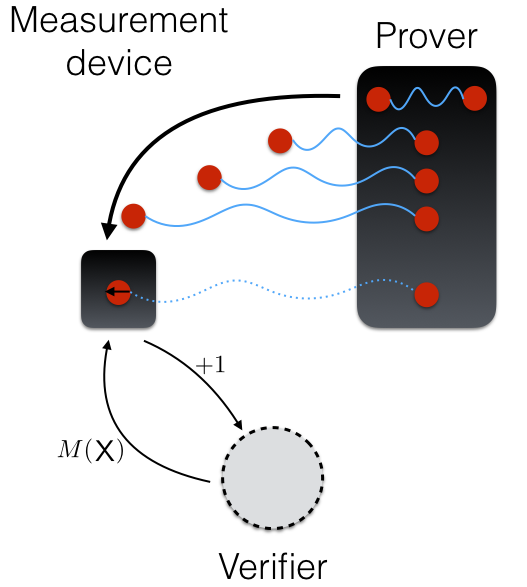}
        \caption{Preparing a qubit}
        \label{fig:prep}
    \end{subfigure}
    \caption{Verified preparation}\label{fig:ver_prep}
\end{figure}

The protocol, therefore, has a two-step structure:
\begin{enumerate}\addtolength{\itemsep}{+0.5\baselineskip}
\item[\textbf{(1)}] \textbf{Verified preparation.} This part is akin to the state tomography subprotocol of RUV. The verifier is trying to certify the correct preparation of states $\{ \ket{+_{\theta}} \}_\theta$ and $\ket{0}$, $\ket{1}$, where $\theta \in \{0, \pi/4, ..., 7\pi/4 \}$. Recall that these are the states used in VUBQC. We shall refer to them as the \emph{resource states}. This is done by self-testing a tensor product of Bell pairs and the observables of the two provers using CHSH games and the rigidity result of Theorem~\ref{thm:rigidity}\footnote{In fact, what is used here is a more general version of Theorem~\ref{thm:rigidity} involving an extended CHSH game. See the appendix section of \cite{ruv}.}. As in the RUV protocol, the verifier will play multiple CHSH games with the provers. This time, however, each game will be an extended CHSH game (as defined in \cite{ruv}) in which the verifier will ask each prover to measure an observable from the set $\{ \xgate, \ygate, \zgate, (\xgate \pm \zgate)/\sqrt{2}, (\ygate \pm \zgate)/\sqrt{2}, (\xgate \pm \ygate)/\sqrt{2}  \}$. Alternatively, this can be viewed as the verifier choosing to play one of $6$ possible CHSH games defined by the observables in that set\footnote{For instance, one game would involve Alice measuring either $\xgate$ or $\ygate$, whereas Bob should measure $(\xgate + \ygate)/\sqrt{2}$ or $(\xgate - \ygate)/\sqrt{2}$. Similar games can be defined by suitably choosing observables from the given set.}
These observables are sufficient for obtaining the desired resource states. In particular, measuring the $\xgate$, $\ygate$, and $(\xgate \pm \ygate) / \sqrt{2}$ observables on the Bell pairs will collapse the entangled qubits to states of the form $\{ \ket{+_{\theta}} \}_\theta$, while measuring $\zgate$ will collapse them to $\ket{0}$, $\ket{1}$.
The verifier accepts if the provers win a fraction $(1-\epsilon)cos^2(\pi/8)$ of the CHSH games, where $\epsilon = poly(\delta, 1/|\mathcal{C}|)$, and $\delta > 0$ is the desired trace distance between the reduced state on the receiver's side and the ideal state consisting of the required resource states in tensor product, up to a local isometry ($\epsilon \rightarrow 0$ as $\delta \rightarrow 0$ or $|\mathcal{C}| \rightarrow \infty$). 
The verifier will also instruct the sender prover to perform additional measurements so as to carry out the remote preparation on the receiver's side. This verified preparation is illustrated in Figure~\ref{fig:ver_prep}.

\item[\textbf{(2)}] \textbf{Verified computation.} This part involves verifying the actual quantum computation, $\mathcal{C}$. Once the resource states have been prepared on the receiver's side, the verifier will perform the VUBQC protocol with that prover as if she had sent him the resource states. She accepts the outcome of the computation if all trap measurements succeed, as in VUBQC.
\end{enumerate}

Note the essential difference, in terms of the provers' requirements, between this protocol and the RUV protocol. In the RUV protocol, both provers had to perform entangling measurements on their side. However, in the GKW protocol, the sender prover is required to only perform single qubit measurements. This means that the sender prover can essentially be viewed as an untrusted measurement device, whereas the receiver is the only universal quantum computer.
For this reason, the GKW protocol is also described as a \emph{device-independent} \cite{barrett2005no, acin2007device} verification protocol. This stems from comparing it to VUBQC or the receive-and-measure protocols, of Section~\ref{sect:recvmeas}, where the verifier had a trusted preparation or measurement device. In this case, the verifier essentially has a measurement device (the sender prover) which is untrusted.

Of course, performing the verified preparation subprotocol and combining it with VUBQC raises some questions. For starters, in the VUBQC protocol, the state sent to the prover is assumed to be an ideal state (i.e. an exact tensor product of states of the form $\Ket{+_{\theta}}$ or $\Ket{0}$, $\Ket{1}$). However, in this case the preparation stage is probabilistic in nature and therefore the state of the receiver will be $\delta$-close to the ideal tensor product state, for some $\delta > 0$. How is the completeness-soundness gap of the VUBQC protocol affected by this? Stated differently, is VUBQC robust to deviations in the input state?
A second aspect is that, since the resource state is prepared by the untrusted sender, even though it is $\delta$-close to ideal, it can, in principle, be correlated with the receiving prover's system. Do these initial correlations affect the security of the protocol? 

Both of these issues are addressed in the proofs of the GKW protocol. Firstly, assume that in the VUBQC protocol the prover receives a state which is $\delta$-close to ideal and uncorrelated with his private system. Any action of the prover can, in the most general sense, be modelled as a CPTP map. This CPTP map is of course distance preserving and so the output of this action will be $\delta$-close to the output in the ideal case. It follows from this that the probabilities of the verifier accepting a correct or incorrect result change by at most $O(\delta)$. As long as $\delta > 1/poly(|\mathcal{C}|)$ (for a suitably chosen polynomial), the protocol remains a valid $\cQPIP$ protocol.

Secondly, assume now that the $\delta$-close resource state is correlated with the prover's private system, in VUBQC. It would seem that the prover could, in principle, exploit this correlation in order to convince the verifier to accept an incorrect outcome. 
However, it is shown that this is, in fact, not the case, as long as the correlations are small.
Mathematically, let $\rho_{VP}$ be the state comprising of the resource state and the prover's private system. In the ideal case, this state should be a product state of the form $\rho_{V} \otimes \rho_{P}$, where $\rho_{V} = \ket{\psi_{id}}\bra{\psi_{id}}$ is the ideal resource state and $\rho_{P}$ the prover's system. However, in the general case the state can be entangled. In spite of this, it is known that:
\begin{equation}
TD( Tr_{P}(\rho_{VP}), \ket{\psi_{id}}\bra{\psi_{id}}) \leq \delta
\end{equation}
Using a result known as the \emph{gentle measurement lemma} \cite{ruv}, one can show that this implies:
\begin{equation}
TD( \rho_{VP}, \ket{\psi_{id}}\bra{\psi_{id}} \otimes Tr_{V}(\rho_{VP})) \leq O(\sqrt{\delta})
\end{equation}
In other words, the joint system of resource states and the prover's private memory is $O(\sqrt{\delta})$-close to the ideal system. Once again, as long as $\delta > 1/poly(|\mathcal{C}|)$ (for a suitably chosen polynomial), the protocol is a valid $\cQPIP$ protocol.

These two facts essentially show that the GKW protocol is a valid entanglement-based protocol, as long as sufficient tests are performed in the verified preparation stage so that the system of resource states is close to the ideal resource states. As with the RUV protocol, this implies a large communication overhead, with the communication complexity being of the order $O(|\mathcal{C}|^c)$, where $c > 2048$. 
One therefore has:
\begin{theorem}
The GKW protocol is an $\mathsf{MIP^*}$ protocol achieving an inverse polynomial gap between completeness and soundness.
\end{theorem}

Before concluding this section, we describe the steering-based VUBQC protocol that we referenced in Section~\ref{sect:recvmeas}. As mentioned, the GKW protocol can be viewed as a protocol involving a verifier with an untrusted measurement device interacting with a quantum prover. In a subsequent paper, Gheorghiu, Wallden and Kashefi addressed the setting in which the verifier's device becomes trusted \cite{gwk}. They showed that one can define a self-testing game for Bell states which involves \emph{steering correlations} \cite{sch} as opposed to non-local correlations. Steering correlations arise in a two-player setting in which one of the players is trusted to measure certain observables. This extra piece of information allows for the characterisation of Bell states with comparatively fewer statistics than in the non-local case.
The steering-based VUBQC protocol, therefore, has exactly the same structure as the GKW protocol. First, the verifier uses this steering-based game, between her measurement device and the prover, to certify that the prover prepared a tensor product of Bell pairs. She then measures some of the Bell pairs so as to remotely prepare the resource states of VUBQC on the prover's side and then performs the trap-based verification.
As mentioned in Section~\ref{sect:recvmeas}, the protocol has a communication complexity of $O(|\mathcal{C}|^{13} log(|\mathcal{C}|))$ which is clearly an improvement over $O(|\mathcal{C}|^{2048})$. This improvement stems from the trust added to the measurement device. However, the overhead is still too great for any practical implementation.

\subsubsection{HPDF protocol.}
Independently from the GKW approach, Hajdu{\v{s}}ek, P{\'e}rez-Delgado and Fitzsimons developed a protocol which also combines the CHSH rigidity result with the VUBQC protocol. This protocol, which we refer to as the \emph{HPDF protocol} has the same structure as GKW in the sense that it is divided into a verified preparation stage and a verified computation stage.
The major difference is that the number of non-communicating provers is on the order $O(poly(|\mathcal{C}|))$, where $\mathcal{C}$ is the computation that the verifier wishes to delegate.
Essentially, there is one prover for each Bell pair that is used in the verified preparation stage. This differs from the previous two approaches in that the verifier knows, a priori, that there is a tensor product structure of states. She then needs to certify that these states are close, in trace distance, to Bell pairs.
The advantage of assuming the existence of the tensor product structure, instead of deriving it through the RUV rigidity result, is that the overhead of the protocol is drastically reduced. Specifically, the total number of provers, and hence the total communication complexity of the protocol is of the order $O(|\mathcal{C}|^4 log( |\mathcal{C}| ))$.

We now state the steps of the HPDF protocol. We will refer to one of the provers as the verifier's untrusted measurement device. This is akin to the sender prover in the GKW protocol. The remaining provers are the ones which will ``receive'' the states prepared by the verifier and subsequently perform the quantum computation.
\begin{enumerate}\addtolength{\itemsep}{+0.5\baselineskip}
\item[\textbf{(1)}] \textbf{Verified preparation.} The verifier is trying to certify the correct preparation of the resource states $\{ \ket{+_{\theta}} \}_\theta$ and $\ket{0}$, $\ket{1}$, where $\theta \in \{0, \pi/4, ..., 7\pi/4 \}$.  The verifier instructs each prover to prepare a Bell pair and send one half to her untrusted measurement device. For each received state, she will randomly measure one of the following observables $\{ \xgate, \ygate, \zgate, (\xgate + \zgate)/\sqrt{2}, (\ygate + \zgate)/\sqrt{2}, (\xgate + \ygate)/\sqrt{2}, (\xgate - \ygate)/\sqrt{2}  \}$. Each prover is either instructed to randomly measure an observable from the set $\{ \xgate, \ygate, \zgate \}$ or to not perform any measurement at all. The latter case corresponds to the qubits which are prepared for the computation stage.
The verifier will compute correlations between the measurement outcomes of her device and the provers and accept if these correlations are above some threshold parametrized by $\epsilon = poly(\delta, 1/|\mathcal{C}|)$ ($\epsilon \rightarrow 0$ as $\delta \rightarrow 0$ or $|\mathcal{C}| \rightarrow \infty$), where $\delta > 0$ is the desired trace distance between the reduced state on the receiving provers' sides and the ideal state consisting of the required resource states in tensor product, up to a local isometry.
\item[\textbf{(2)}] \textbf{Verified computation.} Assuming the verifier accepted in the previous stage, she instructs the provers that have received the resource states to act as a single prover. The verifier then performs the VUBQC protocol with that prover as if she had sent him the resource states. She accepts the outcome of the computation if all trap measurements succeed, as in VUBQC.
\end{enumerate}

In their paper, Hajdu{\v{s}}ek et al have proved that the procedure in the verified preparation stage of their protocol constitutes a self-testing procedure for Bell states. This procedure self-tests individual Bell pairs, as opposed to the CHSH rigidity theorem which self-tests a tensor product of Bell pairs. In this case, however, the tensor product structure is already given by having the $O(|\mathcal{C}|^4 log( |\mathcal{C}| ))$ non-communicating provers.
The correctness of the verified computation stage follows from the robustness of the VUBQC protocol, as mentioned in the previous section.
One therefore has the following:
\begin{theorem}
The HPDF protocol is an $\mathsf{MIP^*[poly]}$ protocol achieving an inverse polynomial gap between completeness and soundness.
\end{theorem}

\subsection{Verification based on self-testing graph states} \label{subsect:mckague}
We saw, in the HPDF protocol, that having multiple non-communicating provers presents a certain advantage in characterising the shared state of these provers, due to the tensor product structure of the provers' Hilbert spaces. This approach not only leads to simplified proofs, but also to a reduced overhead in characterising this state, when compared to the CHSH rigidity Theorem~\ref{thm:rigidity}, from \cite{ruv}. 

Another approach which takes advantage of this tensor product structure is the one of McKague from \cite{mckague}. In his protocol, 
as in HPDF, the verifier will interact with $O(poly(|\mathcal{C}|))$ provers.
Specifically, there are multiple groups of $O(|\mathcal{C}|)$ provers, each group jointly sharing a graph state $\ket{G}$. In particular, each prover should hold only one qubit from $\ket{G}$. The central idea is for the verifier to instruct the provers to measure their qubits to either test that the provers are sharing the correct graph state or to perform an MBQC computation of $\mathcal{C}$. 
This approach is similar to the stabilizer measurement-only protocol of Subsection~\ref{subsect:monly} and, just like in that protocol or the Test-or-Compute or RUV protocols, the verifier will randomly alternate between tests and computation. 

\begin{figure}[htbp!]
\centering
\includegraphics[scale=0.5]{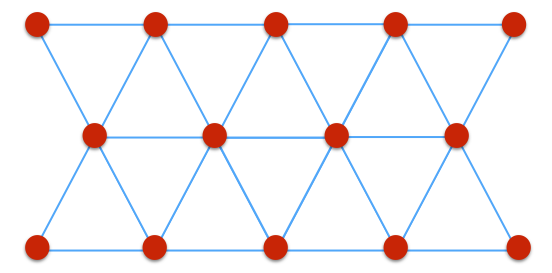}
\caption{Triangular lattice graph}
\label{fig:lattice}
\end{figure}

Before giving more details about this verification procedure, we first describe the type of graph state that is used in the protocol and the properties which allow this state to be useful for verification.
McKague considers $\Ket{G}$ to be a triangular graph state, which is a type of universal cluster state. What this means is that the graph $G$ on which the state is based on, is a triangular lattice (a planar graph with triangular faces). An example is shown in Figure~\ref{fig:lattice}. For each vertex $v$ in $G$ we have that:
\begin{equation}
X_{v} Z_{N(v)} \Ket{G} = \Ket{G}
\end{equation}
Where $N(v)$ denotes the neighbors of the vertex $v$. In other words, $\Ket{G}$ is stabilized by the operators $K_v = X_{v} Z_{N(v)}$, for all vertices $v$. This is important for the testing part of the protocol as it means that measuring the observable $S_v$ will always yield the outcome $1$. Another important property is:
\begin{equation}
X_{\tau} Z_{N(\tau)} \Ket{G} =  -\Ket{G}
\end{equation}
where $\tau$ is a set of $3$ neighboring vertices which comprise a triangle in the graph $G$ (and $N(\tau)$ are the \emph{odd} neighbors of those vertices\footnote{In other words, $N(\tau)$ consists of those vertices that are connected to an odd number of vertices from $\tau$.}). This implies that measuring $T_\tau = X_{\tau} Z_{N(\tau)}$ produces the outcome $-1$.
Triangular graph states are universal for quantum computation, as explained in \cite{mhalla2013graph,mckague}, by performing local measurements (with corrections) on the vertex qubits using the observables $\mathsf{R}(\theta) = cos(\theta) \xgate + sin(\theta) \zgate$, where $\theta \in \{0, \pi/4, ..., 7\pi/4 \}$.

\begin{figure}[htbp!]
\centering
\includegraphics[scale=0.45]{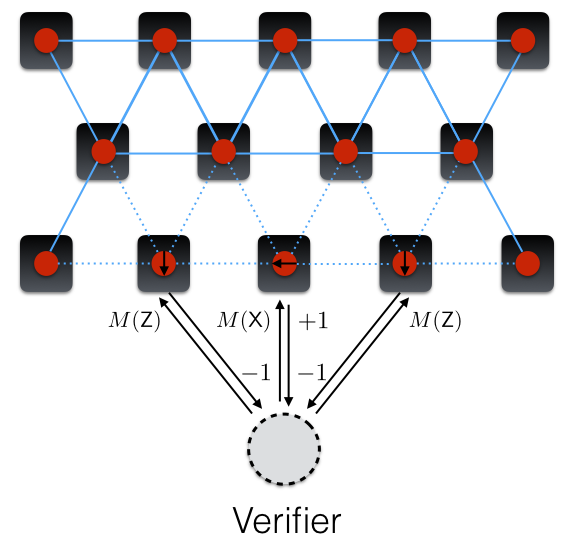}
\caption{Verifier instructing some of the provers to perform measurements in McKague's protocol}
\label{fig:mckague}
\end{figure}

We now have the necessary elements to describe McKague's protocol. The verifier considers a triangular graph state $\Ket{G}$ for the computation she wishes to verify. Let $n = O(|\mathcal{C}|)$ denote the number of vertices in $G$. In the ideal case, there will be multiple groups of $n$ provers and, in each group, every prover should have one of the qubits of this graph (entangled with its neighbors).
Denote $T$ as the number of triangles (consisting of $3$ neighboring vertices) in $G$ and $N_G = 3n + T$.
The protocol's setting is shown in Figure~\ref{fig:mckague}.

The verifier will choose one of the $n$ groups of provers at random to perform the computation $\mathcal{C}$. The computation is performed in an MBQC fashion. In other words, the verifier will pick appropriate measurement angles $\{\theta_v\}_{v \in V(G)}$, for all vertices in $G$, as well as a partial order for the vertices. To perform the computation $\mathcal{C}$, the verifier instructs the provers to measure the qubits of $\ket{G}$ with the observables $\mathsf{R}(\theta_v)$, defined above.
The partial order establishes the temporal ordering of these measurements. Additionally, the $\theta_v$ angles, for the $\mathsf{R}(\theta_v)$ measurements, should be updated so as to account for corrections arising from previous measurement outcomes. In other words, the angles $\{\theta_v\}_{v \in V(G)}$, which we shall refer to as \emph{computation angles}, are the ideal angles assuming no corrections. See Subsection~\ref{subsect:mbqc} for more details about measurement-based quantum computations.

The remaining groups of provers are used to test that the correct graph state, $\ket{G}$, was prepared.
This testing phase consists in the verifier randomly choosing to run one of the following sub-protocols:
\begin{itemize}\addtolength{\itemsep}{+0.5\baselineskip}
\item \textbf{Vertex test}, chosen with probability $\frac{n}{N_G}$. The verifier randomly picks a vertex $v$ of the graph. She instructs the prover having the qubit corresponding to $v$ to measure it in the $\xgate$ basis. At the same time, she asks the provers having the neighboring qubits, $N(v)$, to measure them in the $\zgate$ basis. The $\pm 1$ outcomes of the provers are multiplied by the verifier and she checks to see whether this product is $1$ (since she is essentially measuring the $S_v$ stabilizer). If this is the case, she accepts, otherwise she rejects.
\item \textbf{Triangle test}, chosen with probability $\frac{T}{N_G}$. Similar to the previous test, the verifier picks a random triangle in $G$, consisting of the vertices $\tau$. She instructs the provers having these vertices to measure them in the $\xgate$ basis, while their neighbors ($N(\tau)$) are instructed to measure in the $\zgate$ basis.  The verifier checks if the product of their outcomes is $-1$ and if so, she accepts, otherwise she rejects.
\item $\bm{R(\theta)}$ \textbf{test}, chosen with probability $\frac{2n}{N_G}$. In this case the verifier first randomly picks a vertex $v$ of the graph, a neighbor $u$ of $v$ (so $u \in N(v)$) and $t$ in $\{-1, +1\}$. She then randomly picks $\xgate$ with probability $p = \frac{cos(\theta_v)}{cos(\theta_v) + |sin(\theta_v)|}$ or $\zgate$ with probability $1-p$, where $\theta_v$ is the computation angle associated with $v$. If she chose $\xgate$, then she queries the prover holding $v$ to measure $\mathsf{R}(t \theta_v)$, and his neighbors ($N(v)$) to measure $\zgate$. She accepts if the product of their replies is $+1$. If the verifier instead chose $\zgate$, then she instructs the prover holding $v$ to measure $t \mathsf{R}(t \theta_v)$, the prover holding $u$ to measure $\xgate$ and the neighbors of $u$ and $v$ to measure $\zgate$. She accepts if the product of their outcomes is $+1$.
\end{itemize}

Together, these three tests are effectively performing a self-test of the graph state $\ket{G}$ and the prover's observables. Specifically, McKague showed the following:
\begin{theorem}
For a triangular graph $G$, having $n$ vertices, suppose that $n$ provers, performing the strategy $S = (\ket{\tilde{\psi}}, \{\tilde{O}_i^j\}_{ij})$ succeed in the test described above with probability $1 - \epsilon$, where $\epsilon = poly(\delta, 1/n)$ for some $\delta > 0$ and $\epsilon \rightarrow 0$ as $\delta \rightarrow 0$ or $n \rightarrow \infty$.
The strategy involves sharing the state $\ket{\tilde{\psi}}$ and measuring the observables $\{\tilde{O}_i^j\}_{ij}$, where each prover, $i$, has observables $\{\tilde{O}_i^j\}_{j}$.
Then there exists a local isometry $\Phi = \bigotimes_{i=1}^{n} \Phi_i$ and a state $\ket{junk}$ such that:
\begin{equation}
TD(\Phi(\ket{\tilde{\psi}}), \ket{junk} \ket{G}) \leq \delta
\end{equation}
and for all $j$:
\begin{equation}
TD \left( \Phi \left( \bigotimes_{i=1}^{n} \tilde{O}_i^j \ket{\tilde{\psi}} \right), \ket{junk} \bigotimes_{i=1}^{n} O_i^j \ket{G} \right) \leq \delta
\end{equation}
where for all $i$, $O_i^j \in \{ \mathsf{R}(\theta) | \; \theta \in \{0, \pi/4, ... , 7\pi/4\} \}$\footnote{The measurement angles need not be restricted to this set, however, as in VUBQC, this set of angles is sufficient for performing universal MBQC computations.}.
\end{theorem}
Note that the verifier will ask the provers to perform the same types of measurements in both the testing phase and the computation phase of the protocol. This means that, at the level of each prover, the testing and computation phases are indistinguishable. Moreover, the triangular state $\ket{G}$, being a universal cluster state, will be the same for computations of the same size. Therefore, the protocol is blind in the sense that each prover, on its own, is unaware of what computation is being performed.
In summary, the protocol consists of the verifier choosing to perform one of the following:
\begin{itemize}\addtolength{\itemsep}{+0.5\baselineskip}
\item \textbf{Computation.} In this case, the verifier instructs the provers to perform the MBQC computation of $\mathcal{C}$ on the graph state $\ket{G}$, as described above.
\item \textbf{Testing }$\ket{G}$\textbf{.} In this case, the verifier will randomly choose between one of the three tests described above accepting if an only if the test succeeds.
\end{itemize}

\noindent It is therefore the case that:
\begin{theorem}
McKague's protocol is an $\mathsf{MIP^*}[poly]$ protocol having an inverse polynomial gap between completeness and soundness.
\end{theorem}

As with the previous approaches, the reason for the inverse polynomial gap between completeness and soundness is the use of a self-test with robustness $\epsilon = poly(1/n)$ (and $\epsilon \rightarrow 0$ as $n \rightarrow \infty$). In turn, this leads to a polynomial overhead for the protocol as a whole. Specifically, McKague showed that the total number of required provers and communication complexity, for a quantum computation $\mathcal{C}$, is of the order $O(|\mathcal{C}|^{22})$.
Note, however, that each of the provers must only perform a single-qubit measurement. Hence, apart from the initial preparation of the graph state $\ket{G}$, the individual provers are not universal quantum computers, merely single-qubit measurement devices.

\subsection{Post hoc verification} \label{subsect:entposthoc}
In Subsection~\ref{subsect:posthoc} we reviewed a protocol by Morimae and Fitzsimons for post hoc verification of quantum computation. Of course, that protocol involved a single quantum prover and a verifier with a measurement device. In this section, we review two post hoc protocols for the multi-prover setting having a classical verifier. We start with the first post hoc protocol by Fitzsimons and Hajdu{\v{s}}ek.

\subsubsection{FH protocol.}
Similar to the 1S-Post-hoc protocol from Subsection~\ref{subsect:posthoc}, the protocol of Fitzsimons and Hajdu{\v{s}}ek, which we shall refer to as the \emph{FH protocol}, also makes use of the local Hamiltonian problem stated in Definition~\ref{def:LH}. 
As mentioned, this problem is complete for the class $\cQMA$, which consists of problems that can be decided by a $\cBQP$ verifier receiving a witness state from a prover. Importantly, the size of the witness state is \emph{polynomial} in the size of the input to the problem. However, Fitzsimons and Vidick proposed a protocol for the $k$-local Hamiltonian problem (and hence any $\cQMA$ problem), involving $5$ provers, in which the quantum state received by the verifier is of \emph{constant} size \cite{fv}. That protocol is the basis for the FH protocol and so we start with a description of it.

Suppose that the $k$-local Hamiltonian is of the form $H = \sum_i H_i$, acting on a system of $n$ qubits and each $H_i$ is a $k$-local, $n$-qubit projector. For fixed $a$ and $b$, such that $b-a \geq 1/poly(n)$, the verifier should accept if there exists a state $\ket{\psi}$ such that $\bra{\psi} H \ket{\psi} \leq a$ and reject if for all states $\ket{\psi}$ it is the case that $\bra{\psi} H \ket{\psi} \geq b$.
Suppose we are in the acceptance case and let $\ket{\psi}$ be the witness state. The $5$ provers must share a version of $\ket{\psi}$ encoded in the $5$-qubit error correcting code\footnote{The $5$-qubit code is the smallest error correcting capable of correcting for arbitrary single-qubit errors\cite{fivequbit}.}, denoted $\ket{\psi}_L$. Specifically, for each logical qubit of $\ket{\psi}_L$, each prover should hold one of its constituent physical qubits.
The verifier will then check that the provers are indeed sharing this state, accepting if this is the case and rejecting otherwise. She will also perform an energy measurement on the state, to estimate if it has energy above $b$ or below $a$. To do this she will, with equal probability, choose to either test that the shared state of the provers has energy below $a$ or that the provers share a state encoded in the $5$-qubit code:

\begin{itemize}\addtolength{\itemsep}{+0.5\baselineskip}
\item \textbf{Energy measurement.} In this case, the verifier will pick a random term $H_i$, from $H$, and ask each prover for $k$ qubits corresponding to the logical states on which $H_i$ acts. The verifier will then perform a two-outcome measurement, defined by the operators $\{H_i, I - H_i \}$ on the received qubits. As in the 1S-Post-hoc protocol, this provides an estimate for the energy of $\ket{\psi}$. The verifier accepts if the measurement outcome indicates the state has energy below $a$.
\item \textbf{Encoding measurement.} In this case the verifier will choose at random between two subtests. In the first subtest, she will choose $j$ at random from $1$ to $n$ and ask each prover to return the physical qubits comprising the $j$'th logical qubit. She then measures these qubits to check whether their joint state lies within the code space, accepting if it does and rejecting otherwise. In the second subtest, the verifier chooses a random set, $S$, of $3$ values between $1$ and $n$. She also picks one of the values at random, labelled $j$. The verifier then asks a randomly chosen prover for the physical qubits of the logical states indexed by the values in $S$, while asking the remaining provers for their shares of logical qubit $j$. As an example, if the set contains the values $\{1, 5, 8 \}$, then the verifier picks one of the $5$ provers at random and asks him for his shares (physical qubits) of logical qubits $1$, $5$ and $8$ from $\ket{\psi}$. Assuming that the verifier also picked the random value $8$ from the set, then she will ask the remaining provers for their shares of logical qubit $8$. The verifier then measures logical qubit $j$ (or $8$, in our example) and checks if it is in the code subspace, accepting if it is and rejecting otherwise.
The purpose of this second subtest is to guarantee that the provers respond with different qubits when queried.
\end{itemize}

One can see that when the witness state exists and the provers follow the protocol, the verifier will indeed accept with high probability. On the other hand, Fitzsimons and Vidick show that when there is no witness state, the provers will fail at convincing the verifier to accept with high probability. This is because they cannot simultaneously provide qubits yielding the correct energy measurements and also have their joint state be in the correct code space. 
This also illustrates why their protocol required testing both of these conditions. If one wanted to simplify the protocol, so as to have a single prover providing the qubits for the verifier's $\{H_i, I - H_i \}$ measurement, then it is no longer possible to prove soundness. The reason is that even if there does not exist a $\ket{\psi}$ having energy less than $a$ for $H$, the prover could still find a group of $k$ qubits which minimize the energy constraint for the specific $H_i$ that the verifier wishes to measure. The second subtest prevents this from happening, with high probability, since it forces the provers to consistently provide the requested indexed qubits from the state $\ket{\psi}$.

Note that for a $\cBQP$ computation, defined by the quantum circuit $\mathcal{C}$ and input $x$, the state $\ket{\psi}$ in the Fitzsimons and Vidick protocol becomes the Feynman-Kitaev state of that circuit, as described in Subsection~\ref{subsect:posthoc}.
The FH protocol essentially takes the Fitzsimons and Vidick protocol for $\cBQP$ computations and alters it by making the verifier classical. This is achieved using an approach of Ji \cite{ji} which allows for the two tests to be performed using only classical interaction with the provers.
The idea is based on self-testing and is similar to the rigidity of the CHSH game.
To understand this approach let us first examine the stabilizer generators, $\{ g_i \}_{i=1}^{4}$ for the code space of the $5$-qubit code, shown in Table~\ref{tab:stabs}. Notice that they all involve only Pauli $\xgate$, $\zgate$ or identity operators. In particular, the operator acting on the fifth qubit is always either $\xgate$ or $\zgate$. Ji then considers rotating this operator so that $\xgate \rightarrow \xgate'$ and $\zgate \rightarrow \zgate'$, where $\xgate' = (\xgate + \zgate)/\sqrt{2}$ and $\zgate' = (\xgate - \zgate)/\sqrt{2}$, resulting in the new operators $\{ g'_i \}_{i=1}^{4}$ shown in Table~\ref{tab:rotated}.

\begin{table}
\parbox{.45\linewidth}{
\centering
\begin{tabular}{l|l}
Generator & Name \\
\hline\\[-8pt]
$I \xgate \zgate \zgate \xgate$ & \hspace{2pt} $g_1$ \\[3pt]
$\xgate I \xgate \zgate \zgate$ & \hspace{2pt} $g_2$ \\[3pt]
$\zgate \xgate I \xgate \zgate$ & \hspace{2pt} $g_3$ \\[3pt]
$\zgate \zgate \xgate I \xgate$ & \hspace{2pt} $g_4$ \\[3pt]
\end{tabular}
\caption{Generators for $5$-qubit code}
\label{tab:stabs}
}
\hfill
\parbox{.45\linewidth}{
\centering
\begin{tabular}{l|l}
Generator & Name \\
\hline\\[-8pt]
$I \xgate \zgate \zgate \xgate'$ & \hspace{2pt} $g'_1$ \\[3pt]
$\xgate I \xgate \zgate \zgate'$ & \hspace{2pt} $g'_2$ \\[3pt]
$\zgate \xgate I \xgate \zgate'$ & \hspace{2pt} $g'_3$ \\[3pt]
$\zgate \zgate \xgate I \xgate'$ & \hspace{2pt} $g'_4$ \\[3pt]
\end{tabular}
\caption{Generators with fifth operator rotated}
\label{tab:rotated}
}
\end{table}

The new operators satisfy a useful property. For any state $\ket{\phi}$ in the code space of the $5$-qubit code, it is the case that:
\begin{equation}
\sum_i \bra{\phi} g'_i \ket{\phi} = 4 \sqrt{2}
\end{equation}
This is similar to the CHSH game. In the CHSH game, the ideal strategy involves Alice measuring either $\xgate$ or $\zgate$ and Bob measuring either $\xgate'$ or $\zgate'$, respectively, on the maximally entangled state $\ket{\Phi_+}$. These observables and the Bell state satisfy:
\begin{equation}
\bra{\Phi_+} \xgate \xgate' + \xgate \zgate' + \zgate \xgate' - \zgate \zgate' \ket{\Phi_+} = 2 \sqrt{2}
\end{equation}
It can be shown that having observables which satisfy this relation implies that Alice and Bob win the CHSH game with the (quantum) optimal probability of success $cos^{2}(\pi/8)$.
Analogous to the CHSH game, the stabilizers $\{ g'_i \}_{i=1}^{4}$, viewed as observables, can be used to define a $5$-player non-local game, in which the optimal quantum strategy involves measuring these observables on a state encoded in the $5$-qubit code. Moreover, just like in the CHSH game, observing the players achieve the maximum quantum win-rate for the game implies that, up to local isometry, the players are following the ideal quantum strategy.
We will not detail the game, except to say that it inolves partitioning the $5$ provers into two sets, one consisting of four provers and the other with the remaining prover. Such a bipartition of a state encoded in the $5$-qubit code yields a state which is isometric to a Bell pair. This means that the $5$-player game is essentially self-testing a maximally entangled state, hence the similarity to the CHSH game.
This then allows a classical verifier, interacting with the $5$ provers, to perform the encoding test of the Fitzsimons and Vidick protocol.

We have discussed how a classical verifier can test that the $5$ provers share a state encoded in the logical space of the $5$-qubit code. But to achieve the functionality of the Fitzsimons and Vidick protocol, one needs to also delegate to the provers the measurement of a local term $H_i$ from the Hamiltonian. This is again possible using the $5$-player non-local game. Firstly, it can be shown that, without loss of generality, that each $H_i$, in the $k$-local Hamiltonian, can be expressed as a linear combination of terms comprised entirely of $I$, $\xgate$ and $\zgate$. This means that the Hamiltonian itself is a linear combination of such terms, $H = \sum_i a_i S_i$, where $a_i$ are real coefficients and $S_i$ are $k$-local $\xgate\zgate$-terms. This is akin to the $\xgate\zgate$-Hamiltonian from the 1S-Post-hoc protocol. 

Given this fact, the verifier can measure one of the $S_i$ terms, in order to estimate the energy of the ground state, instead of measuring $\{ H_i, I - H_i \}$. She will pick an $S_i$ term at random and ask the provers to measure the constituent Pauli observables in $S_i$. However, the verifier will also alternate these measurements with the stabilizer measurements of the non-local game, rejecting if the provers do not achieve the maximal non-local value of the game. This essentially forces the provers to perform the correct measurements.

\begin{figure}[htbp!]
\centering
\includegraphics[scale=0.45]{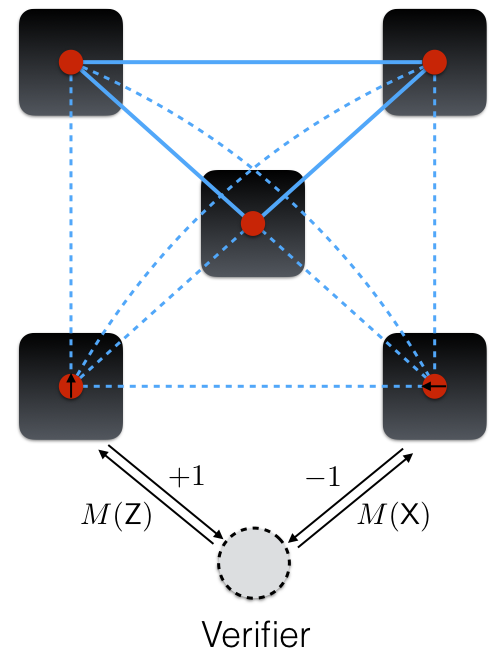}
\caption{Verifier interacting with the $5$ provers}
\label{fig:fh}
\end{figure}

To summarize, the FH protocol is a version of the Fitzsimons and Vidick protocol which restricts the provers to be $\cBQP$ machines and uses Ji's techniques, based on non-local games, to make the verifier classical. The steps of the FH protocol are as follows:
\begin{enumerate}\addtolength{\itemsep}{+0.5\baselineskip}
\item[\textbf{(1)}] The verifier instructs the provers to share the Feynman-Kitaev state, associated with her circuit $\mathcal{C}$, encoded in the $5$-qubit error correcting code, as described above. We denote this state as $\ket{\psi}_L$. The provers are then split up and not allowed to communicate. The verifier then considers a $k$-local Hamiltonian having $\ket{\psi}_L$ as a ground state as well as the threshold values $a$ and $b$, with $b - a > 1/poly(|\mathcal{C}|)$.
\item[\textbf{(2)}] The verifier chooses to either perform the energy measurement or the encoding measurement as described above. For the energy measurement she asks the provers to measure a randomly chosen $\xgate\zgate$-term from the local Hamiltonian. The verifier accepts if the outcome indicates that the energy of $\ket{\psi}_L$ is below $a$. For the encoding measurement the verifier instructs the provers to perform the measurements of the $5$-player non-local game. She accepts if the provers win the game, indicating that their shared state is correctly encoded.
\end{enumerate}

\noindent One therefore has:
\begin{theorem}
The FH protocol is an $\mathsf{MIP^*}$ protocol achieving an inverse polynomial gap between completeness and soundness.
\end{theorem}

There are two significant differences between this protocol and the previous entanglement-based approaches. The first is that the protocol does not use self-testing to enforce that the provers are performing the correct operations in order to implement the computation $\mathcal{C}$. Instead, the computation is checked indirectly by using the self-testing result to estimate the ground-state energy of the $k$-local Hamiltonian. This then provides an answer to the considered $\cBQP$ computation viewed as a decision problem\footnote{In their paper, Fitzsimons and Hajdu{\v{s}}ek also explain how their protocol can be used to sample from a quantum circuit, rather than solve a decision problem \cite{fh}.}. 
The second difference is that the protocol is not blind. In all the previous approaches, the provers had to share an entangled state which was independent of the computation, up to a certain size. However, in the FH protocol, the state that the provers need to share depends on which quantum computation the verifier wishes to perform.

In terms of communication complexity, the protocol, as described, would involve only $2$ rounds of interaction between the verifier and the provers. However, since the completeness-soundness gap is inverse polynomial, and therefore decreases with the size of the computation, it becomes necessary to repeat the protocol multiple times to properly differentiate between the accepting and rejecting cases. 
On the one hand, the local Hamiltonian itself has an inverse polynomial gap between the two cases of acceptance and rejection. As shown in \cite{hangleiter2017direct,gap}, for the Hamiltonian resulting from a quantum circuit, $\mathcal{C}$, that gap is $1/|\mathcal{C}|^2$. To boost this gap to constant, the provers must share $O(|\mathcal{C}|^2)$ copies of the Feynman-Kitaev state.

On the other hand, the self-testing result has an inverse polynomial robustness. This means that estimating the energy of the ground state is done with a precision which scales inverse polynomially in the number of qubits of the state. More precisely, according to Ji's result, the scaling should be $1/O(N^{16})$, where $N$ is the number of qubits on which the Hamiltonian acts \cite{ji}.
This means that the protocol should be repeated on the order of $O(N^{16})$ times, in order to boost the completeness-soundness gap to constant.

\subsubsection{NV protocol.} \label{subsect:NV}
The second entanglement-based post hoc protocol was developed by Natarajan and Vidick \cite{nv} and we therefore refer to it as the \emph{NV protocol}. The main ideas of the protocol are similar to those of the FH protocol. However, Natarajan and Vidick prove a self-testing result having \emph{constant} robustness and use it in order to perform the energy estimation of the ground state for the local Hamiltonian. 

The statement of their general self-testing result is too involved to state here, so instead we reproduce a corollary to their result (also from \cite{nv}) that is used for the NV protocol. This corollary involves self-testing a tensor product of Bell pairs:
\begin{theorem}
For any integer $n$ there exists a two-player non-local game, known as the Pauli braiding test ($PBT$), with $O(n)$-bit questions and $O(1)$-bit answers satisfying the following:

Let $S = (\ket{\tilde{\psi}}, \tilde{A}(\mathbf{a}), \tilde{B}(\mathbf{b}))$ be the strategy employed by two players (Alice and Bob) in playing the game, where $\ket{\tilde{\psi}}$ is their shared state and $\tilde{A}(\mathbf{a})$ and $\tilde{B}(\mathbf{b})$ are their respective (multi-qubit) observables when given $n$-bit questions $\mathbf{a}$ and $\mathbf{b}$, respectively. 
Suppose Alice and Bob win the Pauli braiding test with probability $\omega^*(PBT) - \epsilon$, for some $\epsilon > 0$ (note that $\omega^*(PBT)=1$). Then there exist $\delta = poly(\epsilon)$, a local isometry $\Phi = \Phi_A \otimes \Phi_B$ and a state $\ket{junk}$ such that:
\begin{equation}
TD(\Phi(\ket{\tilde{\psi}}), \ket{junk} \ket{\Phi_+}^{\otimes n}) \leq \delta
\end{equation}
\begin{equation}
TD \left( \Phi \left( \tilde{A}(\mathbf{a}) \otimes \tilde{B}(\mathbf{b}) \ket{\tilde{\psi}} \right), \ket{junk} \xgate(\mathbf{a}) \otimes \zgate(\mathbf{b}) \ket{\Phi_+}^{\otimes n} \right) \leq \delta
\end{equation}
where $\xgate(\mathbf{a}) = \bigotimes\limits_{i=1}^{n} \xgate^{\mathbf{a(i)}}$ and $\zgate(\mathbf{b}) = \bigotimes\limits_{i=1}^{n} \zgate^{\mathbf{b(i)}}$.
\end{theorem}
This theorem is essentially a self-testing result for a tensor product of Bell states, and Pauli $\xgate$ and $\zgate$ observables, achieving a constant robustness.
The Pauli braiding test is used in the NV protocol in a similar fashion to Ji's result, from the previous subsection, in order to certify that a set of provers are sharing a state that is encoded in a quantum error correcting code. Again, this relies on a bi-partition of the provers into two sets, such that, an encoded state shared across the bi-partition is equivalent to a Bell pair.

Let us first explain the general idea of the Pauli braiding test for self-testing $n$ Bell pairs and $n$-qubit observables. We have a referee that is interacting with two players, labelled Alice and Bob. The test consists of three subtests which are chosen at random by the referee. The subtests are:
\begin{itemize}\addtolength{\itemsep}{+0.5\baselineskip}
\item \textbf{Linearity test.} In this test, the referee will randomly pick a basis setting, $W$, from the set $\{ \xgate, \zgate \}$. She then randomly chooses two strings $\mathbf{a_1}, \mathbf{a_2} \in \{ 0, 1 \}^n$ and sends them to Alice. With equal probability, the referee takes $\mathbf{b_1}$ to be either $\mathbf{a_1}$, $\mathbf{a_2}$ or $\mathbf{a_1} \oplus \mathbf{a_2}$. She also randomly chooses a string $\mathbf{b_2} \in \{ 0, 1 \}^n$ and sends the pair $(\mathbf{b_1}, \mathbf{b_2})$ to Bob\footnote{Note that pair can be either $(\mathbf{b_1}, \mathbf{b_2})$ or $(\mathbf{b_2}, \mathbf{b_1})$, so that Bob does not know which string is the one related to Alice's inputs.}. Alice and Bob are then asked to measure the observables $W(\mathbf{a_1})$, $W(\mathbf{a_2})$ and $W(\mathbf{b_1})$, $W(\mathbf{b_2})$, respectively, on their shared state. We denote Alice's outcomes as $a_1$, $a_2$ and Bob's outcomes as $b_1$, $b_2$. If $\mathbf{b_1} = \mathbf{a_1}$ (or $\mathbf{b_1}=\mathbf{a_2}$, respectively), the referee checks that $b_1 = a_1$ (or $b_1 = a_2$, respectively). If $\mathbf{b_1} = \mathbf{a_1} \oplus \mathbf{a_2}$, she checks that $b_1 = a_1 a_2$. This test is checking, on the one hand, that when Alice and Bob measure the same observables, they should get the same outcome (which is what should happen if they share Bell states). On the other hand, and more importantly, it is checking the commutation and linearity of their operators, i.e. that $W(\mathbf{a_1})W(\mathbf{a_2}) = W(\mathbf{a_2})W(\mathbf{a_1}) = W(\mathbf{a_1} + \mathbf{a_2})$ (and similarly for Bob's operators).
\item \textbf{Anticommutation test.} The referee randomly chooses two strings $\mathbf{x}, \mathbf{z} \in \{ 0, 1 \}^n$, such that $\mathbf{x} \cdot \mathbf{z} = 1 \; mod \; 2$, and sends them to both players. These strings define the observables $\xgate(\mathbf{x})$ and $\zgate(\mathbf{z})$ which are anticommuting because of the imposed condition on $\mathbf{x}$ and $\mathbf{z}$. The referee then engages in a non-local game with Alice and Bob designed to test the anticommutation of these observables for both of their systems. This can be any game that tests this property, such as the CHSH game or the \emph{magic square} game, described in \cite{ms1,ms2}. As an example, if the referee chooses to play the CHSH game, then Alice will be instructed to measure either $\xgate(\mathbf{x})$ or $\zgate(\mathbf{z})$ on her half of the shared state, while Bob would be instructed to measure either $(\xgate(\mathbf{x}) + \zgate(\mathbf{z}))/\sqrt{2}$ or $(\xgate(\mathbf{x}) - \zgate(\mathbf{z}))/\sqrt{2}$. The test is passed if the players achieve the win condition of the chosen anticommutation game. Note that for the case of the magic square game, the condition can be achieved with probability $1$ when the players implement the optimal quantum strategy. For this reason, if the chosen game is the magic square game, then $\omega^*(PBT) = 1$.
\item \textbf{Consistency test.} This test combines the previous two. The referee randomly chooses a basis setting, $W \in \{ \xgate, \zgate \}$ and two strings $\mathbf{x}, \mathbf{z} \in \{ 0, 1 \}^n$. Additionally, let $\mathbf{w} = \mathbf{x}$, if $W = \xgate$ and $\mathbf{w} = \mathbf{z}$ if $W = \zgate$. The referee sends $W$, $\mathbf{x}$ and $\mathbf{z}$ to Alice. With equal probability the referee will then choose to perform one of two subtests. In the first subtest, the referee sends $\mathbf{x}, \mathbf{z}$ to Bob as well and plays the anticommutation game with both, such that Alice's observable is $W(\mathbf{w})$. As an example, if $W = \xgate$ and the game is the CHSH game, then Alice would be instructed to measure $\xgate(\mathbf{x})$, while Bob is instructed to measure either $(\xgate(\mathbf{x}) + \zgate(\mathbf{z}))/\sqrt{2}$ or $(\xgate(\mathbf{x}) - \zgate(\mathbf{z}))/\sqrt{2}$. This subtest essentially mimics the anticommutation test and is passed if the players achieve the win condition of the game.
In the second subtest, which mimics the linearity test, the referee sends $W$, $\mathbf{w}$ and a random string $\mathbf{y} \in \{ 0, 1 \}^n$ to Bob, instructing him to measure $W(\mathbf{w})$ and $W(\mathbf{y})$. Alice is instructed to measure $W(\mathbf{x})$ and $W(\mathbf{z})$. The test if passed if Alice and Bob obtain the same result for the $W(\mathbf{w})$ observable. For instance, if $W = \xgate$, then both Alice and Bob will measure $\xgate(\mathbf{x})$ and their outcomes for that measurement must agree.
\end{itemize}
Having observables that satisfy the linearity conditions of the test as well as the anticommutation condition implies that they are isometric to the actual $\xgate$ and $\zgate$ observables acting on a maximally entangled state. This is what the Pauli braiding test checks and what is proven by the self-testing result of Natarajan and Vidick.

We can now describe the NV protocol.
Similar to the FH protocol, for a quantum circuit, $\mathcal{C}$, and an input, $x$, one considers the associated Feynman-Kitaev state, denoted $\ket{\psi}$. This is then used to construct a $2$-local $\xgate\zgate$-Hamiltonian such that the ground state of this Hamiltonian is $\ket{\psi}$. As before, for some $a$ and $b$, with $b-a > 1/poly(|\mathcal{C}|)$, when $\mathcal{C}$ accepts $x$ we have that $\bra{\psi} H \ket{\psi} < a$, otherwise $\bra{\psi} H \ket{\psi} > b$. The verifier will instruct $7$ provers to share a copy of $\ket{\psi}$ state, encoded in a $7$-qubit quantum error correcting code known as \emph{Steane's code}. The provers are then asked to perform measurements so as to self-test an encoded state or perform an energy measurement on this state. The code space, for Steane's code, is the $7$-qubit subspace stabilized by all operators generated by $\{g_i\}_{i=1}^6$, where the generators are listed in Table~\ref{tab:steane}.

\begin{table}
\centering
\begin{tabular}{l|l}
Generator & Name \\
\hline\\[-8pt]
$I I I \xgate \xgate \xgate \xgate$ & \hspace{2pt} $g_1$ \\[3pt]
$I \xgate \xgate I I \xgate \xgate$ & \hspace{2pt} $g_2$ \\[3pt]
$\xgate I \xgate I \xgate I \xgate$ & \hspace{2pt} $g_3$ \\[3pt]
$I I I \zgate \zgate \zgate \zgate$ & \hspace{2pt} $g_4$ \\[3pt]
$I \xgate \xgate I I \xgate \zgate$ & \hspace{2pt} $g_5$ \\[3pt]
$\zgate I \zgate I \zgate I \zgate$ & \hspace{2pt} $g_6$ \\[3pt]
\end{tabular}
\caption{Generators for Steane's $7$-qubit code}
\label{tab:steane}
\end{table}

The reason Natarajan and Vidick use this specific error correcting code is because it has two properties that are necessary for the application of their self-testing result. The first property is that each stabilizer generator is a tensor product of only the $I$, $\xgate$ and $\zgate$ operators. This, of course, is true for the $5$-qubit code as well. The second property is a symmetry condition: for each index $i \in \{1, ..., 6\}$, there exists a pair of stabilizer generators, $S_{\xgate}$ and $S_{\zgate}$, such that $S_{\xgate}$ consists exclusively of $I$ and $\xgate$ operators and has an $\xgate$ on position $i$, whereas $S_{\zgate}$ is identical to $S_{\xgate}$ but with all $\xgate$ operators replaced with $\zgate$. This property is not satisfied by the $5$-qubit code and will allow the verifier to delegate to the provers measurements of the form $\xgate(\mathbf{x})$ and $\zgate(\mathbf{z})$, where $\mathbf{x}$ and $\mathbf{z}$ are binary strings, as in the Pauli braiding test.

Putting everything together, the protocol works as follows. The verifier instructs the $7$ provers to share an encoded instance of the Feynman-Kitaev state, $\ket{\psi}_L$, such that, for each logical qubit in $\ket{\psi}_L$, each prover will hold one of $7$ the constituent physical qubits. She then chooses at random to perform one of the following:

\begin{itemize}\addtolength{\itemsep}{+0.5\baselineskip}
\item \textbf{Pauli braiding test.} The verifier chooses one of the $7$ provers at random to be Alice, while the remaining provers will take on the role of Bob. The verifier then performs the Pauli braiding test with Alice and Bob in order to self-test the logical qubits in $\ket{\psi}_L$. As mentioned, each logical qubit, encoded in the $7$ qubit code, is equivalent to a Bell pair under the chosen bi-partition. The Pauli braiding test is essentially checking that the provers have correctly encoded each of the qubits in $\ket{\psi}$ and that they are correctly measuring $\xgate$ and $\zgate$ observables. The verifier rejects if the provers do not pass the test.
\item \textbf{Energy test.} In this case, the verifier will pick an $\xgate\zgate$-term, $S$, from $H$, at random, and instruct the provers to measure this term on their shared state. 
Note that $S$ consists of \emph{logical} $\xgate$ and $\zgate$ operators. This means that each prover will need to perform local measurements so that their joint measurement acts as either $\xgate_L$ or $\zgate_L$, respectively. Additionally, $\xgate_L$ and $\zgate_L$, for the $7$ qubit code, are expressed as tensor products of physical $\xgate$ and $\zgate$ operations. This means that each prover will be instructed to measure an operators of the form  $\xgate(\mathbf{x})$ and $\zgate(\mathbf{z})$, on its physical qubits, where $\mathbf{x}, \mathbf{z} \in \{0, 1\}^N$, and $N$ is the total number of logical qubits on which $H$ acts. The product $\xgate(\mathbf{x})\zgate(\mathbf{z})$ is the outcome for that prover's share of $S$.
The verifier then takes all of these $\pm 1$ outcomes and multiplies them together, thus obtaining the outcome of measuring $S$ itself. She accepts if the outcome of the measurement indicates that the estimated energy of $\ket{\psi}$ is below $a$ and rejects otherwise. 
\item \textbf{Energy consistency test.} This test is a combination of the previous two. As in the Pauli braiding test, the provers are bi-partitioned into two sets, one consisting of one prover, denoted Alice, and the other consisting of the other $6$ provers, jointly denoted as Bob. The verifier now performs a test akin to the linearity test from Pauli braiding. She randomly chooses $W \in \{ \xgate, \zgate \}$, and let $\mathbf{w} = \mathbf{x}$, if $W = \xgate$ and $\mathbf{w} = \mathbf{z}$ if $W = \zgate$. She also chooses $\mathbf{x}, \mathbf{z} \in \{0, 1\}^N$ according to the same distribution as in the energy test (i.e. as if she were instructing the provers to measure a random $\xgate\zgate$-term from $H$). The verifier then does one of the following:
\begin{itemize}
\item With probability $1/2$, instructs Alice to measure the observables $\xgate(\mathbf{x})$ and $\zgate(\mathbf{z})$. Additionally, the verifier chooses $\mathbf{y} \in \{0, 1\}^N$ at random and instructs Bob to measure $W(\mathbf{y})$ and $W(\mathbf{y} \oplus \mathbf{w})$. If $W = \xgate$, the verifier accepts if the product of Bob's answers agrees with Alice's answer for the  $\xgate(\mathbf{x})$ observable. If $W = \zgate$, the verifier accepts if the product of Bob's answers agrees with Alice's answer for the  $\zgate(\mathbf{z})$ observable. Note that this is the case since the product of Bob's observables should be $W(\mathbf{w})$ if he is behaving honestly.
\item With probability $1/4$, instructs Alice to measure $W(\mathbf{y})$ and $W(\mathbf{v})$, where $\mathbf{y}, \mathbf{w} \in \{0, 1\}^N$ are chosen at random. Bob is instructed to measure $W(\mathbf{y})$ and $W(\mathbf{y} \oplus \mathbf{w})$. The verifier accepts if the outcomes of Alice and Bob for $W(\mathbf{y})$ agree.
\item With probability $1/4$, instructs Alice to measure $W(\mathbf{y} \oplus \mathbf{w})$ and $W(\mathbf{v})$, where $\mathbf{y}, \mathbf{w} \in \{0, 1\}^N$ are chosen at random. Bob is instructed to measure $W(\mathbf{y})$ and $W(\mathbf{y} \oplus \mathbf{w})$. The verifier accepts if the outcomes of Alice and Bob for $W(\mathbf{y} \oplus \mathbf{w})$ agree.
\end{itemize}
\end{itemize}

The self-testing result guarantees that if these tests succeed, the verifier obtains an estimate for the energy of the ground state. Importantly, unlike the FH protocol, her estimate has constant precision. However, the protocol, as described up to this point, will still have an inverse polynomial completeness-soundness gap given by the local Hamiltonian. Recall that this is because the Feynman-Kitaev state will have energy below $a$ when $\mathcal{C}$ accepts $x$ with high probability, and energy above $b$ otherwise, where $b - a > 1/|\mathcal{C}|^2$.
But one can easily boost the protocol to a constant gap between completeness and soundness by simply requiring the provers to share $M = O(|\mathcal{C}|^2)$ copies of the ground state.
This new state, $\ket{\psi}^{\otimes M}$, would then be the ground state of a new Hamiltonian $H'$\footnote{Note that the state still needs to be encoded in the $7$ qubit code.}. One then runs the NV protocol for this Hamiltonian. It should be mentioned that this Hamiltonian is no longer $2$-local, however, all of the tests in the NV protocol apply for these general Hamiltonians as well (as long as each term is comprised of $I$, $\xgate$ and $\zgate$ operators, which is the case for $H'$). Additionally, the new Hamiltonian has a constant gap.  
The protocol therefore achieves a constant number of rounds of interaction with the provers ($2$ rounds) and we have that:
\begin{theorem}
The NV protocol is an $\mathsf{MIP^*}$ protocol achieving a constant gap between completeness and soundness.
\end{theorem} 
To then boost the completeness-soundness gap to $1-\epsilon$, for some $\epsilon > 0$, one can perform a parallel repetition of the protocol $O(log(1/\epsilon))$ times.

\subsection{Summary of entanglement-based protocols}
We have seen that having non-communicating provers sharing entangled states allows for verification protocols with a classical client. What all of these protocols have in common is that they all make use of self-testing results. These essentially state that if a number of non-communicating players achieve a near optimal win rate in a non-local game, the strategy they employ in the game is essentially fixed, up to a local isometry. The strategy of the players consists of their shared quantum state as well as their local observables. Hence, self-testing results provide a precise characterisation for both.

This fact is exploited by the surveyed protocols in order to achieve verifiability. Specifically, we have seen that one approach is to define a number of non-local games so that by combining the optimal strategies of these games, the provers effectively perform a universal quantum computation. This is the approach employed by the RUV protocol \cite{ruv}. Alternatively, the self-testing result can be used to check only for the correct preparation of a specific resource state. This resource state is then used by the provers to perform a quantum computation. 
How this is done depends on the type of resource state and on how the computation is delegated to the provers. For instance, one possibility is to remotely prepare the resource state used in the VUBQC protocol and then run the verification procedure of that protocol. This is the approach used by the GKW and HPDF protocols \cite{gkw, hpdf}. 
Another possibility is to prepare a cluster state shared among many provers and then have each of those provers measure their states so as to perform an MBQC computation. This approach was used by McKague in his protocol \cite{mckague}.
Lastly, the self-tested resource state can be the ground state of a local Hamiltonian leading to the post hoc approaches employed by the FH and NV protocols.

\begin{table}[htb] 
\centering
\begin{tabular}{l*{5}{c}} 
\toprule
 \bfseries Protocol  \hspace{0.1in} & \bfseries Provers \hspace{0.1in} & \bfseries Qmem provers \hspace{0.1in} & \bfseries Rounds \hspace{0.1in} & \bfseries Communication \hspace{0.1in} & \bfseries Blind \\
\midrule
RUV      & $2$              & $2$ & $O(N^{8192} \cdot log(1/\epsilon))$    & $O(N^{8192} \cdot log(1/\epsilon))$    & Y  \\
McKague  & $O(N^{22} \cdot log(1/\epsilon))$      & $0$ & $O(N^{22} \cdot log(1/\epsilon))$      & $O(N^{22} \cdot log(1/\epsilon))$      & Y  \\
GKW      & $2$              & $1$ & $O(N^{2048}  \cdot log(1/\epsilon))$   & $O(N^{2048}  \cdot log(1/\epsilon))$   & Y  \\
HPDF     & $O(N^4 log(N) \cdot log(1/\epsilon))$  & $O(log(1/\epsilon))$ & $O(N^4 log(N)  \cdot log(1/\epsilon))$ & $O(N^4 log(N)  \cdot log(1/\epsilon))$ & Y  \\
FH       & $5$              & $5$ & $O(N^{16} \cdot log(1/\epsilon))$      & $O(N^{19}  \cdot log(1/\epsilon))$        & N  \\
NV       & $7$              & $7$ & $O(1)$                                 & $O(N^3  \cdot log(1/\epsilon))$        & N  \\ \bottomrule
\end{tabular}
\begin{center}
\caption{Comparison of entanglement-based protocols. We denote $N = |\mathcal{C}|$ to be the size of the delegated quantum computation together with the input to that computation. The listed values are given assuming a completeness-soundness gap of at least $1 - \epsilon$, for some $\epsilon > 0$. For the ``Qmem provers'' column, the numbers indicate how many provers need to have a quantum memory that is not of constant size, with respect to $|\mathcal{C}|$ (if we ignore the preparation of the initial shared entangled state). The ``Rounds'' column quantifies how many rounds of interaction are performed between the verifier and the provers, whereas ``Communication'' quantifies the total amount of communication (number of rounds times the size of the messages). Note that a similar table can be found in \cite{leash}.}
\label{tab:entbased}
\end{center}
\end{table}

We noticed that, depending on the approach that is used, there will be different requirements for the quantum operations of the provers. Of course, all protocols require that collectively the provers can perform $\cBQP$ computations, however, individually some provers need not be universal quantum computers. 
Related to this is the issue of blindness. Again, based on what approach is used some protocols utilize blindness and some do not. In particular, the post hoc protocols are not blind since the computation and the input are revealed to the provers so that they can prepare the Feynman-Kitaev state.

We have also seen that the robustness of the self-testing game impacts the communication complexity of the protocol. Specifically, having robustness which is inverse polynomial in the number of qubits of the self-tested state, leads to an inverse polynomial gap between completeness and soundness. In order to make this gap constant, the communication complexity of the protocol has to be made polynomial. This means that most protocols will have a relatively large overhead, when compared to prepare-and-send or receive-and-measure protocols.
Out of the surveyed protocols, the NV protocol is the only one which utilizes a self-testing result with constant robustness and therefore has a constant completeness-soundness gap.
We summarize all of these facts in Table~\ref{tab:entbased}\footnote{Note that for the HPDF protocol we assumed that there is one prover with quantum memory, comprised of the individual provers that come together in order to perform the MBQC computation at the end of the protocol. Since, to achieve a completeness-soundness gap of $1 - \epsilon$. the protocol is repeated $O(log(1/\epsilon))$ times, this means there will be $O(log(1/\epsilon))$ provers with quantum memory in total.}.

\section{Outlook} \label{sect:outlook}

\subsection{Sub-universal protocols} \label{subsect:others}
So far we have presented protocols for the verification of universal quantum computations, i.e. protocols in which the provers are assumed to be $\cBQP$ machines. In the near future, however, quantum computers might be more limited in terms of the type of computations that they can perform. 
Examples of this include the class of so-called \emph{instantaneous quantum computations}, denoted $\sf{IQP}$, \emph{boson sampling} or the \emph{one-pure qubit model} of quantum computation \cite{iqp, bosonsampling, onepurequbit}. While not universal, these examples are still highly relevant since, assuming some plausible complexity theoretic conjectures hold, they could solve certain problems or sample from certain distributions that are intractable for classical computers.
One is therefore faced with the question of how to verify the correctness of outcomes resulting from these models.
In particular, when considering an interactive protocol, the prover should be restricted to the corresponding sub-universal class of problems and yet still be able to prove statements to a computationally limited verifier. 
We will see that many of the considered approaches are adapted versions of the VUBQC protocol from Subsection~\ref{subsect:fk}. It should be noted, however, that the protocols themselves are not direct applications of VUBQC. In each instance, the protocol was constructed so as to adhere to the constraints of the model.

The first sub-universal verification protocol is for the one-pure (or one-clean) qubit  model. A machine of this type takes as input a state of limited purity (for instance, a system comprising of the totally mixed state and a small number of single qubit pure states), and is able to coherently apply quantum gates. The model was considered in order to reflect the state of a quantum computer with noisy storage.
In \cite{kapourniotis14opq_verification}, Kapourniotis, Kashefi and Datta introduced a verification protocol for this model by adapting VUBQC to the one-pure qubit setting. The verifier still prepares individual pure qubits, as in the original VUBQC protocol, however the prover holds a mixed state of limited purity at all times\footnote{The purity of a $d$-qubit state, $\rho$, is quantified by the \emph{purity parameter} defined in \cite{kapourniotis14opq_verification} as: $\pi(\rho) = log(Tr(\rho^2)) + d$.}. Additionally, the prover can inject or remove pure qubits from his state, during the computation, as long as it does not increase the total purity of the state. The resulting protocol has an inverse polynomial completeness-soundness gap. However, unlike the universal protocols we have reviewed, the constraints on the prover's state do not allow for the protocol to be repeated. This means that the completeness-soundness gap cannot be boosted through repetition.

Another model, for which verification protocols have been proposed, is that of instantaneous quantum computations, or $\sf{IQP}$ \cite{iqp, iqpcol}.
An $\sf{IQP}$ machine is one which can only perform unitary operations that are diagonal in the $\xgate$ basis and therefore commute with each other. The name ``instantaneous quantum computation'' illustrates that there is no temporal structure to the quantum dynamics \cite{iqp}. Additionally, the machine is restricted to measurements in the computational basis.
It is important to mention that $\sf{IQP}$ does not represent a decision class, like $\sf{BQP}$, but rather a \emph{sampling class}. The input to a sampling problem is a specification of a certain probability distribution and the output is a sample from that distribution. The class $\sf{IQP}$, therefore, contains all distributions which can be sampled efficiently (in polynomial time) by a machine operating as described above.
Under plausible complexity theoretic assumptions, it was shown that this class is not contained in the set of distributions which can be efficiently sampled by a classical computer \cite{iqpcol}.

In \cite{iqp}, Shepherd and Bremner proposed a hypothesis test in which a classical verifier is able to check that the prover is sampling from an $\sf{IQP}$ distribution. The verifier cannot, however, check that the prover sampled from the correct distributions. Nevertheless, the protocol serves as a practical tool for demonstrating a quantum computational advantage.
The test itself involves an encoding, or obfuscation scheme which relies on a computational assumption (i.e. it assumes that a particular problem is intractable for $\sf{IQP}$ machines).

Another test of $\sf{IQP}$ problems is provided by the Hangleiter et al approach, from Subsection~\ref{subsect:posthoc} \cite{hangleiter2017direct}. Recall that this was essentially the 1S-Post-hoc protocol for certifying the ground state of a local Hamiltonian. 
Hangleiter et al have the prover prepare multiple copies of a state which is the Feynman-Kitaev state of an $\sf{IQP}$ circuit. 
They then use the post hoc protocol to certify that the prover prepared the correct state (measuring local terms from the Hamiltonian associated with that state) and then use one copy to sample from the output of the $\sf{IQP}$ circuit. This is akin to the measurement-only approach of Subsection~\ref{subsect:monly}.
In a subsequent paper, by Bermejo-Vega et al, they consider a subclass of sampling problems that are contained in $\sf{IQP}$ and prove that this class is also hard to classically simulate (subject to standard complexity theory assumptions). The problems can be viewed as preparing a certain entangled state and then measuring all qubits in a fixed basis. The authors provide a way to certify that the state prepared is close to the ideal one, by giving an upper bound on the trace distance. Moreover, the measurements required for this state certification can be made using local stabilizer measurements, for the considered architectures and settings \cite{bermejovega2017}.

Recently, another scheme has been proposed, by Mills et al \cite{mills2017information}, which again adapts the VUBQC protocol to the $\sf{IQP}$ setting. This eliminates the need for computational assumptions, however it also requires the verifier to have a single qubit preparation device. In contrast to VUBQC, however, the verifier need only prepare eigenstates of the $\ygate$ and $\zgate$ operators.

Yet another scheme derived from VUBQC was introduced in \cite{kapourniotis2017} for a model known as the \emph{Ising spin sampler}. This is based on the \emph{Ising model}, which describes a lattice of interacting spins in the presence of a magnetic field \cite{ising}. The Ising spin sampler is a translation invariant Ising model in which one measures the spins thus obtaining samples from the partition function of the model. Just like with $\sf{IQP}$, it was shown in \cite{gao2017quantum} that, based on complexity theoretic assumptions, sampling from the partition function is intractable for classical computers. 

Lastly, Disilvestro and Markham proposed a verification protocol \cite{disilvestro2017quantum} for \emph{Spekkens' toy model} \cite{spekkens2007evidence}. This is a local hidden variable theory which is phenomenologically very similar to quantum mechanics, though it cannot produce non-local correlations. 
The existence of the protocol, again inspired by VUBQC, suggests that Bell non-locality is not a necessary feature for verification protocols, at least in the setting in which the verifier has a trusted quantum device.

\subsection{Fault tolerance} \label{sect:ft}
The protocols reviewed in this paper have all been described in an ideal setting in which all quantum devices work perfectly and any deviation from the ideal behaviour is the result of malicious provers.
This is not, however, the case in the real world. The primary obstacle, in the development of scalable quantum computers, is noise which affects quantum operations and quantum storage devices. As a solution to this problem, a number of fault tolerant techniques, utilizing quantum error detection and correction, have been proposed. Their purpose is to reduce the likelihood of the quantum computation being corrupted by imperfect gate operations.
But while these techniques have proven successful in minimizing errors in quantum computations, it is not trivial to achieve the same effect for verification protocols.
To clarify, while we have seen the use of quantum error correcting codes in verification protocols, their purpose was to either boost the completeness-soundness gap (in the case of prepare-and-send protocols), or to ensure an honest behaviour from the provers (in the case of entanglement-based post hoc protocols).
The question we ask, therefore, is: how can one design a fault-tolerant verification protocol? Note that this question pertains primarily to protocols in which the verifier is not entirely classical (such as the prepare-and-send or receive-and-measure approaches) or in which one or more provers are assumed to be single-qubit devices (such as the GKW and HPDF protocols). For the remaining entanglement-based protocols, one can simply assume that the provers are performing all of their operations on top of a quantum error correcting code.

Let us consider what happens if, in the prepare-and-send and receive-and-measure protocols, the devices of the verifier and the prover are subject to noise\footnote{Different noise models have been examined when designing fault tolerant protocols, however, a very common model and one which can be considered in our case, is \emph{depolarizing noise} \cite{nc,buhrman2006new}. This can be single-qubit depolarizing noise, which acts as $\mathcal{E}(\rho) = (1 -p) [I] + p/3 ([\xgate] + [\ygate] + [\zgate])$, or two-qubit depolarizing noise, which acts as $\mathcal{E}(\rho) = (1 -p) [I \otimes I] + p/15 ([I \otimes \xgate] + ... [\zgate \otimes \zgate])$, for some probability $p > 0$. The square bracket notation indicates the action of an operator.}.
If, for simplicity, we assume that the errors on these devices imply that each qubit will have a probability, $p$, of producing the same outcome as in the ideal setting, when measured, we immediately notice that the probability of $n$ qubits producing the same outcomes scales as $O(p^n)$. This means that, even if the prover behaves honestly, the computation is very unlikely to result in the correct outcome \cite{gkw}.

Ideally, one would like the prover to perform his operations in a fault tolerant manner. In other words, the prover's state should be encoded in a quantum error correcting code, the gates he performs should result in logical operations being applied on his state and he should, additionally, perform error-detection (syndrome) measurements and corrections. But we can see that this is problematic to achieve. Firstly, in prepare-and-send protocols, the computation state of the prover is provided by the verifier. Who should then encode this state in the error-correcting code, the verifier or the prover?
It is known that in order to suppress errors in a quantum circuit, $\mathcal{C}$, each qubit should be encoded in a logical state having $O(polylog(|\mathcal{C}|))$-many qubits \cite{nc}. 
This means that if the encoding is performed by the verifier, she must have a quantum computer whose size scales poly-logarithmically with the size of the circuit that she would like to delegate. It is preferable, however, that the verifier has a constant-size quantum computer.
Conversely, even if the prover performs the encoding, there is another complication. Since the verifier needs to encrypt the states she sends to the prover, and since her operations are susceptible to noise, the errors acting on these states will have a dependency on her secret parameters. This means that when the prover performs error-detection procedures he could learn information about these secret parameters and compromise the protocol. 

For receive-and-measure protocols, one encounters a different obstacle. While the verifier's measurement device is not actively malicious, if the errors occurring in this device are correlated with the prover's operations in preparing the state, this can compromise the correctness of the protocol.

A number of fault tolerant verification protocols have been proposed, however, they all overcome these limitations by making additional assumptions. 
For instance, one proposal, by Kapourniotis and Datta \cite{kapourniotis2017}, for making VUBQC fault tolerant, uses a topological error-correcting code described in \cite{rhg1, rhg2}. The error-correcting code is specifically designed for performing fault tolerant MBQC computations, which is why it is suitable for the VUBQC protocol.
In the proposed scheme, the verifier still prepares single qubit states, however there is an implicit assumption that the errors on these states are independent of the verifier's secret parameters.
The prover is then instructed to perform a blind MBQC computation in the topological code.
The protocol described in \cite{kapourniotis2017} is used for a specific type of MBQC computation designed to demonstrate a quantum computational advantage. However, the authors argue that the techniques are general and could be applied for universal quantum computations.

A fault-tolerant version of the measurement-only protocol from Subsection~\ref{subsect:monly} has also been proposed in \cite{fujii2016verifiable}. The graph state prepared by the prover is encoded in an error-correcting code, such as the topological lattice used by the previous approaches. As in the `non-fault-tolerant' version of the protocol, the prover is instructed to send many copies of this state which the verifier will test using stabilizer measurements. The verifier also uses one copy in order to perform her computation in an MBQC fashion. 
The protocol assumes that the errors occurring on the verifier's measurement device are independent of the errors occurring on the prover's devices.

More details, regarding the difficulties with achieving fault tolerance in $\cQPIP$ protocols, can be found in \cite{abem}.

\subsection{Experiments and implementations} \label{sect:exp}
Protocols for verification will clearly be useful for benchmarking experiments implementing quantum computations. Experiments implementing quantum computations on a small number of qubits can be verified with brute force simulation on a classical computer. However, as we have pointed out that this is not scalable, in the long-term it is worthwhile to try and implement verification protocols on these devices. As a result, there have been proof of concept experiments that demonstrate the components necessary for verifiable quantum computing. 

Inspired by the prepare-and-send VUBQC protocol, Barz et al implemented a four-photon linear optical experiment, where the four-qubit linear cluster state was constructed from entangled pairs of photons produced through parametric down-conversion \cite{barz13verification_experiment}. Within this cluster state, in runs of the experiment, a trap qubit was placed in one of two possible locations, thus demonstrating some of the elements of the VUBQC protocol. However, it should be noted that the trap qubits are placed in the system through measurements on non-trap qubits within the cluster state, i.e. through measurements made on the the other three qubits. Because of this, the analysis of the VUBQC protocol cannot be directly translated over to this setting, and bespoke analysis of possible deviations is required. In addition, the presence of entanglement between the photons was demonstrated through Bell tests that are performed blindly. This work also builds on a previous experimental implementation of blind quantum computation by Barz et al \cite{barz12verification_experiment}.

With regards to receive-and-measure protocols, and in particular the measurement-only protocol of Subsection~\ref{subsect:monly}, Greganti et al implemented \cite{greganti2016} some of the elements of these protocols with a four-photon experiment, similar to the experiment of Barz et al mentioned above \cite{barz13verification_experiment}. This demonstration builds on previous work in the experimental characterisation of stabiliser states \cite{greganti2015}. In this case, two four-qubit cluster states were generated: the linear cluster state and the star graph state, where in the latter case the only entanglement is between one central qubit and pairwise with every other qubit. In order to demonstrate the elements for measurement-only verification, by suitable measurements made by the client, traps can be placed in the state. Furthermore, the linear cluster state and star graph state can be used as computational resources for implementing single qubit unitaries and an entangling gate respectively. 

Finally, preliminary steps have been taken towards an experimental implementation of the RUV protocol, from Subsection~\ref{subsect:RUV}. Huang et al implemented a simplified version of this protocol using sources of pairs of entangled photons \cite{huang2017}. Repeated CHSH tests were performed on thousands of pairs of photons demonstrating a large violation of the CHSH inequality; a vital ingredient in the protocol of RUV. In between the many rounds of CHSH tests, state tomography, process tomography, and a computation were performed, with the latter being the factorisation of the number $15$. Again, all of these elements are ingredients in the protocol, however, the entangled photons are created ‘on-the-fly’. In other words, in RUV, two non-communicating provers share a large number of maximally entangled states prior to the full protocol, but in this experiment these states are generated throughout.

\section{Conclusions}
The realization of the first quantum computers capable of outperforming classical computers at non-trivial tasks is fast approaching. All signs indicate that their development will follow a similar trajectory to that of classical computers. In other words, the first generation of quantum computers will comprise of large servers that are maintained and operated by specialists working either in academia, industry or a combination of both.
However, unlike with the first super-computers, the Internet opens up the possibility for users, all around the world, to interface with these devices and delegate problems to them.
This has already been the case with the $5$-qubit IBM machine \cite{ibm}, and more powerful machines are soon to follow \cite{ibm16, googlequantum}.
But how will these computationally restricted users be able to verify the results produced by the quantum servers? That is what the field of quantum verification aims to answer.
Moreover, as mentioned before and as is outlined in \cite{aharonov2013quantum}, the field also aims to answer the more foundational question of: how do we verify the predictions of quantum mechanics in the large complexity regime?

In this paper, we have reviewed a number of protocols that address these questions. While none of them achieve the ultimate goal of the field, which is to have a classical client verify the computation performed by a single quantum server, each protocol provides a unique approach for performing verification and has its own advantages and disadvantages. 
We have seen that these protocols combine elements from a multitude of areas including: cryptography, complexity theory, error correction and the theory of quantum correlations. We have also seen that proof-of-concept experiments, for some of these protocols, have already been realized.

What all of the surveyed approaches have in common, is that none of them are based on computational assumptions. In other words, they all perform verification unconditionally. However, recently, there have been attempts to reduce the verifier's requirements by incorporating computational assumptions as well. What this means is that the protocols operate under the assumption that certain problems are intractable for quantum computers. We have already mentioned an example: a protocol for verifying the sub-universal sampling class of $\sf{IQP}$ computations, in which the verifier is entirely classical. 
Other examples include protocols for \emph{quantum fully homomorphic encryption} \cite{qfhe1, qfhe2}. In these protocols, a client is delegating a quantum computation to a server while trying to keep the input to the computation hidden. The use of computational assumptions allows these protocols to achieve this functionality using only one round of back-and-forth communication. However, in the referenced schemes, the client does require some minimal quantum capabilities. A recent modification of these schemes has been proposed in order to make the protocols verifiable as well \cite{qfhever}.
Additionally, an even more recent paper introduces a protocol for quantum fully homomorphic encryption with an entirely classical client (again, based on computational assumptions) \cite{mahadev2017classical}.
We can therefore see a new direction emerging in the field of delegated quantum computations. This recent success in developing protocols based on computational assumptions could very well lead to the first single-prover verification protocol with a classical client.

Another new direction, especially pertaining to entanglement-based protocols, is given by the development of self-testing results achieving constant robustness. This started with the work of Natarajan and Vidick, which was the basis of their protocol from Subsection~\ref{subsect:NV} \cite{nv}. We saw, in Section~\ref{sect:entanglement}, that all entanglement-based protocols rely, one way or another, on self-testing results. Consequently, the robustness of these results greatly impacts the communication complexity and overhead of these protocols. Since most protocols were based on results having inverse polynomial robustness, this led to prohibitively large requirements in terms of quantum resources (see Table~\ref{tab:entbased}).
However, subsequent work by Coladangelo et al, following up on the Natarajan and Vidick result, has led to two entanglement-based protocols, which achieve near linear overhead \cite{leash}\footnote{The result from \cite{leash} appeared on the arxiv close to the completion of this work, which is why we did not review it.}. This is a direct consequence of using a self-testing result with constant robustness and combining it with the Test-or-Compute protocol of Broadbent from Subsection~\ref{subsect:howtoverify}.
Of course, of the two protocols proposed by Coladangelo et al, only one is blind and so an open problem, of their result, is whether the second protocol can also be made blind. Another question is whether the protocols can be further optimized so that only one prover is required to perform universal quantum computations, in the spirit of the GKW protocol from Subsection~\ref{subsect:GKW}.

We conclude by listing a number of other open problems that have been raised by the field of quantum verification. The resolution of these problems is relevant not just to quantum verification but to quantum information theory as a whole.
\begin{itemize}\addtolength{\itemsep}{+0.5\baselineskip}
\item While the problem of a classical verifier delegating computations to a single prover is the main open problem of the field, we emphasize a more particular instance of this problem: can the proof that any problem in $\sf{PSPACE}$\footnote{$\sf{PSPACE}$ is the class of problems which can be solved in polynomial space by a classical computer.} admits an interactive proof system, be adapted to show that any problem in $\sf{BQP}$ admits an interactive proof system with a $\cBQP$ prover?
The proof that $\sf{PSPACE} = \cIP$ (in particular the $\sf{PSPACE} \subseteq \sf{IP}$ direction) uses error-correcting properties of low-degree polynomials to give a verification protocol for a $\sf{PSPACE}$-complete problem \cite{shamir1992ip}. We have seen that the Poly-QAS VQC scheme, presented in Subsection~\ref{subsect:polyqas}, also makes use of error-correcting properties of low-degree polynomials in order to perform quantum verification (albeit, with a quantum error correcting code and a quantum verifier). Can these ideas lead to a classical verifier protocol for $\cBQP$ problems with a $\cBQP$ prover?
\item In all existing entanglement-based protocols, one assumes that the provers are not allowed to communicate during the protocol. However, this assumption is not enforced by physical constraints. Is it, therefore, possible to have an entanglement-based verification protocol in which the provers are \emph{space-like separated}\footnote{In an experiment, two regions are space-like separated if the time it takes light to travel from one region to the other is longer than the duration of the experiment. Essentially, according to relativity, this means that there is no causal ordering between events occurring in one region and events occurring in the other.}? Note, that since all existing protocols require the verifier to query the two (or more) provers adaptively, it is not directly possible to make the provers be space-like separated.
\item What is the optimal overhead (in terms of either communication complexity, or the resources of the verifier) in verification protocols? For all types of verification protocols we have seen that, for a fixed completeness-soundness gap, the best achieved communication complexity is linear. For the prepare-and-send case is it possible to have a protocol in which the verifier need only prepare a poly-logarithmic number of single qubits (in the size of the computation)? For the entanglement-based case, can the classical verifier send only poly-logarithmic sized questions to the provers? This latter question is related to the quantum $\sf{PCP}$ conjecture \cite{qpcp}.
\item Are there other models of quantum computation that are suitable for developing verification protocols? We have seen that the way in which we view quantum computations has a large impact on how we design verification protocols and what characteristics those protocols will have.
Specifically, the separation between classical control and quantum resources in MBQC lead to VUBQC, or the $\cQMA$-completeness of the local Hamiltonian problem lead to the post hoc approaches.
Of course, all universal models are equivalent in terms of the computations which can be performed, however each model provides a particular insight into quantum computation which can prove useful when devising new protocols. Can other models of quantum computation, such as the adiabatic model, the anyon model etc, provide new insights?
\item We have seen that while certain verification protocols employ error-correcting codes, these are primarily used for boosting the completeness-soundness gap. Alternatively, for the protocols that do in fact incorporate fault tolerance, in order to cope with noisy operations, there are additional assumptions such as the noise in the verifier's device being uncorrelated with the noise in the prover's devices. Therefore, the question is: can one have a fault tolerant verification protocol, with a minimal quantum verifier, in the most general setting possible? By this we mean that there are no restrictions on the noise affecting the quantum devices in the protocol, other than those resulting from the standard assumptions of fault tolerant quantum computations (constant noise rate, local errors etc). This question is addressed in more detail in \cite{abem}.
Note that the question refers in particular to prepare-and-send and receive-and-measure protocols, since entanglement-based approaches are implicitly fault tolerant (one can assume that the provers are performing the computations on top of error correcting codes).
\end{itemize}

\section*{Acknowledgements}
The authors would like to thank Petros Wallden, Alex Cojocaru, Thomas Vidick for very useful comments and suggestions for improving this work, and Dan Mills for \TeX \hspace{0.5pt} support. AG would also like to especially thank Matty Hoban for many helpful remarks and comments and Vivian Uhlir for useful advice in improving the figures in the paper.
EK acknowledges funding through EPSRC grant EP/N003829/1 and EP/M013243/1. TK acknowledges funding through EPSRC grant EP/K04057X/2.

\section{Appendix} \label{sect:preliminaries}
\subsection{Quantum information and computation} \label{subsect:qinfo}
In this section, we provide a few notions regarding the basics of quantum information and quantum computation and refer the reader to the appropriate references for a more in depth presentation \cite{nc, watrousintro, watrous2009quantum}.

\subsubsection{Basics of quantum mechanics.}
A quantum state (or a quantum register) is a unit vector in a complex Hilbert space, $\mathcal{H}$. We denote quantum states, using standard Dirac notation, as $\Ket{\psi} \in \mathcal{H}$, called a `ket' state. The dual of this state is denoted $\Bra{\psi}$, called a `bra', and is a member of the dual space $\mathcal{H}^{\perp}$.
We will only be concerned with finite-dimensional Hilbert spaces.
Qubits are states in two-dimensional Hilbert spaces. Traditionally, one fixes an orthonormal basis for such a space, called \emph{computational basis}, and denotes the basis vectors as $\Ket{0}$ and $\Ket{1}$. Gluing together systems to express the states of multiple qubits is achieved through \emph{tensor product}, denoted $\otimes$. The notation $\Ket{\psi}^{\otimes n}$ denotes a state comprising of $n$ copies of $\Ket{\psi}$. If a state $\Ket{\psi} \in \mathcal{H}_1 \otimes \mathcal{H}_2$ cannot be expressed as $\ket{a} \otimes \ket{b}$, for any $\ket{a} \in \mathcal{H}_1$ and any $\ket{b} \in \mathcal{H}_2$, we say that the state is \emph{entangled}. As a shorthand, we will sometimes write $\Ket{a}\ket{b}$ instead of $\ket{a} \otimes \ket{b}$.
As a simple example of an entangled state one can consider the \emph{Bell state}:
\begin{equation}
\ket{\Phi_+} = \frac{\ket{00} + \ket{11}}{\sqrt{2}}
\end{equation}

Quantum mechanics postulates that there are two ways to change a quantum state: \emph{unitary evolution} and \emph{measurement}. Unitary evolution involves acting with some unitary operation $U$ on $\Ket{\psi}$, thus producing the mapping $\Ket{\psi} \rightarrow U \Ket{\psi}$. Note that any such operation is reversible through the application of the hermitian conjugate of $U$, denoted $U^{\dagger}$, since $UU^{\dagger} = U^{\dagger}U = I$.

Measurement, in its most basic form, involves expressing a state $\ket{\psi}$ in a particular orthonormal basis, $\mathcal{B}$, and then choosing one of the basis vectors as the state of the system post-measurement. The index of that vector is the classical outcome of the measurement. The post-measurement vector is chosen at random and the probability of obtaining a vector $\ket{v} \in \mathcal{B}$ is given by $| \braket{v|\psi} |^2$. 

More generally, a measurement involves a collection of operators $\{ M_i \}_i$ acting on the state space of the system to be measured and satisfying:
\begin{equation}
\sum_i M_i^{\dagger} M_i = I
\end{equation}
The label $i$ indicates a potential measurement outcome. Given a state $\ket{\psi}$ to be measured, the probability of obtaining outcome $i$ is $p(i) = \bra{\psi} M_i^{\dagger} M_i \ket{\psi}$ and the state of the system after the measurement will be $M_i \ket{\psi} / \sqrt{p(i)}$.
If we are only interested in the probabilities of the different outcomes and not in the post-measurement state then we can denote $E_i = M_i^{\dagger} M_i$ and we will refer to the set $\{ E_i \}_i$ as a \emph{positive-operator valued measure} (POVM).
When performing a measurement in an orthonormal basis $\mathcal{B} = \{ \ket{i} \}_i$, we are essentially choosing $M_i = \ket{i}\bra{i}$. This is known as a \emph{projective measurement} and in general consists of operators $M_i$ satisfying the property that $M_i^2 = M_i$.

Lastly, when discussing measurements we will sometimes use \emph{observables}. These are hermitian operators which define a measurement specified by the diagonal basis of the operator. Specifically, for some hermitian operator $O$, we know that there exists an orthonormal basis $\mathcal{B} = \{ \ket{i} \}_i$ such that:
\begin{equation}
O = \sum_i \lambda_i \ket{i}
\end{equation}
where $\{ \lambda_i \}_i$ is the set of eigenvalues of $O$. Measuring the $O$ observable on some state $\ket{\psi}$ is equivalent to performing a projective measurement of $\ket{\psi}$ in the basis $\mathcal{B}$\footnote{Note that if the operator is degenerate (i.e. has repeating eigenvalues) then the projectors for degenerate eigenvalues will correspond to projectors on the subspaces spanned by the associated eigenvectors.}. When using observables, one takes the measurement outcomes to be the eigenvalues of $O$, rather than the basis labels. In other words, if when measuring $O$ the state is projected to $\ket{i}$, then the measurement outcome is taken to be $\lambda_i$.

\subsubsection{Density matrices.}
States denoted by kets are also referred to as \emph{pure states}.
Quantum mechanics tells us that for an isolated quantum system the complete description of that system is given by a pure state\footnote{It should be noted that this is the case provided that quantum mechanics is a \emph{complete} theory in terms of its characterisation of physical systems. See \cite{harrigan2010einstein} for more details.}.
This is akin to classical physics where pure states are points in phase space, which provide a complete characterisation of a classical system. However, unlike classical physics where knowing the pure state uniquely determines the outcomes of all possible measurements of the system, in quantum mechanics measurements are probabilistic even given the pure state.
It is also possible that the state of a quantum system is specified by a probability distribution over pure states.
This is known as a \emph{mixed state} and can be represented using \emph{density matrices}. These are positive semidefinite, trace one, hermitian operators. 

The density matrix of a pure state $\ket{\psi}$ is $\rho = \Ket{\psi} \bra{\psi}$.
For an ensemble of states $\{ \ket{\psi_i} \}_i$, each occurring with probability $p_i$, such that $\sum_i p_i = 1$, the corresponding density matrix is:
\begin{equation}
\rho = \sum_i p_i \Ket{\psi_i} \Bra{\psi_i}
\end{equation}
It can be shown that if $\rho$ corresponds to a pure state then $Tr(\rho^2) = 1$, whereas when $\rho$ is a mixed state $Tr(\rho^2) < 1$.
One of the most important mixed states, which we encounter throughout this review, is the \emph{maximally mixed state}. The density matrix for this state is $I/d$, where $I$ is the identity matrix and $d$ is the dimension of the underlying Hilbert space. As an example, the maximally mixed state for a one qubit system is $I/2$.
This state represents the state of maximal uncertainty about quantum system. What this means is that for any basis $\{\ket{v_i}\}_i$ of the Hilbert space of dimension $d$, the maximally mixed state is:
\begin{equation}
\frac{I}{d} = \frac{1}{d} \sum\limits_{i=1}^d  \Ket{v_i} \Bra{v_i}
\end{equation}
Equivalently, any non-degenerate projective measurement, specified by an orthonormal basis $\mathcal{B}$, of the maximally mixed state will have all outcomes occurring with equal probability.
We will denote the set of all density matrices over some Hilbert space $\mathcal{H}$ as $\mathcal{D}(\mathcal{H})$.

When performing a measurement on a state $\rho$, specified by operators $\{ M_i \}_i$, the probability of outcome $i$ is given by $p(i) = Tr( M_i^{\dagger} M_i \rho)$ and the post-measurement state will be $M_i \rho M_i^{\dagger} / p(i)$.

\subsubsection{Purification.}
An essential operation concerning density matrices is the \emph{partial trace}. This provides a way of obtaining the density matrix of a subsystem that is part of a larger system. Partial trace is linear, and is defined as follows. Given two density matrices $\rho_1$ and $\rho_2$ with Hilbert spaces $\mathcal{H}_1$ and $\mathcal{H}_2$, we have that:
\begin{equation}
\rho_1 = Tr_2(\rho_1 \otimes \rho_2) \; \; \; \; \; \; \; \rho_2 = Tr_1(\rho_1 \otimes \rho_2)
\end{equation}
In the first case one is `tracing out' system $2$, whereas in the second case we trace out system $1$. This property together with linearity completely defines the partial trace. For if we take any general density matrix, $\rho$, on $\mathcal{H}_1 \otimes \mathcal{H}_2$, expressed as:
\begin{equation}
\rho = \sum_{i,i',j, j'} a_{ii'jj'} \Ket{i}_1\Bra{i'}_1 \otimes \Ket{j}_2 \Bra{j'}_2
\end{equation}
where $\{\ket{i}\}$ ($\{\ket{i'}\}$) and $\{\ket{j}\}$ ($\{\ket{j'}\}$) are orthonormal bases for $\mathcal{H}_1$ and $\mathcal{H}_2$,
if we would like to trace out subsystem $2$, for example, we would then have:
\begin{equation}
Tr_2(\rho) = Tr_2 \left( \sum_{i,i',j, j'} a_{ii'jj'} \Ket{i}_1\Bra{i'}_1 \otimes \Ket{j}_2 \Bra{j'}_2 \right) =
\sum_{i,i',j} a_{ii'jj} \Ket{i}_1\Bra{i'}_1
\end{equation}

An important fact, concerning the relationship between mixed states and pure states, is that any mixed state can be \emph{purified}. In other words, for any mixed state $\rho$ over some Hilbert space $\mathcal{H}_1$ one can always find a pure state $\ket{\psi} \in \mathcal{H}_1 \otimes \mathcal{H}_2$ such that $dim(\mathcal{H}_1) = dim(\mathcal{H}_2)$\footnote{One could allow for purifications in larger systems, but we restrict attention to same dimensions.} and:
\begin{equation}
Tr_2(\Ket{\psi}\Bra{\psi}) = \rho
\end{equation}
Moreover, the purification $\ket{\psi}$ is not unique and so another important result is the fact that if $\ket{\phi} \in \mathcal{H}_1 \otimes \mathcal{H}_2$ is another purification of $\rho$ then there exists a unitary $U$, acting only on $\mathcal{H}_2$ (the additional system that was added to purify $\rho$) such that:
\begin{equation}
\ket{\phi} = (I \otimes U) \ket{\psi}
\end{equation}
We will refer to this as the \emph{purification principle}.

\subsubsection{CPTP maps and isometries.}
All operations on quantum states can be viewed as maps from density matrices on an input Hilbert space to density matrices on an output Hilbert space, $\mathcal{O} : \mathcal{D}(\mathcal{H}_{in}) \rightarrow \mathcal{D}(\mathcal{H}_{out})$, which may or may not be of the same dimension.
Quantum mechanics dictates that such a map, must satisfy three properties:
\begin{enumerate}
\item \textbf{Linearity}: $\mathcal{O}(a\rho_1 + b\rho_2) = a \mathcal{O}(\rho_1) + b \mathcal{O}(\rho_2)$.
\item \textbf{Complete positivity}: the map $\mathcal{O} \otimes I : \mathcal{D}(\mathcal{H}_{in} \otimes \mathcal{H}_E) \rightarrow \mathcal{D}(\mathcal{H}_{out} \otimes \mathcal{H}_E))$ takes positive states to positive states, for all extensions $\mathcal{H}_E$.
\item \textbf{Trace preserving}: $Tr(\mathcal{O}(\rho)) = Tr(\rho)$.
\end{enumerate}
For this reason, such maps are referred to as \emph{completely positive trace-preserving} (CPTP) maps. It can be shown that any CPTP map can be equivalently expressed as:
\begin{equation}
\mathcal{O}(\rho) = \sum_i K_i \rho K_i^{\dagger}
\end{equation}
for some set of linear operators $\{ K_i \}_i$, known as \emph{Kraus operators}, satisfying:
\begin{equation}
\sum_i K_i^{\dagger} K_i = I
\end{equation}
CPTP maps are also referred to as \emph{quantum channels}. 

Let us also define \emph{isometries}. First, let $\Phi : \mathcal{H}_{in} \rightarrow \mathcal{H}_{out}$ be a bounded linear map. The adjoint of $\Phi$, denoted $\Phi^{\dagger}$ is the unique linear map $\Phi^{\dagger} : \mathcal{H}_{out} \rightarrow \mathcal{H}_{in}$ such that for all $\ket{\psi} \in \mathcal{H}_{in}$, $\ket{\phi} \in \mathcal{H}_{out}$:
\begin{equation}
\braket{\Phi(\psi) | \phi} = \braket{\psi | \Phi^{\dagger} (\phi)}
\end{equation}
An isometry is a bounded linear map, $\Phi : \mathcal{H}_{in} \rightarrow \mathcal{H}_{out}$ such that:
\begin{equation}
\Phi^{\dagger} \circ \Phi = id
\end{equation}
where $id$ is the identity map (on $\mathcal{H}_{in}$).

\subsubsection{Trace distance.} \label{subsubsect:td}
We will frequently be interested in comparing the ``closeness'' of quantum states. To do so we will use the notion of \emph{trace distance} which generalizes \emph{variation distance} for probability distributions. Recall that if one has two probability distributions $p(x)$ and $q(x)$, over a finite sample space, the variation distance between them is defined as:
\begin{equation}
D(p, q) = \frac{1}{2} \sum_x | p(x) - q(x) |
\end{equation}
Informally, this represents the largest possible difference between the probabilities that the two distributions can assign to some even $x$. The quantum analogue of this, for density matrices, is:
\begin{equation}
TD(\rho_1, \rho_2) = \frac{1}{2} Tr \left( \sqrt{(\rho_1 - \rho_2)^{2}} \right)
\end{equation}
One could think that the trace distance simply represents the variation distance between the probability distributions associated with measuring $\rho_1$ and $\rho_2$ in the same basis (or using the same POVM). However, there are infinitely many choices of a measurement basis. So, in fact, the trace distance is the \emph{maximum} over all possible measurements of the variation distance between the corresponding probability distributions.

Similar to variation distance, the trace distance takes values between $0$ and $1$, with $0$ corresponding to identical states and $1$ to perfectly distinguishable states. Additionally, like any other distance measure, it satisfies the triangle inequality.

\subsubsection{Quantum computation.} \label{subsubsect:qcomp}
Quantum computation is most easily expressed in the \emph{quantum gates model}. In this framework, gates are unitary operations which act on groups of qubits. As with classical computation, universal quantum computation is achieved by considering a fixed set of quantum gates which can approximate any unitary operation up to a chosen precision. The most common universal set of gates is given by:
\begin{equation}
\xgate = \begin{bmatrix}
0 & 1 \\
1 & 0 \\
\end{bmatrix} \;
\zgate = \begin{bmatrix}
1 & 0 \\
0 & -1 \\
\end{bmatrix} \;
\hgate = \frac{1}{\sqrt{2}} \begin{bmatrix}
1 & 1 \\
1 & -1 \\
\end{bmatrix} \;
\tgate = \begin{bmatrix}
1 & 0 \\
0 & e^{i\pi/4} \\
\end{bmatrix} \;
\cnot = \begin{bmatrix}
		1 & 0 & 0 & 0 \\
		0 & 1 & 0 & 0 \\
		0 & 0 & 0 & 1 \\
		0 & 0 & 1 & 0 \\		
		\end{bmatrix}
\end{equation}
In order, the operations are known as Pauli $\xgate$ and Pauli $\zgate$, Hadamard, the $\tgate$-gate and controlled-NOT.
 Note that general controlled-$U$ operations are operations performing the mapping $\Ket{0}\ket{\psi} \rightarrow \ket{0}\ket{\psi}$, $\ket{1}\ket{\psi} \rightarrow \ket{1} U\ket{\psi}$. The first qubit is known as a \emph{control qubit}, whereas the second is known as \emph{target qubit}. The matrices express the action of each operator on the computational basis.
A classical outcome for a particular quantum computation can be obtained by measuring the quantum state resulting from the application of a sequence of quantum gates. Another gate, which we will encounter, is the \emph{Toffoli} gate, or the controlled-controlled-NOT gate, described by the matrix:
\begin{equation}
\sf{CCNOT} = \begin{bmatrix}
1 & 0 & 0 & 0 & 0 & 0 & 0 & 0 \\
0 & 1 & 0 & 0 & 0 & 0 & 0 & 0 \\
0 & 0 & 1 & 0 & 0 & 0 & 0 & 0 \\
0 & 0 & 0 & 1 & 0 & 0 & 0 & 0 \\
0 & 0 & 0 & 0 & 1 & 0 & 0 & 0 \\
0 & 0 & 0 & 0 & 0 & 1 & 0 & 0 \\
0 & 0 & 0 & 0 & 0 & 0 & 0 & 1 \\
0 & 0 & 0 & 0 & 0 & 0 & 1 & 0 \\
\end{bmatrix}
\end{equation}
The effect of this gate is to apply an $\xgate$ on a target qubit if both control qubits are in the $\ket{1}$ state.

We also mention an important class of quantum operations known as \emph{Clifford operations}. To define them, consider first the $n$-qubit Pauli group:
\begin{equation}
\mathds{P}_n = \{ \alpha \; \sigma_1 \otimes ... \otimes \sigma_n | \alpha \in \{+1, -1, +i, -i\}, \sigma_i \in \{ I, \xgate, \ygate, \zgate \} \}
\end{equation}
As a useful side note, the $n$-qubit Pauli group forms a basis for all $2^n \times 2^n$ matrices.
The \emph{Clifford group} is then defined as follows:
\begin{equation}
\mathfrak{C}_n = \{ U \in U(2^n) | \sigma \in \mathds{P}_n \implies U \sigma U^{\dagger} \in \mathds{P}_n \}
\end{equation}
Where $U(2^n)$ is the set of all $2^n \times 2^n$ unitary matrices. Clifford operations, therefore, are operations which leave the Pauli group invariant under conjugation (in other words, they normalise the Pauli group). Operationally they can be obtained through combinations of the Pauli gates together with $\hgate$, $\cnot$ and $\sgate = \tgate^2$, in which case they are referred to as \emph{Clifford circuits}.
We note that the $\tgate$ and Toffoli gates are not Clifford operations. However, Clifford circuits combined with either of these two gates gives a universal set of quantum operations.

\subsubsection{Bloch sphere.}
\begin{figure}[htbp!]
\centering
\includegraphics[scale=0.27]{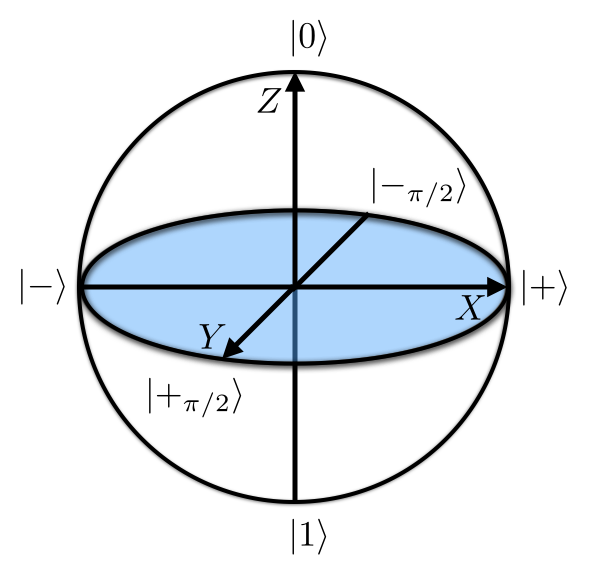}
\caption{Bloch sphere}
\label{fig:bloch}
\end{figure}

The final aspect we mention is the \emph{Bloch sphere}, which offers a useful geometric picture for visualizing single qubit states. Any such state is represented as a point on the surface of the sphere. In Figure~\ref{fig:bloch}, one can see a visualization of the Bloch sphere together with the states $\Ket{0}, \Ket{1}$, the eigenstates of $\zgate$, as well as $\Ket{+} = \frac{1}{\sqrt{2}}(\Ket{0} + \Ket{1}), \Ket{-} = \frac{1}{\sqrt{2}}(\Ket{0} - \Ket{1})$, the eigenstates of $\xgate$ and $\Ket{+_{\pi/2}} = \frac{1}{\sqrt{2}}(\Ket{0} + i\Ket{1}), \Ket{-_{\pi/2}} = \frac{1}{\sqrt{2}}(\Ket{0} - i\Ket{1})$, the eigenstates of $\ygate$.
All of the previously mentioned single-qubit operations can be viewed as rotations on this sphere. The Pauli $\xgate, \ygate, \zgate$ gates correspond to rotations by $\pi$ radians around the corresponding $\xgate, \ygate, \zgate$ axes. The Hadamard gate, which can be expressed as $\hgate = \frac{1}{\sqrt{2}}(\xgate + \zgate)$ acts as a rotation by $\pi$ radians around the $\xgate + \zgate$ axis. Lastly, the $\tgate$ gate, corresponds to a rotation by $\pi/4$ radians around the $\zgate$ axis.

We will frequently mention the states $\ket{+_{\phi}} = \frac{1}{\sqrt{2}}(\Ket{0} + e^{i\phi}\Ket{1})$ and $\ket{-_{\phi}} = \frac{1}{\sqrt{2}}(\Ket{0} - e^{i\phi}\Ket{1})$ which all lie in the $\xgate\ygate$-plane of the Bloch sphere, represented in blue in the above figure. These states can be viewed as rotations of the $\ket{+}, \ket{-}$ states by $\phi$ radians around the $\zgate$ axis. For example, the $\ket{+_{\pi/2}}, \ket{-_{\pi/2}}$ states are rotations by $\pi/2$ around the $\zgate$ axis of the $\ket{+}, \ket{-}$ states.
One can also consider measurements in the $\xgate\ygate$-plane. Any two diametrically opposed states in this plane form a basis for a one-qubit Hilbert space and therefore define a projective measurement. Suppose we choose the basis $(\ket{+_{\phi}}, \ket{-_{\phi}})$ and wish to measure the state $\ket{+_{\theta}}$. It can be shown that the probability of the state being projected to $\ket{+_{\phi}}$ is $cos^2((\phi - \theta)/2)$, whereas the probability of it being projected to $\ket{-_{\phi}}$ is $sin^2((\phi - \theta)/2)$. In other words, the probabilities only depend on the \emph{angle difference} between $\phi$ and $\theta$. This fact will prove very useful later on.

\subsubsection{Quantum error correction}
One important consideration, when discussing quantum protocols, is that any implementation of quantum operations will be subject to noise stemming from interactions with the external environment. For this reason, one needs a \emph{fault tolerant} way of performing quantum computation.
This is achieved using protocols for quantum error detection and correction, for which we give a simplified description. 

Suppose we have a $k$-qubit quantum state $\ket{\psi}$ on which we want to perform some quantum gate $G$. The quantum memory storing $\ket{\psi}$ as well as the implementation of $G$ are subject to noise. This means that if we were to apply $G$ directly on $\ket{\psi}$ the result would be $\mathcal{E}(G \ket{\psi})$, where $\mathcal{E}$ is a CPTP error map associated with the noisy application of $G$. Using the Kraus decomposition, the action of $\mathcal{E}$ can be expressed as:
\begin{equation}
\mathcal{E}(G \ket{\psi}) = \sum_j E_j \; G \ket{\psi}\bra{\psi} G^{\dagger} \; E_j^{\dagger}
\end{equation}
where $\{ E_j \}_j$ is a set of Kraus operators. If one can correct for all $E_j$'s then one can correct for $\mathcal{E}$ as well \cite{gottesman2009introduction}.

To detect and correct for errors from the set $\{ E_j \}_j$, one first performs an encoding procedure on $\ket{\psi}$ mapping it to a so-called \emph{logical state} $\ket{\psi}_L$ on $n$ qubits, where $n > k$. 
This procedure involves the use of $n - k$ auxiliary qubits known as \emph{ancilla} qubits. If we denote the state of these $n - k$ ancillas as $\ket{anc}$ we then have the encoding procedure $Enc(\ket{\psi}\ket{anc}) \rightarrow \ket{\psi}_L$.
This state is part of a $2^k$-dimensional subspace of the $2^n$-dimensional Hilbert space of all $n$ qubits, denoted $\mathcal{H}$. The subspace is usually referred to as the \emph{code space} of the error correcting code.
One way to represent this space is by giving a set of operators such that the code space is the intersection of the $+1$ eigenspaces of all the operators. 

As an example, consider the $3$-qubit \emph{flip code}. We will take $k=1$ and $n=3$, so that one qubit is encoded in $3$ qubits. The code is able to detect and correct for Pauli $\xgate$ errors occurring on \emph{a single} qubit.
The encoding procedure for a state $\ket{\psi} = a\ket{0} + b\ket{1}$ maps it to the state $\ket{\psi}_L = a\ket{000} + b\ket{111}$. The code space is therefore defined by $span(\ket{000}, \ket{111})$. It is also the unique $+1$ eigenspace of the operators $g_1 = Z \otimes Z \otimes I$ and $g_2 = I \otimes Z \otimes Z$\footnote{These are known as \emph{stabilizer} operators for the states in the code spaces. We also encounter these operators in Subsection~\ref{subsect:mbqc}. The operators form a group under multiplication and so, when specifying the code space, it is sufficient to provide the generators of the group.}.
All valid operations on $\ket{\psi}_L$ must be invariant on this subspace, whereas any error from the set $\{ E_j \}_j$ should map the state to a different subspace.
In this case, valid operations, or \emph{logical operations}, are the analogues of the single-qubit unitaries that map $\ket{\psi} \rightarrow \ket{\phi} = U\ket{\psi}$. Thus, a logical operation $U_L$ would map $\ket{\psi}_L \rightarrow \ket{\phi}_L$.
The error set simply consists of $\{ \xgate \otimes I \otimes I, I \otimes \xgate \otimes I, I \otimes I \otimes \xgate \}$. 
We can see that any of these errors will map a state inside $span(\ket{000}, \ket{111})$ to a state outside of this code space.
One then defines a projective measurement in which the projectors are associated with each of the $2^{n - k}$ subspaces of $\mathcal{H}$. This is called a \emph{syndrome measurement}. Its purpose is to detect whether an error has occurred and, if so, which error. Knowing this, the effect of the error can be undone by simply applying the inverse operation.
For the $3$-qubit code, there are $2^{3-1} = 4$ possible subspaces in which the state can be mapped to, meaning that we need a $4$-outcome measurement. The syndrome measurement is defined by jointly measuring the observables $g_1$ and $g_2$. An outcome of $+1$ for both observables indicates that the state is in the correct subspace, $span(\ket{000}, \ket{111})$. Conversely, if either of the two observables produces a $-1$ outcome, then this corresponds to one of the $3$ possible errors.
For instance, an outcome of $+1$ for the first observable and $-1$ for the second, indicates that the state is in the subspace $span(\ket{001}, \ket{110})$, corresponding to an $\xgate$ error on the third qubit.
The error is corrected by applying another $\xgate$ operation on that qubit.

Since Kraus operators can be expressed in terms of Pauli matrices acting on the individual qubits, one often speaks about the \emph{weight} of an error correcting code. If the code can correct non-identity Pauli operations on at most $w$ qubits, then $w$ is the weight of the code.

The smallest error correcting code which can correct for \emph{any} single-qubit error is the \emph{$5$-qubit code} (i.e. one qubit is encoded as $5$ qubits) \cite{fivequbit}. This code is used Subsection~\ref{subsect:entposthoc}.

\subsection{Measurement-based quantum computation} \label{subsect:mbqc}
Since some of the protocols we review are expressed in the model of \emph{measurement-based quantum computation} (MBQC), defined in \cite{mbqc1, mbqc2, raussendorf2003measurement}, we provide a brief description of this model.

Unlike the quantum gates model of computation, in MBQC a given computation is performed by measuring qubits from a large entangled state.
Traditionally, this state consists of qubits prepared in the state $\Ket{+} = \frac{1}{\sqrt{2}}(\Ket{0} + \Ket{1})$, entangled using the $\cz$ (controlled-$\zgate$) operation, where:
\[
\mathsf{CZ} = \begin{bmatrix}
		1 & 0 & 0 & 0 \\
		0 & 1 & 0 & 0 \\
		0 & 0 & 1 & 0 \\
		0 & 0 & 0 & -1 \\		
		\end{bmatrix}
\]
They are then measured in the basis $(\Ket{+_{\phi}}, \Ket{-_{\phi}})$.
These measurements are denoted as $M(\phi)$, and depending on the value of $\phi$ chosen for each qubit one can perform universal quantum computation. For this to work, the entangled qubits need to form a \emph{universal graph state}. 
A graph state, denoted $\ket{G}$, is one in which the qubits have been entangled according to the structure of a graph $G$.
Given some fixed constant $k$, a universal graph state is a family of graph states, denoted $\{ \ket{G_N} \}_N$, with $N>0$, and having $kN$ qubits, such that, for any quantum circuit $\mathcal{C}$, consisting of $N$ gates, there exists a measurement pattern\footnote{A measurement pattern is simply a tuple consisting of the measurement angles, for the qubits in $\ket{G_N}$, and the partial ordering of these measurements.} on $\ket{G_N}$ that implements $\mathcal{C}\ket{00..0}$.
In other words, for each quantum circuit of size $N$, there is an MBQC computation using $\ket{G_N}$ that performs that circuit.

\begin{figure}[htbp!]
\centering
\includegraphics[scale=1.2]{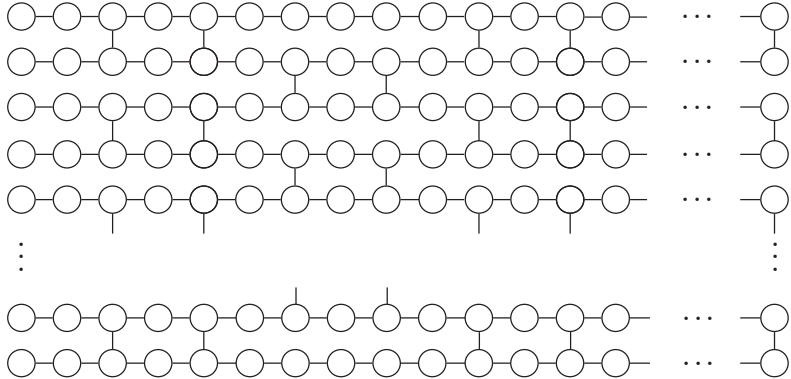}
\caption{Brickwork state, reproduced from \cite{ubqc}}
\label{fig:brickwork}
\end{figure}

An example of such a state is the \emph{brickwork state}, defined in \cite{ubqc} from which we illustrate Figure~\ref{fig:brickwork}. To be more precise, suppose we would like to perform some quantum computation described by a circuit consisting of $N$ gates. The corresponding MBQC computation consists of the following steps:
\begin{enumerate}
\item \textbf{Initialization}. Prepare $O(N)$ qubits, each in the state $\Ket{+}$.
\item \textbf{Entanglement}. Entangle the qubits according to some universal graph state structure, such as the brickwork state.
\item \textbf{Measurement}. Measure each qubit, $i$ using $M(\phi_i)$, for some angle $\phi_i$ determined based on the computation we would like to perform. The angles $\phi_i$ are referred to as the \emph{computation angles}.
\item \textbf{Correction}. Apply appropriate corrections (Pauli $\xgate$ and $\zgate$ operations) to the qubits, based on the measurement outcomes.
\end{enumerate}

\noindent The last two steps can be performed together. This is because if we would like to apply a Pauli $\xgate$ correction to a qubit, $i$, before measuring it, we can simply measure it using $M(-\phi_i)$. Similarly, if we would like to apply a Pauli $\zgate$ correction to that same qubit we measure it using $M(\phi_i + \pi)$. Therefore, the general measurement performed on a particular qubit will be $M((-1)^s \phi_i + r \pi)$, where $s, r \in \{0, 1\}$ are determined by previous measurement outcomes.

One element concerning graph states, which we will encounter in some protocols, is the representation of these states using \emph{stabilizers}. A stabilizer state for a unitary hermitian operator, $O$, is some state $\ket{\psi}$ such that $O\ket{\psi} = \ket{\psi}$. $O$ is referred to as a stabilizer of $\ket{\psi}$. It is possible to specify a state, $\ket{\psi}$, by giving a set of operators, such that $\ket{\psi}$ is the unique state (up to global phase) which is stabilized by all the operators in the set.
As an example, the state $\ket{\Phi_+} = (\ket{00} + \ket{11})/\sqrt{2}$ is uniquely stabilized by the set $\{ \xgate \otimes \xgate, \zgate \otimes \zgate \}$. Note that the set of all stabilizers for a state forms a group, since if $O_1 \ket{\psi} = \ket{\psi}$ and $O_2 \ket{\psi} = \ket{\psi}$, then clearly $O_1 O_2 \ket{\psi} = \ket{\psi}$. So, it is sufficient to specify a set of generators for that group in order to describe the stabilizer group for a particular state.

To specify the generators for the stabilizer group of a graph state $\ket{G}$, let us first denote $V(G)$ as the set of vertices in the graph $G$ and $N_G(v)$ as the set of neighbouring vertices for some vertex $v$ (i.e. all vertices in $G$ that are connected to $v$ through an edge). Additionally, for some operator $O$, when we write $O_v$ we mean that $O$ is acting on the qubit from $\ket{G}$ associated with vertex $v$ in $G$.
The generators for the stabilizer group of $\ket{G}$ are then given by:
\begin{equation}
K_v = \xgate_v \prod_{w \in N_G(v)} \zgate_w
\end{equation}
for all $v \in V(G)$.

As a final remark, it should be noted that one can translate quantum circuits into MBQC patterns in a canonical way. For instance, the universal gate set mentioned in the previous subsection, and hence any quantum circuit comprising of those gates, can be translated directly into MBQC. See for example \cite{raussendorf2003measurement, ubqc} for more details.

\subsection{Complexity theory} \label{subsect:complexity}
As mentioned in the introduction, the questions regarding verification of quantum computation can be easily expressed in the language of complexity theory. To that end, we provide definitions for the basic complexity classes used in this paper.
We let $\{0, 1\}^*$ denote the set of all binary strings of finite length and $\{0, 1\}^n$ the set of all binary strings of length $n$. We use standard complexity theory notation and assume familiarity with the concepts of Turing machines and uniform circuits. For a more general introduction into the subject we refer the reader to \cite{zoo, arorabarak}.

\begin{definition}
A language $L \subseteq \{0, 1\}^*$ belongs to $\sf{BPP}$ if there exists a polynomial $p$ and a probabilistic Turing machine $M$, whose running time on inputs of size $n$ is bounded by $p(n)$, such that for any $x \in \{0, 1\}^{n}$ the following is true:
\begin{itemize}
\item when $x \in L$, $M(x)$\footnote{The notation $M(x)$ means running the Turing machine $M$ on input $x$.} accepts with probability at least $c$,
\item when $x \not\in L$, $M(x)$ accepts with probability at most $s$,
\end{itemize}
where $c - s \geq 1/p(n)$.
\end{definition}
Here, and in all subsequent definitions, $c$ is referred to as \emph{completeness} and $s$ is referred to as \emph{soundness}. Traditionally, one takes $c = 2/3$ and $s = 1/3$, however, in full generality, the only requirement is that there exists an inverse polynomial gap between $c$ and $s$.

\begin{definition}
A language $L \subseteq \{0, 1\}^*$ belongs to $\sf{BQP}$ if there exists a polynomial $p$ and a uniform quantum circuit family $\{C_{n}\}_{n}$, where each circuit has at most $p(n)$ gates, such that for any $x \in \{0, 1\}^{n}$ the following is true:
\begin{itemize}
\item when $x \in L$, $C_{n}(x)$ accepts with probability at least $c$,
\item when $x \not\in L$, $C_{n}(x)$ accepts with probability at most $s$,
\end{itemize}
where $c - s \geq 1/p(n)$. 
\end{definition}
For the quantum circuit $C_n$, acceptance can be defined as having one of its output qubits outputting $1$ when measured in the computational basis.

\begin{definition}
A language $L \subseteq \{0, 1\}^*$ belongs to $\sf{MA}$ if there exists a polynomial $p$ and a probabilistic Turing machine $V$, whose running time on inputs of size $n$ is bounded by $p(n)$, such that for any $x \in \{0, 1\}^{n}$ the following is true:
\begin{itemize}
\item when $x \in L$, there exists a string $w \in \{0,1\}^{\leq p(n)}$, such that $V(x, w)$ accepts with probability at least $c$,
\item when $x \not\in L$, for all strings $w \in \{0,1\}^{\leq p(n)}$, $V(x,w)$ accepts with probability at most $s$,
\end{itemize}
where $c - s \geq 1/p(n)$.
\end{definition}
For this class, $V$ is traditionally referred to as the verifier (or Arthur), whereas $w$, which is the witness string, is provided by the prover (or Merlin). Essentially, the verifier and is tasked with checking a purported proof that $x \in L$, provided by the prover. There is also a quantum version of this class:

\begin{definition}
A language $L \subseteq \{0, 1\}^*$ belongs to $\sf{QMA}$ if there exists a polynomial $p$ and a uniform quantum circuit family $\{V_{n}\}_{n}$ taking $x$ and a quantum state $\ket{\psi}$ as inputs, such that for any $x \in \{0, 1\}^n$ the following are true:
\begin{itemize}
\item when $x \in L$, there exists a quantum state $\ket{\psi} \in \mathcal{H}$, such that $V_{n}(x, \ket{\psi})$ accepts with probability at least $c$, and
\item when $x \not\in L$, for all quantum states $\ket{\psi} \in \mathcal{H}$, $V_{n}(x, \ket{\psi})$ accepts with probability at most $s$,
\end{itemize}
where $dim(\mathcal{H}) \leq 2^{p(|x|)}$ and $c - s \geq 1/p(|x|)$.
\end{definition}

For $\cQMA$ we also provide the definition of a complete problem\footnote{A problem, $P$, is complete for the complexity class $\cQMA$ if $P \in \cQMA$ and all problems in $\cQMA$ can be reduced in quantum polynomial time to $P$.} since this will be referenced in some of the protocols we review. The specific problem we state was defined by Kitaev et al and is known as the \emph{k-local Hamiltonian problem} \cite{kitaev2002classical}.
A $k$-local Hamiltonian, acting on a system of $n$ qubits, is a hermitian operator $H$ that can be expressed as $H = \sum_{i} H_i$, where each $H_i$ is a hermitian operator which acts non-trivially on at most $k$ qubits.
We give the definition of the $k$-local Hamiltonian problem from \cite{qpcp}:
\begin{definition}[The $k$-local Hamiltonian (LH) problem] \ 
\label{def:LH}
  \begin{itemize}
    \item  \textbf{Input:} $H_1,\ldots,H_m$, a set of $m$ Hermitian matrices
      each acting on $k$ qubits  
      out of an $n$-qubit system and
      satisfying $\|H_i\|\le 1$. Each matrix entry is specified by
      $poly(n)$-many bits. Apart from the $H_i$ we are also given
      two real numbers, $a$ and $b$ (again, with polynomially many
      bits of precision) such that $\Gamma = b-a>1/poly(n)$.
      $\Gamma$ is referred to as the \emph{absolute promise gap} of
      the problem.

    \item \textbf{Output:} Is the smallest eigenvalue 
      of $H=H_1+H_2+...+H_m$ smaller than $a$ or are all its
      eigenvalues larger than $b$?
  \end{itemize} 
\end{definition}
Essentially, for some language $L \in \sf{QMA}$, and given $a$ and $b$, one can construct a $k$-local Hamiltonian such that, whenever $x \in L$, its smallest eigenvalue is less than $a$ and whenever $x \not\in L$, all of its eigenvalues are greater than $b$.
The witness $\ket{\psi}$, when $x \in L$, is the eigenstate of $H$ corresponding to its lowest eigenvalue (or one such eigenstate if the Hamiltonian is degenerate). The uniform circuit family $\{V_n\}_n$ represents a $\cBQP$ verifier, whereas the state $\ket{\psi}$ is provided by a prover. The verifier receives this witness from the prover and measures one of the local terms $H_i$ (which is an observable) on that state. This can be done with a polynomial-size quantum circuit and yields an estimate for measuring $H$ itself. Therefore, when $x \in L$ and the prover sends $\ket{\psi}$, with high probability the verifier will obtain the corresponding eigenvalue of $\ket{\psi}$ which will be smaller than $a$. Conversely, when $x \not\in L$, no matter what state the prover sends, with high probability, the verifier will measure a value above $b$.
The constant $k$, in the definition, is not arbitrary. In the initial construction of Kitaev, $k$ had to be at least $5$ for the problem to be $\sf{QMA}$-complete. Subsequent work has shown that even with $k=2$ the problem remains $\sf{QMA}$-complete \cite{kempe2006complexity}.

\begin{definition}
A language $L \subseteq \{0, 1\}^*$ belongs to $\sf{IP}$ if there exists a polynomial $p$ and a probabilistic Turing machine $V$, whose running time on inputs of size $n$ is bounded by $p(n)$, such that for any $x \in \{0, 1\}^{n}$ the following is true:
\begin{itemize}
\item when $x \in L$, there exists a prover $P$ which exchanges at most $p(n)$ messages (of length at most $p(n)$) with $V$ and makes $V$ accept with probability at least $c$,
\item when $x \not\in L$, any prover $P$ which exchanges at most $p(n)$ messages (of length at most $p(n)$) with $V$, makes $V$ accept with probability at most $s$,
\end{itemize}
where $c - s \geq 1/p(n)$.
\end{definition}

While the previous are fairly standard complexity classes, we now state the definition of a more non-standard class, which first appeared in \cite{abe}:
\begin{definition}\label{def:QPIP}
A language $L \subseteq \{0, 1\}^*$ belongs to $\sf{QPIP}$ if there exists a polynomial $p$, a constant $\kappa$ and a probabilistic Turing machine $V$, whose running time on inputs of size $n$ is bounded by $p(n)$, and which is augmented with the ability to prepare and measure groups of $\kappa$ qubits, such that for any $x \in \{0, 1\}^{n}$ the following is true:
\begin{itemize}
\item when $x \in L$, there exists a $\cBQP$ prover $P$ which exchanges at most $p(n)$ classical or quantum messages (of length at most $p(n)$) with $V$ and makes $V$ accept with probability at least $c$,
\item when $x \not\in L$, any $\cBQP$ prover $P$ which exchanges at most $p(n)$ classical or quantum messages (of length at most $p(n)$) with $V$, makes $V$ accept with probability at most $s$,
\end{itemize}
where $c - s \geq 1/p(n)$.
\end{definition}
Some clarifications are in order. The class $\cQPIP$ differs from $\cIP$ in two ways. Firstly, while computationally the verifier is still restricted to the class $\cBPP$, operationally it has the additional ability of preparing or measuring groups of $\kappa$ qubits. Importantly, $\kappa$ is a constant which is independent of the size of the input. This is why this extra ability does not add to the verifier's computational power, since a constant-size quantum device can be simulated in constant time by a $\cBPP$ machine.
Secondly, unlike $\cIP$, in $\cQPIP$ the prover is restricted to $\cBQP$ computations. This constraint on the prover is more in line with Problem~\ref{prob:verification} and it also has the direct implication that $\cQPIP \subseteq \cBQP$.

As we will see, all the protocols in Section~\ref{sect:prepsend} and Section~\ref{sect:recvmeas} are $\cQPIP$ protocols. And since these protocols allow for the delegation of arbitrary $\cBQP$ problems, it follows that $\cQPIP = \cBQP$.

\noindent We now proceed to the multi-prover setting and define the multi-prover generalization of $\cIP$:
\begin{definition}
A language $L \subseteq \{0, 1\}^*$ belongs to $\sf{MIP[k]}$ if there exists a polynomial $p$ and a probabilistic Turing machine $V$, whose running time on inputs of size $n$ is bounded by $p(n)$, such that for any $x \in \{0, 1\}^{n}$ the following is true:
\begin{itemize}
\item when $x \in L$, there exists a $k$-tuple of provers $(P_1, P_2, ... P_k)$ which are not allowed to communicate and which exchange at most $p(n)$ messages (of length at most $p(n)$) with $V$ and make $V$ accept with probability at least $c$,
\item when $x \not\in L$, any $k$-tuple of provers $(P_1, P_2, ... P_k)$ which are not allowed to communicate and which exchange at most $p(n)$ messages (of length at most $p(n)$) with $V$, make $V$ accept with probability at most $s$,
\end{itemize}
where $c - s \geq 1/p(n)$.
\end{definition}
Note that $\mathsf{MIP[1]} = \cIP$ and it was shown that for all $k > 2$, $\mathsf{MIP[k]} = \mathsf{MIP[2]}$ \cite{bgk}. The latter class is simply denoted $\mathsf{MIP}$. If the provers are allowed to share entanglement then we obtain the class:
\begin{definition}
A language $L \subseteq \{0, 1\}^*$ belongs to $\sf{MIP^*[k]}$ if there exists a polynomial $p$ and a probabilistic Turing machine $V$, whose running time on inputs of size $n$ is bounded by $p(n)$, such that for any $x \in \{0, 1\}^{n}$ the following is true:
\begin{itemize}
\item when $x \in L$, there exists a $k$-tuple of provers $(P_1, P_2, ... P_k)$ which can share arbitrarily many entangled qubits, are not allowed to communicate and which exchange at most $p(n)$ messages (of length at most $p(n)$) with $V$ and make $V$ accept with probability at least $c$,
\item when $x \not\in L$, any $k$-tuple of provers $(P_1, P_2, ... P_k)$ which can share arbitrarily many entangled qubits, are not allowed to communicate and which exchange at most $p(n)$ messages (of length at most $p(n)$) with $V$, make $V$ accept with probability at most $s$,
\end{itemize}
where $c - s \geq 1/p(n)$.
\end{definition}
As before it is the case that $\mathsf{MIP^*[k]} = \mathsf{MIP^*[2]}$ and this class is denoted as $\mathsf{MIP^*}$ \cite{cleve2004consequences}. It is not known whether $\mathsf{MIP} = \mathsf{MIP^*}$, however, it is known that both classes contain $\mathsf{BQP}$. Importantly, for $\mathsf{MIP^*}$ protocols, if the provers are restricted to $\mathsf{BQP}$ computations, the resulting complexity class is equal to $\mathsf{BQP}$ \cite{ruv}. Most of the protocols presented in Section~\ref{sect:entanglement} are of this type.

Note that while the protocols we review can be understood in terms of the listed complexity classes, we will often give a more fine-grained description of their functionality and resources than is provided by complexity theory. To give an example, for a $\cQPIP$ protocol, from the complexity theoretic perspective, we are interested in the verifier's ability to delegate arbitrary $\cBQP$ decision problems to the prover by interacting with it for a polynomial number of rounds.
In practice, however, we are interested in a number of other characteristics of the protocol such as:
\begin{itemize}
\item whether the verifier can delegate not just decision problems, but also sampling problems (i.e. problems in which the verifier wishes to obtain a sample from a particular probability distribution and is able to certify that, with high probability, the sample came from the correct distribution),
\item whether the prover can receive a particular quantum input for the computation or return a quantum output to the verifier,
\item having minimal quantum communication between the verifier and the prover,
\item whether the verifier can ``hide'' the input and output of the computation from the prover.
\end{itemize}

\bibliographystyle{splncs}
\bibliography{bibliography}

\end{document}